\documentclass[a4paper,10pt]{article}
\usepackage[utf8]{inputenc}
\usepackage[T1]{fontenc}
\usepackage[english]{babel}%
\RequirePackage{fix-cm}
\usepackage{geometry}
\usepackage{amsmath, amsfonts, amssymb, amsthm, mathrsfs}
\usepackage{epsfig,graphicx,color}
\usepackage{dsfont}
\usepackage{natbib}
\usepackage{hyperref}%
\usepackage{url}
\usepackage{mathtools}
\usepackage{environ}
\usepackage[position=top,font={sf,md,up}]{subfig}
\usepackage[right]{lineno}

\usepackage{bbm}
\usepackage{hyperref}
\hypersetup{
    colorlinks,
    linkcolor={gray!50!black},
    citecolor={blue!50!black},
    urlcolor={blue!80!black}
}
\usepackage{verbatim}
\usepackage{fancyhdr}
\usepackage{setspace}
\usepackage{orcidlink}
\usepackage[all,line,pdf]{xy}
\newcommand{\mengi}[1]{\ensuremath{\left\{ #1 \right\}}}
\newcommand{\svigi}[1]{\ensuremath{\left( #1 \right)}}
\newcommand{\uN}{{\ensuremath{\zeta(N)}}}

\newcommand{\N}{\mathds N}
\newcommand{\IN}{\ensuremath{\mathds N}}%

\newcommand{\one}[1]{\ensuremath{\mathds{1}_{\left\{ #1 \right\}}}}%
\newcommand{\prob}[1]{\ensuremath{\mathds{P}\left( #1 \right)}}%
\newcommand{\EE}[1]{\ensuremath{\mathds{E}\left[ #1 \right]}}%
\newcommand{\ao}{\ensuremath{{\alpha_1}}}%
\newcommand{\breyta}[1]{\textcolor{black}{#1}}
\newcommand{\be}{\begin{equation}}
\newcommand{\ee}{\end{equation}}%
\newcommand{\bd}{\begin{displaymath}}%
\newcommand{\ed}{\end{displaymath}}%
\newcommand{\norm}[1]{\ensuremath{\left\lVert #1 \right\rVert } }
\newcommand{\set}[1]{\ensuremath{\left\{ #1 \right\}}}
\newtheorem{thm}{Theorem}

\newtheorem{propn}{Proposition}[section]
\newtheorem{lemma}[propn]{Lemma}%

\newtheorem{defn}[propn]{Definition}
\newtheorem{remark}[propn]{Remark}

\newtheorem{corollary}[propn]{Corollary}

\usepackage{tocbasic}
\DeclareTOCStyleEntry[numwidth=2em,indent=0pt]{tocline}{part}
\usepackage{expl3}
\ExplSyntaxOn
  \clist_map_inline:nn
  {part,chapter,section,subsection,subsubsection,paragraph,subparagraph}
  {\DeclareTOCStyleEntry[
    linefill=\enskip,
    raggedpagenumber,
    pagenumberbox=\mbox
  ]{tocline}{#1}}
\ExplSyntaxOff

\makeatletter
\renewcommand{\maketitle}{\bgroup\setlength{\parindent}{0pt}
\begin{flushleft}
  \textbf{\textsf{\Large\@title}}
\end{flushleft}
\begin{center}
 {\textsc{ \small \@author} }
 \end{center}
\egroup} \makeatother

\title{Gene genealogies in diploid populations evolving according to
sweepstakes reproduction}
\author{Bjarki Eldon\orcidlink{0000-0001-9354-2391}
~\footnote{
 Corresponding author:
\href{mailto:beldon11@gmail.com}{beldon11@gmail.com}}
}

\begin{document}
\setcounter{page}{0}
\pagenumbering{arabic}
\setcounter{page}{1}
\maketitle

\renewcommand{\abstractname}{\vspace{-\baselineskip}}
\vspace{-\baselineskip}

\begin{abstract} Recruitment dynamics, or the distribution
of the number of offspring among individuals, is central for
understanding ecology and evolution.~\breyta{Sweepstakes reproduction
(when the  offspring number distribution has a heavy right-tail)
 may characterize  the recruitment dynamics of 
 highly fecund natural populations.}  Sweepstakes reproduction can induce jumps in type
frequencies,  and multiple mergers in gene genealogies of sampled gene
copies.~\breyta{Here, we consider gene genealogies in 
diploid panmictic populations evolving in a random environment. The
heavy-tailed offspring number distribution is  generated
by mechanisms not involving natural selection,
such as in chance matching of broadcast spawning with favourable
environmental conditions.  Our model of  sweepstakes
reproduction  extends the one  considered by
~\cite{schweinsberg03} by applying an upper bound to the number of
potential offspring of any given parent pair.  Depending on the stated
bound, the gene genealogies are in the domain of attraction of the
Kingman coalescent, or specific familes of continuous-time Beta- or 
Poisson-Dirichlet simultaneous multiple-merger coalescents. The gene
genealogies in a large population are viewed on  a  timescale  proportional to at least
$N/\log N$ generations;  $N$ is proportional to the population
size (when constant).  Incorporating deterministic  population size changes  leads to
 time-changed coalescents;   the time-change is independent of the
  skewness of the offspring-number distribution.   Using simulations,   we show
  that   gene genealogies in finite populations  are not well
  approximated  by the coalescent trees.  Simulation results also
  indicate that quenched (conditioned on the population ancestry)
  and annealed gene genealogies in finite populations  are not in good agreement whenever the
  skewness of the offspring number distribution is increased. }
\end{abstract}

\tableofcontents

\bigskip


\bigskip

{\bf Significance Statement}

\breyta{It has been suggested that highly fecund natural populations
characterised by Type III survivorship may evolve according to
sweepstakes reproduction, meaning high skewness in the offspring
number distribution.  Sweepstakes reproduction may affect the
evolution of the population in a number of ways. Here, we consider
gene genealogies in diploid panmictic populations with a heavy-tailed
offspring number distribution, and evolving in a random environment.
We derive simultaneous multiple-merger coalescents, time-changed when
the population size varies in time.  We use simulations to study
patterns of genetic variation as predicted by the resulting
coalescents, and to compare quenched (conditioning on the ancestral
relations of the gene copies in the population) and annealed gene
genealogies. Our results are relevant for inferring sweepstakes.}

\bigskip

\section*{Declarations}

\subsection*{Ethical Statement}

The author has no conflict of interest to declare that are relevant to
the content of this article

\subsection*{Funding}

Funded in part  by DFG Priority Programme SPP 1819 `Rapid Evolutionary
Adaptation' DFG grant Projektnummer 273887127 through SPP 1819 grant
STE 325/17 to Wolfgang Stephan;  Icelandic Centre of Research (Rann\'is)
through the Icelandic Research Fund (Ranns\'oknasj\'o{\dh}ur) Grant of
Excellence no.\ 185151-051 with Einar \'Arnason, Katr\'in
Halld\'orsd\'ottir, Alison Etheridge, and Wolfgang Stephan, and DFG
SPP1819 Start-up module grants with Jere Koskela and Maite Wilke
Berenguer, and with Iulia Dahmer.

\subsection*{Data availability}

The code  (C/C++) to generate  the numerical results is
freely available at \\
\url{https://github.com/eldonb/gene_genealogies_diploid_pops_sweepstakes}

\clearpage\pagebreak\newpage

\section{Introduction}\label{sec:intro}
\linenumbers

Inferring evolutionary histories of natural populations is one of the
main aims of population genetics. Inheritance, or the transfer of gene
copies from a parent to an offspring, is the characteristic of
organisms that makes inference possible.  Inheritance leaves a 
`trail' of ancestral relations.~\breyta{The 
structure of the trail shapes  the pattern of genetic
variation observed in a sample.}  By modeling the random (unknown)
ancestral relations of sampled gene copies, one hopes to be able to
distinguish between evolutionary histories by comparing model
predictions to population genetic data.  This sample-based 
  approach to inference forms a framework for deriving
efficient inference methods~\citep{Donnelly1995,Bertoin2006,W07,B09}.

Recruitment dynamics, or individual recruitment success (the offspring
number distribution) is central to ecology and evolution
\citep{Eldon2020,HP11}.~\breyta{In the absence of}  natural selection and complex demography
the offspring number distribution is a deciding factor in how the
sample trees look like, and therefore for predictions about data.
Models  such as the Wright-Fisher model
\citep{wright31:_evolut_mendel,fisher22}, in which large families (with
numbers of offspring proportional to the population size) occur only
with negligible probability in an arbitrarily large population, are
commonly used as offspring number distributions.   However,
such `small family' models may be a  poor choice for highly fecund populations
\citep{HP11,Arnasonsweepstakes2022}.


Highly fecund natural populations are diverse and widely found
\citep{Eldon2020}.  By `high fecundity' we mean that individuals have
the capacity to produce numbers (at least) proportional to the
population size of `potential' offspring \breyta{(the surviving
offspring are sampled from the pool of potential offspring).}  It has been suggested that the evolution of highly
fecund populations may be characterised by high variance in the
offspring number distribution
\citep[sweepstakes reproduction;][]{A88,PW90,Li1998,H82,H94,B94,A04,AH2015,Vendrami2021,Arnasonsweepstakes2022}.
We will use the term `sweepstakes reproduction' for when there
occasionally (randomly occurring, \breyta{and without  involving
natural selection})  is an increased chance of large families with 
numbers of surviving offspring proportional to the population size.

The evolution of populations evolving according to sweepstakes
reproduction may be different from the evolution of populations not
characterised by sweepstakes.~\breyta{
Forward-in-time processes (in the form of Fleming-Viot measure-valued
diffusions \citep{fleming79:_some_markov,ethier93:_flemin_viot})
describing the evolution of type frequencies in populations
characterised by sweepstakes admit \linelabel{jumps}  jumps \citep{BB09}.
Gene genealogies  describing the
random ancestral relations   of gene copies sampled from populations evolving
according to sweepstakes are characterised by multiple mergers, where
a  random number of ancestral lineages 
merges in one or more groups   whenever mergers occur
\citep{DK99,P99,S99,S00,S03,MS01,MS03,HM11b,HM12}.}
Strong positive selection inducing recurrent selective sweeps, 
 recurrent strong bottlenecks, and recurrent extinction
and recolonisation  in subdivided populations, 
 are additional examples of mechanisms
generating multiple-mergers \breyta{in gene genealogies}~\citep{DS05,BBMST09,EW06,TV09}.  Loosely speaking, sweepstakes introduce
jumps to the evolution of the population, where `jumps' refer to
multiple mergers of \breyta{ancestral lineages},  and discontinuous changes in
type frequencies.  We are only beginning to understand \textit{(i)} if
one can distinguish between the mechanisms producing jumps using
population genetic
data~\citep{BBMST09,Hoban2013,freund2023interpreting,Johri2022,Koskela2018,Koskela2019,Freund2021},  and \textit{(ii)} what sweepstakes reproduction
may mean for the ecology and evolution of natural populations
\citep{A04,Hedrick2005,HP11,JFB:JFB13143,Eldon2023mec,AH2015,Arnasonsweepstakes2022,Eldon2016,Eldon2020,Vendrami2021}.

\breyta{Our}  results are relevant for inferring sweepstakes
reproduction using   data sampled from diploid populations.~\breyta{A diploid population is one where each individual carries a pair of
chromosomes. For our purposes  the
population can be seen as  consisting  of pairs of (homologous) chromosomes.}  We will consider gene genealogies of a
single contiguous non-recombining segment of a chromosome in a diploid
panmictic population.  The population evolves \breyta{without}  selfing and
according to sweepstakes reproduction.  The evolution over a single
generation is seen as occurring in two stages.  In the first stage the
current individuals (randomly paired) produce potential offspring
according to a given law. In the second stage a given number of the
potential offspring (conditional on there being enough of them) is
then sampled uniformly and without replacement to survive to maturity
and replace the current individuals; if the potential offspring are
too few (fewer than the population size) the population remains
unchanged over the generation.

\breyta{A coalescent is a Markov process  describing the random
ancestral relations of sampled gene copies in a large population~\citep{K82,K82b,K82c,MS01,P99,S99,DK99,S00}.
Multiple-merger coalescents describe gene genealogies in (arbitrarily large) populations
evolving according to sweepstakes 
\citep{schweinsberg03,S99,EW06,SW08,MS01,HM11b,Der2012,Der2014,Gnedin2014,Matuszewski2018,freund2020cannings,Eldonhaploid2026}.}
A simultaneous multiple-merger coalescent (Xi-, $\Xi$-coalescent)  is a coalescent where
ancestral lineages may merge in two or more groups simultaneously
\citep{S00,MS01}.
Diploidy intuitively induces simultaneous multiple mergers in gene
genealogies   from a diploid
population\linelabel{evolves} \breyta{evolving} according to sweepstakes
\citep{MS03,BBE13,BLS15}.  One should distinguish between haploidy and
diploidy when it comes to comparing multiple-merger
coalescents to data;~\breyta{$\Xi$-coalescents 
  can predict patterns of genetic variation different from the ones
admitting only  asynchronous mergers} 
\citep{BBE2013a,Blath2016,SKS2016,Eldonhaploid2026}.
The~\breyta{coalescents derived here}  can be seen as mixtures of the Kingman coalescent
\citep{K82,K82b} and specific families of  $\Xi$-coalescents.
~\breyta{Such mixtures}
also arise, for example,  from  models of strong positive selection
\citep[Proposition~3.2]{DS05}.

\breyta{Incorporating diploidy in studies of  gene genealogies   involves tracing ancestral lineages through
homologous chromosomes of  diploid
individuals.}   Thus, a pair of ancestral lineages can be
found in the same diploid individual without the lineages having
merged.  Viewed on the timescale applied when deriving a limit, such
states occur over infinitesimally short periods of time, but prevent
convergence in the $J_{1}$-Skorokhod topology \citep{Skorokhod1956};
one nevertheless has convergence in finite-dimensional distributions
\citep{BLS15}.  A topology, which can be seen as an extension of the
Skorokhod topology, has been proposed for convergence of Markov chains
with such states \citep{Landim2015}.  However, we will follow
~\cite{BLS15} in proving convergence in the space of
\linelabel{cadlag}~\breyta{càdlàg}
paths
for a process where the instantaneous states (occurring over an
infinitesimal length of time in the limit of an arbitrarily large
population) are simply ignored.

The layout of the paper: \S~\ref{back} contains  a brief background to
coalescent processes,  and to models of sweepstakes reproduction. In 
\S~\ref{mathresults} we state our main  
results in Theorems~\ref{thm:diploid-alpha-random-all}, 
~\ref{thm:diploid-alpha-random-one-environment}, and 
~\ref{thm:time-varyingXi}; 
\S~\ref{sec:incr-sample-size} contains numerical  examples comparing
functionals of gene genealogies and coalescents; \S~\ref{sec:proofs}
holds  
proofs of Theorems~\ref{thm:diploid-alpha-random-all} and 
~\ref{thm:diploid-alpha-random-one-environment}.~\breyta{Appendix~\ref{sec:appr-eer_-omega} considers    an 
 example  of a  specific $\Xi$-coalescent};     Appendices~\ref{sec:estimatingERin}--\ref{sec:estimatingERiNA}
contain  brief descriptions of the simulation algorithms.

\section{Background}\label{back}

For ease of reference we state notation used throughout.
\begin{defn}[Standard notation]\label{def:notation}

\breyta{For any  $i\in \set{0,1,2,\ldots}$ write  $\N_{i} \equiv 
\set{i,i+1,\ldots}$, $\IN \equiv \IN_{1}$.}  Let $N\in \N$  be fixed with $2N$
being  the total number of diploid individuals (when the
population size is constant). 

Asymptotic relations are seen to hold as the population becomes
arbitrarily large,  unless otherwise noted.

We let $C,c,c^{\prime},K$ denote  positive constants.

Write $[n] := \set{1,2,\ldots, n}$ for any $n \in \N$.

Let $\mathcal E_{n}$ denote the set of partitions on $[n]$ for any
$n\in \N$ ($\mathcal E_{1} = \set{  \set{ \set{1} } }$,
$\mathcal E_{2} = \set{  \set{ \set{1}, \set{2} }, \set{[2]} }$, etc.)

We let ${(x)}_{m}$ denote the  falling factorial;  for any real $x$,  and
any $m\in \N_{0}$,  
\begin{equation}
\label{eq:1}
 {(x)}_m := x(x-1)\cdots (x-m+1),\quad  {(x)}_0 := 1
\end{equation}

For positive sequences ${(x_n)}_{n\in
\N}$ and ${(y_n)}_{n\in \N}$ with  $(y_{n})$   bounded away from zero we will write   
\begin{equation}
\label{eq:8}
x_n =   O( y_n)
\end{equation}
if   $\limsup_{n\to \infty} x_n/y_n < \infty$, and 
\begin{equation}
\label{simseq}
   x_n \sim  y_n 
\end{equation}
 if    $\lim_{n\to \infty}x_n/ y_n = 1$.  We will also write
\begin{equation}
\label{eq:simc}
x_{n}\overset{c}{\sim} y_{n}
\end{equation}
if  $\lim_{n\to \infty} x_{n}/y_{n} = c$ for some (unspecified)
constant $c > 0$;~\eqref{eq:simc} is of course equivalent to $x_{n}
\sim c y_{n}$.~\breyta{Given $x_{n}$ and $y_{n}$ we  will use~\eqref{eq:simc}   
when we want to emphasize the conditions  under which~\eqref{eq:simc} holds.}

Define
\begin{equation}
\label{eq:one}
\one{E} := 1
\end{equation}
when  a given condition/event  $E$ holds, and
take $\one{E} = 0$ otherwise.

Take  $|A|$ to be the number of elements in  a given finite   set $A$. 

The abbreviation i.i.d.\ will stand for independent and identically
distributed (random variables).
\end{defn}

The introduction of the \breyta{(annealed)}  coalescent \citep{H83b,T83,K2000} marks a
milestone in mathematical and empirical population genetics, as it is
a rigorous probabilistic description of the random ancestral relations
of sampled gene copies (chromosomes)~\citep{K82,K82b,K82c,K1978}.  A
coalescent $\set{\xi} \equiv \{\xi(t); t\ge 0 \}$ is a Markov chain
\breyta{(a Markov process with a countable state space and evolving in
continuous time)} taking values in the partitions of $\IN$, such that the restriction
$\set{\xi^{n}} \equiv \{\xi^{n}(t); t\ge 0 \} $ to the partitions of  $[n]$ for a fixed
$n\in \IN_{2}$  is still Markov and takes values in $\mathcal{E}_{n}$, the
set of partitions of $[n]$.  For a partition
$\xi^{n} \in \mathcal{E}_{n}$    write
$\xi^{n} = \set{\xi_{1}^{n}, \ldots, \xi_{b}^{n}}$, where  $\xi^{n}_{j} \subset [n]$,  
$b = | \xi^{n}|$ is the number of blocks in $\xi^{n}$,
$\xi_{i}^{n}\cap \xi_{j}^{n} = \emptyset$ for $i \neq j$, and
$\cup_{i=1}^{b} \xi_{i}^{n} = [n]$.  The only transitions are the
merging of blocks of the current partition each time (we exclude
elements such as recombination,  population structure, or natural
selection).      Each block in a partition represents an
ancestor to the elements (the leaves) in each of the \linelabel{block}\breyta{blocks}, in the
sense that distinct leaves (corresponding to sampled gene copies) $i$
and $j$ are in the same block at time $t \ge 0$ if and only if they
share a common ancestor at time $t$ in the past.  At time zero,
$\xi^{n}(0) = \{ \{1\}, \ldots, \{n\}\}$, and the time
$\inf\{t \ge 0 : \xi^{n}(t) = \{ [n]\} \}$, where the partition
$\{[n]\}$ contains only the block $[n]$, is when all the $n$ leaves
have found a common ancestor.

Here,  we focus on diploid populations.~\breyta{Our  aim is to}  
describe the random ancestral relations of $2n$  gene copies
from  $n$ \breyta{randomly sampled} diploid individuals.
Following~\cite{BLS15} and~\cite{MS03} we define the state space
\begin{equation}
\label{eq:Sn}
\mathcal{S}_{n} \equiv  \set{  \set{ \{\xi_{1}^{n}, \xi_{2}^{n}\}, \ldots, \set{\xi_{2x-1}^{n},\xi_{2x}^{n}}, \xi_{2x+1}^{n},\ldots, \xi_{b}^{n}} : b\in [2n], x \in \set{0,1, \ldots, \lfloor b/2 \rfloor }, \set{             \xi_{1}^{n}, \ldots,                   \xi_{b}^{n}} \in \mathcal{E}_{2n} }
\end{equation}
The elements of $\mathcal{S}_{n}$ when $x = 0$ ($x$ is the number of
diploid individuals carrying two ancestral blocks) are precisely the
elements of $\mathcal{E}_{n}$, so that
$\mathcal{E}_{n} \subset \mathcal{S}_{n}$.  The elements of
$\mathcal{S}_{n}$ corresponding to $x > 0$ have $2x$ blocks paired
together in diploid individuals.  A block in a partition
$\xi^{n}\in \mathcal S_{n}$ is `ancestral' in the sense that it is an
ancestor of the sampled gene copies (arbitrarily labelled) whose
labels are contained in the block.~\breyta{Let} 
$\set{ \xi^{n,N}} \equiv \{ \xi^{n,N}(\tau); \tau \in  \N_{0} \}$ be a Markov
sequence \linelabel{Markov} (a Markov process with a countable state space and evolving
in discrete time) with values in $\mathcal S_{n}$.  We will refer to
$\set{\xi^{n,N}}$ as the  \emph{annealed ancestral process};
leaves (sampled gene copies) labelled $i$ and $j$ are in the same
block at time $\tau$ if and only if they share a common ancestor at time
$\tau$  in the past.  We measure time~\breyta{going  from the present into the
past,}  we take
$ \xi^{n,N}(0) = \{ \{1,2\}, \ldots, \{2n-1, 2n\}\}$, and the only
transitions are the merging of blocks of the current partition (we
exclude further elements such as recombination or population
structure).~\breyta{We will also refer to the process keeping track
of the block sizes  of a  given ancestral process resp.\ coalescent
as an ancestral process resp.\  coalescent.}


A central quantity in deriving limits of \set{\xi^{n,N}} is the
coalescence probability \citep{S99}.
\begin{defn}[The coalescence probability]\label{def:cNhapl}
Define $c_{N}$ as the probability that two given   gene copies
in separate diploid individuals from a given generation derive
from the same parent gene copy
\end{defn}
We will show that $c_{N}\to 0$, and that $1/c_{N}$ is proportional to
(at least) $N/\log N$ for the models we will consider.  It holds that
$1/c_{N}$ is the correct scaling of time of the ancestral process for
proving convergence~\cite[Equation~1.4]{S03}.~\breyta{If
$c_{N}\to 0$,   the limiting coalescent evolves in
continuous time, with one unit of time corresponding to
$\lfloor 1/c_{N}\rfloor$ generations \citep{schweinsberg03,MS01}.}

\breyta{For each $N\in \IN$}  let $\nu_{1}, \ldots, \nu_{N}$ denote the random number of surviving
offspring from the $N$ current parent pairs at some arbitrary time.
We will assume that
$(\nu_{1},\ldots, \nu_{N})$ are independent  across generations \breyta{(and
identically distributed when the population size is constant)}.
Definition~\ref{def:cNhapl} then gives, with $2N$ being   the population size
(recall the notation from Definition~\ref{def:notation}),  
\begin{equation}
\label{eq:6}
   c_N = \frac 14 N\frac{\EE{\nu_1(\nu_1 - 1)} }{2N(2N-1) } =   \frac{\EE{\nu_1(\nu_1 - 1)} }{8(2N-1) } \sim \frac{\EE{\nu_1(\nu_1 - 1)} }{16N } \overset{c}{\sim} \frac 1N \EE{\nu_1(\nu_1 - 1)}
\end{equation}

\subsection{The population model}\label{sec:population-model}

In this section we define more
precisely the evolution of a diploid population \breyta{evolving without selfing}.

\begin{defn}[Evolution of a diploid population]\label{dschwpop}%
Consider a diploid panmictic population.~\breyta{In each generation
the current individuals arbitrarily form  parent pairs,  each pair
is made up of distinct individuals,  each individual shows up in at
most 1 pair (individuals are
monogamous),   and selfing is prohibited.}  The parent pairs thus formed
independently produce random numbers of diploid potential offspring
according to some given law.  Each offspring receives two chromosomes
(gene copy),
one chromosome from each of its two parents, with each inherited 
chromosome sampled independently and uniformly at random from
among the two parent chromosomes.  If the total number of potential
offspring is at least some given number \breyta{($2N$ when the
population size is constant)}, we sample \breyta{the given number of}
the potential offspring uniformly at random without replacement to
survive to maturity and replace the current individuals; otherwise we
assume the population is unchanged over the generation (all the
potential offspring perish before reaching maturity).
\end{defn}

\begin{remark}[Illustrating Definition~\ref{dschwpop}]\label{rm:illdschwpop}
The mechanism described in Definition~\ref{dschwpop} is illustrated
below, where $\set{a,b}$ denotes a diploid individual carrying gene
copies $a,b$,   and $\set{\set{a,b},\set{c,d}}$ denotes a pair of diploid
individuals (a parent pair).  Here we have arbitrarily labelled the
gene copies \breyta{in order to illustrate} the evolution over one
generation.~\breyta{In this example the population size  stays constant at $2N$ diploid individuals.}  
\begin{displaymath}
\begin{split}
\text{stage} &\quad   \text{individuals involved} \\
1 & \quad  \set{\{a,b\} , \{c,d\}}, \ldots,  \set{\set{w,x}, \set{y,z}} \quad \text{:  $N$ parent pairs} \\
2 & \quad \underset{X_{1}}{\underbrace{\set{a,d},\set{b,d}, \ldots, \set{a,c}}}, \ldots, \underset{X_{N}}{ \underbrace{ \set{w,y}, \ldots, \set{x,z}} } \quad \text{: $X_{1} + \cdots +  X_{N}$ potential offspring}  \\
3 & \quad    \set{b,d},\ldots, \set{x,y} \quad \text{: $2N$ surviving offspring (whenever $X_{1}+ \cdots +X_{N} \ge 2N$)}
\end{split}
\end{displaymath}
In stage~1 above, the current $2N$ diploid individuals randomly form
$N$ pairs; in stage~2 the $N$ pairs formed in stage~1 independently
produce random numbers $X_{1}, \ldots, X_{N}$ of potential offspring,
where each diploid offspring receives one gene copy (sampled uniformly at random)
from each of its two parents; in the third stage $2N$ of the
$X_{1} + \cdots + X_{N}$ potential offspring (conditional on there
being at least $2N$ of them) are sampled uniformly and without
replacement to survive to maturity and replace the parents.
\end{remark}

\begin{remark}[Relation to~\cite{BLS15}]\label{remark:relationbls15}
The reproduction mechanism described in Definition~\ref{dschwpop} is a
special case of the one studied~\breyta{by}~\cite{BLS15}.~\breyta{They} 
consider an array $\left(V_{i,j}^{(m)} \right)$ of exchangeable
offspring numbers where $V_{i,j}^{(m)}$ is the random number of
\emph{surviving} offspring produced in generation $m$ by individuals
$i$ and $j$ (arbitrarily labelled, and with $V_{i,i}^{(m)} = 0$). The
idea is that any individual may produce offspring with more than one
individual in the same generation (promiscuous mating).  In principle
it should be possible to use the framework of~\cite{BLS15}. However,
as we are interested in comparing a given ancestral process to the
limiting coalescent, we follow a simpler framework where \breyta{in
any given generation each
diploid individual produces offspring in collaboration with exactly 
one other individual (monogamous mating).   We elect to  leave an  extension
of~\citep{BLS15} to random environments to future work. A
$\Xi$-coalescent with an atom at zero is briefly discussed
in~\citep[Section~1.1]{BLS15}. 
 Here,  we specifically
consider  scenarios where the limiting  $\Xi$-coalescent does have an
atom  at zero. In our case  the time-scaling used
for viewing gene genealogies in a large population is affected.    The concept
of time-scaling is relevant for explaining genetic variation (see
 \S~\ref{sec:conclusion}). }
\end{remark}

\breyta{Due to diploidy (see Definition~\ref{dschwpop} and
Remark~\ref{rm:illdschwpop}) the ancestral process may enter 
 states} where two gene copies (ancestral to the sampled ones)
reside in the same diploid individual.
\breyta{This is}  why in Definition~\ref{def:cNhapl} we require the two gene
copies to be in distinct diploid individuals.  It should also be clear
from the illustration in Remark~\ref{rm:illdschwpop} why one could
expect to see simultaneous mergers in the genealogy of a sample from a
diploid population where large families regularly
occur.~\linelabel{simmergers}~\breyta{For family $i$ we sample $2X_{i}$
times (with replacement)  from the 4 parental chromosomes
involved. Then, when a parent pair  produces  
of order $\mathcal O(N)$ number  of offspring,  we can have up to 4 groups  of
offspring chromosomes of order $\mathcal O(N)$ belonging to the same
tetrad (quadruplet) of parent chromosomes,  which translates to simultaneous
mergers of ancestral lineages. }

\breyta{The following population
model was initially introduced for a haploid population.   Suppose $\alpha, C > 0$
are fixed and $X$ is the   random number  of
potential offspring produced by an arbitrary  individual 
(gene copy), and that~\cite[Equation~11]{schweinsberg03}
\begin{equation}
\label{eq:4}
\lim_{x \to \infty}Cx^{\alpha}\prob{X \ge x}  = 1
\end{equation}
From the pool of potential offspring produced at the same time  $N$ of them (conditional on there
being at least $N$ potential offspring) are sampled uniformly at
random and without replacement to survive to maturity and replace the
parents (recall Definition~\ref{dschwpop}).~\breyta{When $1\le \alpha
< 2$~\eqref{eq:4} gives rise to the (asynchronous)
Beta$(2-\alpha,\alpha)$-coalescent~\cite[Theorem~4c]{schweinsberg03}, a
 family of 
multiple-merger coalescents studied to some extent}
~\citep{BBC05,DAHMER2014,BBS07,BBS08,Kersting2014,Birkner2024}.  }
We will consider extensions of
~\eqref{eq:4} applied to diploid populations evolving as described in
Definition~\ref{dschwpop}.

\subsection{Simultaneous  multiple-merger coalescents}\label{sec:mult-merg-coals}

\breyta{We will be 
concerned with  simultaneous   multiple-merger coalescents
($\Xi$-coalescents,Xi-coalescents)   where ancestral lineages
may merge in  several groups simultaneously~\citep{S00,MS01,S03,MS03}.}
\breyta{Sweepstakes reproduction, recurrent strong bottlenecks, or 
strong positive selection,  are examples of
mechanisms generating  $\Xi$-coalescents}
\citep{SW08,Eldonhaploid2026,BLS15,BBE13,MS03,BBMST09,DS05,SD2005,TV09}.

Define 
\begin{equation}\label{eq:simplex}
\Delta_{+} := \set{ (x_1, x_2, \ldots ) :   x_1 \ge x_2 \ge \ldots  \ge 0, \quad 
x_{1} + x_{2} + \cdots   \le 1 } \setminus \{(0,\ldots ) \}.
\end{equation}
\breyta{Suppose the merging measure $\Xi$ of a $\Xi$-coalescent on
$\Delta_{+}\cup\set{(0,\ldots)}$  is of the form ($c,c^{\prime} \ge
0$ fixed)
\begin{equation}
\label{eq:34}
 \Xi(dx) =  c\delta_{0} +  c^{\prime}\int_{\Delta_{+}}\delta_{\varphi(x) } \Xi_{+}(dx) 
\end{equation}
with   $\Xi_+$  a finite measure on $\Delta_{+}$, and  $\varphi:
\Delta_{+} \to  \Delta_{+}$ a given map.}
Then, with $n\ge 2$ active blocks in the current partition,  $k_1,
\ldots,  k_r \ge 2$,  $r \in \N $, and $s = n - \sum_{j=1}^r k_j \ge 0$, the
rate at which $k_{1} + \cdots + k_{r} \in \{2,\ldots, n\}$  blocks merge in $r$ groups with group $j$ of 
size $k_{j}$  is given by ($c,c^{\prime} \ge 0$ fixed, \breyta{and $\varphi$
the identity map})
\begin{equation}
\label{eq:xirate}
\begin{split}
\lambda_{n; k_1, \ldots, k_r;s} & =  c\one{r=1, k_1=2} \\
& +  c^{\prime} \int_{\Delta_{+}} \sum_{\ell = 0}^s \sum_{i_1 \neq \ldots \neq i_{r+\ell} } \binom{s}{\ell} x_{i_1}^{k_1} \cdots  x_{i_r}^{k_r}x_{i_{r+1}}\cdots x_{i_{r+\ell}}{\left( 1 -    \sum_{j=1}^\infty x_j  \right)}^{s - \ell} \frac{1}{\sum_{j=1}^\infty x_j^2 } \Xi_+ (dx) 
\end{split}
\end{equation}
\citep{S00,MS01}. The constant $c$ in~\eqref{eq:xirate} is the mass  the
merging measure $\Xi$ assigns to $\set{(0,\ldots)}$,
 and $c^{\prime}$ the mass assigned to  $\Delta_{+}$.

The application of~\eqref{eq:4} to a diploid panmictic population of
constant size evolving according to Definition~\ref{dschwpop}   leads to a $\Xi$-coalescent without an atom at zero
(corresponding to $c= \Xi\set{(0,\ldots)} =  0$ in~\eqref{eq:xirate}) and driving measure
$\Xi_{+}$ of the form
\begin{equation}
\label{eq:35}
\Xi_{+}(dx) =  \int_{(0,1]}\delta_{\left( \tfrac x4, \tfrac x4, \tfrac x4, \tfrac x4, 0,   \ldots  \right) }\text{Beta}(2-\alpha,\alpha)(dx)
\end{equation}
\breyta{where Beta$(2-\alpha,\alpha)$ stands for the
beta density   with parameters $2-\alpha$ and $\alpha$ for $1 <\alpha
< 2$}~\cite[Proposition~2.5(2); Equation~29]{BLS15}.~\breyta{The form
of $\Xi_{+}$ in~\eqref{eq:35}  says that the measure $\Xi_{+}$
is of the form
$\Xi_{+} = \text{Beta}(2-\alpha,\alpha) \circ \varphi^{-1}$, where
\linelabel{varphi} $\varphi : (0,1]\to \Delta_{+}$ takes values
$\varphi(x) \equiv (x/4,x/4,x/4,x/4,0,\ldots)$. }

The $x$ in
~\eqref{eq:35} can be seen as the
fraction of surviving offspring produced by an arbitrary 
 parent pair (a pair of pairs of gene copies). 
 \breyta{The}  ancestral lines belonging to the same  family are then
split (uniformly at random and with replacement) among the four
parental chromosomes,~\breyta{and the ones assigned to the same
chromosome are merged.}  We will consider extensions of~\eqref{eq:4} 
that, when combining the results, allow us to take $0< \alpha < 2$.
Moreover, we will consider a truncated (incomplete) version of the
Beta$(2-\alpha,\alpha)$-coalescent.

When $0 < \alpha < 1$ one
obtains, using~\eqref{eq:4} for a haploid panmictic population of
constant size, a discrete-time $\Xi$-coalescent associated with the
Poisson-Dirichlet distribution with parameter $(\alpha,0)$
\citep[Theorem~4(d)]{schweinsberg03}.
The Poisson-Dirichlet distribution \citep{Kingman1975} has found wide
applicability, including in population genetics 
\citep{feng2010poisson,Bertoin2006,S03,Eldonhaploid2026}.  We will be concerned with
the two-parameter Poisson-Dirichlet$(\alpha, \theta)$ distribution 
for $0 < \alpha < 1$ and
$\theta > -\alpha$ restricted to $\theta = 0$.

\begin{defn}[A Poisson-Dirichlet$(\alpha,0)$-coalescent; \citep{schweinsberg03}]\label{def:poisson-dirichlet}
Let $0 < \alpha < 1$ be fixed, recall the simplex $\Delta_{+}$ in
~\eqref{eq:simplex}, write $x = (x_1, x_2, \ldots)$ for
$x \in \Delta_{+}$, and  $(x,x) \equiv  \sum_{j=1}^{\infty} x_{j}^{2}$.   Let $F_\alpha$ be a
probability measure on $\Delta_{+}$ associated with the
Poisson-Dirichlet$(\alpha,0)$-distribution, and $\Xi_\alpha$ a measure
on $\Delta_{+}$ given by
\begin{displaymath}
\Xi_\alpha(dx) \equiv   \svigi{x,x}  F_\alpha( dx)
\end{displaymath}
A Poisson-Dirichlet$(\alpha,0)$-coalescent is a discrete-time
$\Xi$-coalescent with $\Xi$-measure $\Xi_\alpha$ and no atom at zero.
The transition probability of merging blocks in $r$ groups of size
$k_{1}, \ldots, k_{r} \ge 2$ with current number of blocks
$b \ge k_{1} + \cdots + k_{r}$,  and
$s \equiv b - k_{1} - \cdots - k_{r}$,  is given by (recall
~\eqref{eq:1}),
\begin{equation}
\label{eq:24}
p_{b; k_{1}, \ldots, k_{r}; s} =  \frac{\alpha^{r+s-1}(r+s-1)! }{(b-1)!}\prod_{i=1}^{r} {(k_{i}-1 - \alpha)}_{k_{i}-1}
\end{equation}\end{defn}
(cf.\ \cite[Equation~13]{schweinsberg03},
~\cite[Equation~12]{Eldonhaploid2026}).  The $\Xi_{+}$-measure (recall
~\eqref{eq:xirate}) of the 
Poisson-Dirichlet$(\alpha,0)$-coalescent  is of the form 
\begin{displaymath}
\Xi_{+}(dx)  =   \int_{\Delta_{+}}\delta_{(x_{1},  x_{2},\ldots)}\Xi_{\alpha}(dx) 
\end{displaymath}

\subsection{Two $\Omega$-$\Xi$-coalescents}\label{sec:two-omega-xi-coals}

\breyta{We define two $\Xi$-coalescents, the
$\Omega$-$\delta_{0}$-Beta$(\gamma,2-\alpha,\alpha)$-coalescent, and  the
$\Omega$-$\delta_{0}$-Poisson-Dirichlet$(\alpha,0)$-coalescent (see
Appendix~\ref{sec:appr-eer_-omega} for the
$\Omega$-Beta$(2-\beta,\beta)$-Poisson-Dirichlet$(\alpha,0)$-coalescent)
that we will need for stating  our results. We use the prefix  $\Omega$ to denote a $\Xi$-coalescent where the ancestral lines
involved in each group of mergers are assigned (uniformly at random and with replacement)
into four subgroups, and the lines assigned to the same subgroup are
merged; an $\Omega$-$\Xi$-coalescent is a  $\Xi$-coalescents with the merging  measure $\Xi$
 as given in~\eqref{eq:34},   with
\begin{displaymath}
 \varphi((x_{1}, x_{2},\ldots)) = \left( \frac {x_{1}}4, \frac  {x_{1}}4,  \frac  {x_{1}}4,  \frac  {x_{1}}4,  \frac  {x_{2}}4,  \frac  {x_{2}}4,  \frac  {x_{2}}4,  \frac  {x_{2}}4,      \ldots \right).
\end{displaymath}}
\begin{defn}[The $\Omega$-$\delta_{0}$-Beta$(\gamma,2-\alpha,\alpha)$-coalescent]\label{def:xi-kingman-beta}
Let $\mathbf{k} \equiv (k_{1}, \ldots, k_{r})$ for $r \in [4]$, where
$2 \le k_{i} \le n$, $k \equiv k_{1} + \cdots + k_{r} \le n$, and
$s \equiv n - k$ for $n \ge 2$.  The
$\Omega$-$\delta_{0}$-Beta$(\gamma, 2-\alpha,\alpha)$-coalescent, for
$0 < \gamma \le 1$ and $0 < \alpha < 2$, is a continuous-time
$\Xi$-coalescent (recall~\eqref{eq:xirate}) on the partitions of
$[n]$, where the only transitions are the merging of blocks of the
current partition.  The transition  rate, at which $r\in [4]$  (simultaneous
when $r \ge 2$) mergers occur with the merger sizes given by
$k_{1},\ldots, k_{r}$ and $2 \le \sum_{i}k_{i} \le n$, is ($c,c^{\prime} > 0$ fixed)
\begin{equation}
\label{eq:7}
\begin{split}
\lambda_{n;\mathbf{k}; s}  &  =  c\one{r=1,k_{1}=2} +   c^{\prime}\sum_{\ell = 0}^{s\wedge (4-r) }\binom{s}{\ell}\frac{ {(4)}_{r+\ell}}{4^{k+\ell}}\int_{0}^{1}\one{0 < x \le \gamma} x^{k+\ell-2}{(1-x)}^{n-k-\ell}\Lambda_{+}(dx) 
\end{split}
\end{equation}
  The finite measure  $\Lambda_{+}$~\eqref{eq:7}   on
\breyta{$(0,1]$} is given by,  \breyta{for $0<\gamma \le 1$ and  $0 <
\alpha < 2$ both fixed,}
\begin{equation}
 \label{eq:3}
\text{d} \Lambda_+(x) =  \frac{1}{B(\gamma,  2-\alpha,\alpha)}  \one{0 < x \le \gamma } x^{1-\alpha} {(1-x)}^{\alpha - 1} \text{d}x
\end{equation}
In~\eqref{eq:3}   $B(\gamma, a,b) \equiv   \int_{0}^{1}\one{0 < t \le \gamma } t^{a-1}{(1-t)}^{b-1}dt$ for
$a,b > 0$, and $0 < \gamma \le 1$, is the   beta function. Then,
$\lambda_{n;\mathbf{k};s}$ in~\eqref{eq:7} takes the form  \breyta{($x\wedge y$
is the smaller of $x$ and $y$)}
\begin{displaymath}
\lambda_{n; \mathbf{k};s} =    c\one{r=1,k_{1}=2} +   \frac{c^{\prime}}{B(\gamma, 2-\alpha,\alpha) } \sum_{\ell = 0}^{s\wedge (4-r) }\binom{s}{\ell}\frac{ {(4)}_{r+\ell}}{4^{k+\ell}}B(\gamma, k+\ell - \alpha, n-k-\ell+\alpha)
\end{displaymath}
\end{defn}
To see that the sum in~\eqref{eq:7} follows from~\eqref{eq:xirate}
when $\Xi_{+}$ is as given, recall that $\set{\xi^{n}}$ is restricted
to $\mathcal E_{n}$ \breyta{(recall Definition~\ref{def:notation})}.  It then
suffices to see that given $k$ ancestral lines merging in $r$ groups,
we can have up to $s \wedge (4-r)$ additional lines each assigned to
one `free' parental chromosome (to which none of the $k$ lines are
assigned) and  there are ${(4)}_{r+\ell}$ equivalent ways of
ordering the $r+\ell$ parent chromosomes receiving an ancestral line~\cite[Equation~27]{BBE13}.
The merging measure \breyta{of the
$\Omega$-$\delta_{0}$-Beta$(\gamma,2-\alpha,\alpha)$-coalescent
defined in Definition~\ref{def:xi-kingman-beta}}  is   of the form (recall~\eqref{eq:35})
\begin{equation}
\label{eq:XideltaBetameasure}
\Xi(dx) = \delta_{0} +   \int_{(0,1]} \delta_{ \left( \frac x4,  \frac x4,  \frac x4,  \frac x4, 0, 0, \ldots  \right) } \text{Beta}(\gamma, 2-\alpha, \alpha) dx
\end{equation}
where Beta$(\gamma,2-\alpha,\alpha)$   is the law on
$(0,1]$ with density~\eqref{eq:3}.

\breyta{The $\delta_{0}$-Poisson-Dirichlet$(\alpha,0)$-coalescent
is a continuous-time    Poisson-Dirichlet$(\alpha,0)$-coalescent 
~\citep{schweinsberg03}   with an
atom at zero; it  describes gene
genealogies in haploid populations evolving in a random environment
similar to the ones we will consider~\citep{Eldonhaploid2026}.}
\begin{defn}[The $\delta_{0}$-Poisson-Dirichlet$(\alpha,0)$-coalescent]\label{def:ndeltaPD}
The $\delta_{0}$-Poisson-Dirichlet$(\alpha,0)$-coalescent is a
continuous-time $\Xi$ coalescent (recall~\eqref{eq:xirate}) taking
values in $\mathcal E_{n}$ with $\Xi$-measure
$\Xi = \delta_{0} + \Xi_{+}$, where $\Xi_{+} = \Xi_{\alpha}$ and
$\Xi_{\alpha}$ is as in Definition~\ref{def:poisson-dirichlet}.  The
only transitions are the merging of blocks of the current partition.
Given $n$ blocks in a partition, $k_{1},\ldots, k_{r}\ge 2$ with
$2 \le k_{1} + \cdots + k_{r}\le n$ denoting the merger sizes of $r$
(simultaneous when $r \ge 2$) mergers, $s = n - k_{1} - \cdots - k_{r}$, a
$(k_{1},\ldots, k_{r})$-merger occurs at rate (where $c,c^{\prime} >0$
fixed)
\begin{displaymath}
\lambda_{n;k_{1},\ldots, k_{r}; s} = c\one{r=1, k_{1}=2}  + c^{\prime} p_{n;k_{1},\ldots, k_{r}; s}
\end{displaymath}
where $ p_{n;k_{1},\ldots, k_{r}; s} $ is as in~\eqref{eq:24}. 
\end{defn}



\begin{defn}[A $\Omega$-$\delta_{0}$-Poisson-Dirichlet$(\alpha,0)$-coalescent]\label{def:diploid-kingman-poisson-dirichlet-coalescent}
A $\Omega$-$\delta_{0}$-Poisson-Dirichlet$(\alpha,0)$-coalescent is
a  $\delta_{0}$-Poisson-Dirichlet$(\alpha,0)$-coalescent (recall
Definition~\ref{def:ndeltaPD}) where the blocks in each group of
merging blocks 
are split among four subgroups independently and uniformly at random
and with replacement, and the blocks assigned to the same subgroup are merged.
\end{defn}
The merging measure of  the  $\Omega$-$\delta_{0}$-Poisson-Dirichlet$(\alpha,0)$-coalescent in Definition~\ref{def:diploid-kingman-poisson-dirichlet-coalescent}  is given by
\begin{displaymath}
\Xi(dx)  =  \delta_{0} +   \int_{\Delta_{+}} \delta_{\svigi{ \frac{x_{1}}{4}, \frac{x_{1}}{4}, \frac{x_{1}}{4}, \frac{x_{1}}{4},  \frac{x_{2}}{4},  \frac{x_{2}}{4},  \frac{x_{2}}{4},  \frac{x_{2}}{4},   \ldots     }  } \Xi_{\alpha}(dx)
\end{displaymath}
where $\Xi_{\alpha}$ is as in Definition~\ref{def:poisson-dirichlet}
and $\Delta_{+}$ as in~\eqref{eq:simplex}.

\begin{remark}[Extending~\citep{Koskela2019}]\label{remark:relKWB}
The
$\Omega$-$\delta_{0}$-Poisson-Dirichlet$(\alpha,0)$-coalescent extends
the coalescents considered by~\cite{Koskela2019}, who consider
coalescents based on population models where at most one large family
(with an offsping number proportional to the  population size) occurs
with non-negligible probability at any given time in a large
population~\cite[Equations~9 and 10]{Koskela2019}, thus leading to
$\Xi$-coalescents with measure of the form as in~\eqref{eq:35} or
~\eqref{eq:XideltaBetameasure}.~\breyta{In our model behind the
$\Omega$-$\delta_{0}$-Poisson-Dirichlet$(\alpha,0)$-coalescent, when
using~\eqref{eq:PXiJ} with $a = \alpha \in (0,1)$,  we expect to  see of
order $\mathcal O(N^{1-\alpha})$ families  with at least $N$ potential
offspring.  We expect, during the history of a sample,  
   to use $\alpha$   on the order of  $\mathcal O(1)$ number of
  times  for sampling numbers of offspring. }
\end{remark}

\subsection{Equivalent convergence conditions}\label{sec:equiv-conv-cond}

Equivalent conditions for convergence of $\set{\xi^{n,N}}$ to
$\Xi$-coalescents are summarised by~\cite{BLS15}.  Convergence, in the
sense of convergence of finite-dimensional distributions  of
$\set{\xi^{n,N}}$ to a $\Xi$-coalescent,  depend on  the existence of the
limits \breyta{(recall $c_{N}$ from Definition~\ref{def:cNhapl} and~\eqref{eq:6})}
\begin{equation}
\label{xmoments}
\phi_{n; k_1, \ldots, k_r} :=  \lim_{N\to \infty}  \frac{1 }{c_N} \frac{ \EE{ {(\nu_1)}_{k_1} \cdots {(\nu_r)}_{k_r}  }}{2^{k_{1} + \cdots + k_{r}}  N^{k_{1} + \cdots + k_{r} - r} }
\end{equation}
for all  $r\in \N$,  $k_{1}, \ldots, k_{r} \ge 2$ are the merger sizes, 
$2 \le k_{1}+ \cdots + k_{r} \le n$,  and $n$ is the current number of
blocks \citep{MS01,BLS15}.  See~\cite[Theorem~A.5]{BLS15} for a summary
of equivalent conditions for convergence to a $\Xi$-coalescent.
Existence of the limits in~\eqref{xmoments}  is equivalent to
\begin{equation}
\label{eq:vagueconv}
\frac{1}{c_{N}} \Phi_{N}(dx) \to \frac{1 }{\sum_{i}x_{i}^{2} }\Xi^{\prime} (dx) 
\end{equation}
vaguely on $\Delta_{+}$ as $\N \to \infty$;  $\Xi^{\prime}$ is a
probability measure on $\Delta_{+}$,  and 
\begin{displaymath}
\Phi_{N} \equiv   \mathscr L \svigi{  \frac{  \nu_{(1)}  }{2N}, \ldots,   \frac{  \nu_{(N)}  }{2N}, 0,  \ldots    }
\end{displaymath}
is the law of the ranked \linelabel{iw} \breyta{individual-wise}  offspring frequencies
$\nu_{(1)}/(2N) \ge \nu_{(2)}/(2N) \ge \ldots \ge \nu_{(N)}/(2N) $.   Write 
\begin{displaymath}
\Delta_{r} \equiv   \set{ (x_{1},\ldots, x_{r}) :   x_{1} \ge  \ldots
\ge x_{r} \ge 0 \text{,  and } x_{1} + \cdots + x_{r} \le 1 }.
\end{displaymath}
\breyta{Let}  $F_{r}$ be a symmetric measure on   $\Delta_{r}$ for $r\in \N$.   
Equivalent to
equivalent conditions~\eqref{xmoments} and~\eqref{eq:vagueconv} are
the two conditions~\cite[Condition~II in Appendix~A]{BLS15} (see also
~\cite[Equations~21 and 22]{MS01})  
\begin{subequations}
\begin{align}
& \lim_{N\to \infty} \frac{1}{2^{2r} N^{r} c_{N} }\EE{ \svigi{\nu_{1} }_{2} \cdots  \svigi{\nu_{r} }_{r}  }  =   F_{r}( \Delta_{r}) \label{eq:14} \\ 
& \lim_{N\to \infty} \frac{N^{r} }{c_{N}} \prob{\nu_{1} > 2N x_{1}, \ldots, \nu_{r} > 2N x_{r}  } =   \int_{x_{1}}^{1} \cdots \int_{x_{r}}^{1} \frac{ F_{r}\svigi{ d y_{1} \cdots d y_{r}  }  }{ y_{1}^{2} \cdots y_{r}^{2}   } \label{eq:15}
\end{align}
\end{subequations}
with the limits in~\eqref{eq:14} holding for all $r \in \N$, and
~\eqref{eq:15} holding for points of continuity for $F_{r}$.

\section{Mathematical results}\label{mathresults}%

\breyta{In \S~\ref{sec:law-number-potential} we present  the model
behind the main results.  
Theorems~\ref{thm:diploid-alpha-random-all} and 
~\ref{thm:diploid-alpha-random-one-environment} record the results for
constant population
size (\S~\ref{sec:const-popul-size}),   and
Theorem~\ref{thm:time-varyingXi} for (deterministically)  varying population size  (\S~\ref{sec:vary-popul-size}).}

\subsection{The law for the number of potential offspring}\label{sec:law-number-potential}

\breyta{In this section we present the law for the random  number  of
potential offspring underlying the main results.}  For ease of
reference we first state key notation.
\begin{defn} (Notation)\label{def:not}
\breyta{For each  $N\in \N$,  we let } $X_1, \ldots, X_N$,  resp.\
$\nu_{1},\ldots, \nu_{N}$, 
denote the random number of potential resp.\  surviving  offspring
produced in an arbitrary generation by the current  $N$ parent pairs
(\breyta{arbitrarily formed};  recall Definition~\ref{dschwpop}). Recall that $2N$ is the population
size, the number of diploid individuals (pairs of gene copies) at any
time when the population size is constant,  and then  $\nu_{1} +
\cdots + \nu_{N} = 2N$ \breyta{for each $N\in \IN$}.  The  $X_{1}, \ldots, X_{N}$ are  independent. Write, \breyta{for
each $N\ge 2$,}
\begin{subequations}
\begin{align}
S_N &  := X_1 + \cdots + X_N,  \label{SN} \\
  \widetilde{S}_N &  := X_2 + \cdots + X_N; \label{SN2}
\end{align}
\end{subequations}
 $S_{N}$ is the total number of diploid potential offspring
produced in a given generation.  Write
\begin{equation}
\label{eq:5}
\begin{split}
m_{N} &  := \EE{X_{1}} \to m_{\infty} 
\end{split}
\end{equation}
for the expected value of $X_{1}$, with $m_{\infty}$ denoting the
\breyta{(assumed fixed)} mean
as $N \to \infty$ (\linelabel{m} we will identify conditions for $m_{\infty}$ to
\breyta{be finite};  see Lemmas~\ref{lma:finitemN} and~\ref{lm:finitemarandone}).

Write, with $\uN$ a positive  function of $N$ (see
~\eqref{eq:PXiJ}),
\begin{equation}
\label{eq:26}
\frac{ \uN}{N} \gneqq    0
\end{equation}
\linelabel{bbound}\breyta{if  either  $\uN / N$ converges to a strictly
positive  constant, or  $\uN / N$ diverges,  as $N\to \infty$.}

We write, with   $\kappa \ge 2$  a given  positive  constant
\breyta{(recalling~$\one{\cdot}$ from~\eqref{eq:one})},
\begin{equation}
\label{eq:CNmap}
 C_\kappa^{N} :=  \one{\kappa > 2}N  +  \one{\kappa = 2}N/ \log N
\end{equation}
\end{defn}

Now we state the distribution for the number of potential offspring
~\breyta{underlying our results}.
Suppose $X$ represents the random number of potential offspring
produced by an arbitrary  pair of diploid individuals (parent pair).
 We write, with $a>0$ fixed,  and $\zeta(N)$ a
positive deterministic function of $N$, with $a$ and $\zeta(N)$ as
given \breyta{in each case,  }
\begin{equation}
\label{eq:28}
X \vartriangleright \mathds L(a, \uN)
\end{equation} 
when the probability  mass function of the law of $X$ is bounded by,
for  $k \in \set{2,3,\ldots, \uN}$~\breyta{and each $N\in \N$},  
\begin{equation}
\label{eq:PXiJ}
g_{a}(k) \left( \frac{1}{k^a} - \frac{1}{ {(1+k)}^a} \right)  \leq
\prob{X = k} \leq f_{a}(k) \left( \frac{1}{k^a} -
\frac{1}{ {(1+k)}^a} \right) 
\end{equation}
\breyta{We assign the remaining mass (outside $\set{2,3,\ldots, \zeta(N)}$) to
$\{X \in \{0,1\}\}$; how the mass is distributed between  0 and 1 will
become irrelevant in the limit (see Lemma~\ref{lm:PSNltwoN}). }
We take $\set{f_{a}}, \set{g_{a}}$ to be families of bounded positive
functions on $\N$ such that~\linelabel{cutoff} $\prob{X \le \uN} = 1$ for any $a > 0$
\breyta{and $N\in \N$}.
We will identify conditions on $g_{a}$ and
$f_{a}$ such that the ancestral process converges to a non-trivial
limit.  The law~\eqref{eq:PXiJ} is an extension of the one in
~\eqref{eq:4}; with $X$ distributed as in~\eqref{eq:4} we have
$\prob{X = k} = \prob{X \ge k} - \prob{X \ge k+1} \overset{c}{\sim}
k^{-\alpha} - {(k+1)}^{-\alpha}$ as $k\to
\infty$.~\linelabel{bounds}\breyta{The bounds  on the mass function in
~\eqref{eq:PXiJ} are the same for each $N\in \N$. We can equivalently 
define  $X^{(N)} \equiv 
X\one{X\in [\zeta(N)]} + \one{X > \zeta(N)}$,  and
take $X_{1},\ldots,X_{N}$ to be independent copies of $X^{(N)}$ (see
Remark~\ref{rm:sequence}) for each $N \ge 2$.}
See~\citep{Eldon2026,Eldonhaploid2026} for 
~\eqref{eq:PXiJ}  in the context of  haploid populations. 


\breyta{
\begin{remark}[A sequence of independent random variables]\label{rm:sequence}
The law~\eqref{eq:PXiJ} is seen to hold for the random numbers
$X_{1},\ldots,X_{N}$ of  potential offspring of the current parent
pairs ($N$ pairs when the population size is $2N$) for each $N\in \IN$.   Equivalently,  one
may consider a sequence  ${(X_{i}^{\prime})}_{i\in \N}$ of independent  random variables,
where $X_{i}^{\prime} = X\one{X \in [\zeta(N)]} +  Y_{i}\one{X >
\zeta(N)}$,   ${(Y_{i})}_{i\in \N}$ is a sequence of independent
random variables suitably defined,  and $X\vartriangleright
\mathds L(a,\zeta(N))$ (recall~\eqref{eq:28}). We will work with such
sequences when we  prove almost sure   convergence of
$S_{N}/(Nm_{N})$  (recall~\eqref{SN} and~\eqref{eq:5} in \S~\ref{def:not}).
\end{remark}}

We assume $g_{a}(k) \le f_{a}(k)$ for all $k$.   We define, for any $a,g_{a}^{(\infty)},f_{a}^{(\infty)} >   0$ fixed,
\begin{equation}
\label{eq:isfg}
\begin{split}
\underline{f_{a}}(k)  \equiv \inf_{i\ge k } f_{a}(i), & \quad \overline{f_{a}}(k)  \equiv \sup_{i\ge k } f_{a}(i), \\
\underline{g_{a}}(k)  \equiv \inf_{i\ge k } g_{a}(i), & \quad \overline{g_{a}}(k)  \equiv \sup_{i\ge k } g_{a}(i), \\
\lim_{k\to \infty}g_{a}(k) \equiv  g_{a}^{(\infty)}, &  \quad \lim_{k\to \infty}f_{a}(k) \equiv  f_{a}^{(\infty)}.
\end{split}
\end{equation}
We then have, for $k \in \{2,3,\ldots, \uN\}$, with the law of  $X$ as
in~\eqref{eq:PXiJ}  \breyta{for each $N\ge 2$,}
\begin{equation}
\label{eq:12}
\left( \frac{1}{k^a} - \frac{1}{ {(1 + \zeta(N))}^a} \right)\underline{g_{a}}(k) \le  \prob{X  \ge  k} \le \left( \frac{1}{k^a} - \frac{1}{ {(1 + \zeta(N))}^a} \right)\overline{f_{a}}(k).
\end{equation}

The functions $f_{a}$ and $g_{a}$ in~\eqref{eq:PXiJ} can be chosen such that
$\prob{X = k-1} \ge \prob{X = k}$ for $k\in \set{3,\ldots, \uN}$, and that $\EE{X} > 2$  
for all $N$ large enough.  The monotonicity requirement is reasonable
(but not necessary).  The requirement that $\EE{X_{i}} > 2$ for all
$i \in [N]$ \breyta{and $N\in \N$}  results in
$\prob{X_{1} + \cdots + X_{N} <  2N}$ decreasing exponentially in $N$ (Lemma~\ref{lm:PSNltwoN}),
where $X_1, \ldots, X_N$ are as in Definition~\ref{def:not} \breyta{(recall Remark~\ref{rm:sequence}).}

\begin{remark}[Assumption on $g_{a}$ and  $f_{a}$ in~\eqref{eq:PXiJ}]\label{rm:assumptionmN2+}
With $X \vartriangleright \mathds L(a,\uN)$ (recall~\eqref{eq:28}) we
will assume that the functions $f_{a}$ and $g_{a}$ in~\eqref{eq:PXiJ} are such
that $\EE{X} > 2$ \breyta{for all $N\in \N$} (recall that parent pairs
produce offspring), \breyta{and that the constant  $m_{\infty} > 2$ (recall~\eqref{eq:5}).}
\end{remark}

\subsection{Constant population size}\label{sec:const-popul-size}

\breyta{In this section we define the random environments, and state
the convergence results for constant population size.  Results for
when the population size varies in time are stated in
\S~\ref{sec:vary-popul-size}. }

The coalescent in Definitions~\ref{def:xi-kingman-beta} and 
~\ref{def:diploid-kingman-poisson-dirichlet-coalescent} arises in
specific cases when the population evolves  as follows. 
\begin{defn}[A random environment]\label{dfn:alpharandomall}
Suppose a diploid panmictic population evolves  as in
Definition~\ref{dschwpop}.  Fix $0 < \alpha < 2$,  and $2 \le \kappa$.  Recall
$X_{1},\ldots, X_{N}$ from Definition~\ref{def:not}.  Write $E$ for the event
 when $X_{i} \vartriangleright \mathds L(\alpha,\zeta(N)) $ for all 
$i \in [N] $ (recall~\eqref{eq:28}),   and  $E^{\sf c}$ for the event  when $\kappa$ replaces $\alpha$ in
$E$ ($X_{i} \vartriangleright \mathds L(\kappa, \zeta(N))$ for all
$i\in [N]$).  Let ${(\varepsilon_{N})}_{N\in \N}$ be a positive sequence with
$0 < \varepsilon_{N} < 1$ for all $N$.~\breyta{It} may hold that
$\varepsilon_{N}\to 0$ as $N\to \infty$.~\breyta{Take, for each $N\in \IN$,}
\begin{displaymath}
\prob{ E  } = \varepsilon_{N}, \quad \prob{E^{\sf c}} = 1 - \varepsilon_{N}
\end{displaymath}
\end{defn}
Definition~\ref{dfn:alpharandomall} says that the
$X_{1},\ldots, X_{N}$ are i.i.d.\ copies of $X$ where
$X \vartriangleright \mathds L\left( \one{E}\alpha + \one{E^{\sf
c}}\kappa, \uN \right)$.  The idea represented by
Definition~\ref{dfn:alpharandomall} is that most of the time (with
probability $1 - \varepsilon_{N}$)  small (relative to the
population size) numbers of offspring \breyta{per family}  are produced with high
probability through $\kappa$ (event $E^{\sf c}$); occasionally,
however,  
(with probability $\varepsilon_{N}$) environmental conditions are
favourable for producing an increased number of offspring through
$\alpha$ (event $E$), and then the current parent pairs will all~\breyta{(independently)} 
produce  a random number 
of potential  offspring~\breyta{using~\eqref{eq:PXiJ} with  $a=\alpha$.}

\breyta{The coalescents resulting from
Definition~\ref{dfn:alpharandomall} are described in
Theorem~\ref{thm:diploid-alpha-random-all}; \S~\ref{sec:proofthmall}
contains a proof of Theorem~\ref{thm:diploid-alpha-random-all}.}

\begin{thm}[Coalescents under Definition~\ref{dfn:alpharandomall}]\label{thm:diploid-alpha-random-all}
Suppose a diploid population evolves according to
Definitions~\ref{dschwpop} and~\ref{dfn:alpharandomall}, and that 
Assumption~\ref{rm:assumptionmN2+} holds.   Then
$\{ \xi^{n,N}(\lfloor t/c_{N} \rfloor); t \ge 0\} $ converges, as
$N\to \infty$, in the sense of convergence of finite-dimensional
distributions,   to \breyta{a continuous-time coalescent}
$\{\xi^{n}\} \equiv \{ \xi^{n}(t); t \ge 0\}$ as specified in each
case.~\breyta{When $1 \le \alpha < 2$ we take  \linelabel{c1}  $\varepsilon_{N} =  cN^{\alpha - 2}\left(\one{\kappa >
2} + \one{\kappa = 2}\log N\right)$  with $c >
0$ fixed.  }
\begin{enumerate}
\item~\label{item:1} The limiting coalescent $\set{\xi^{n} }$ is the Kingman coalescent when one of
\begin{enumerate}
\item $0<\alpha<1$,  $\uN^{3-\alpha} / N^{2} \to 0$ as $N\to \infty$,  and $\limsup_{N\to \infty} \varepsilon_{N}/c_{N} < \infty $; 
\item $1 \le \alpha < 2$, \breyta{$\varepsilon_{N}$ is as stated
above,}  and    $\uN/N \to 0$ as $N\to \infty$; 
\end{enumerate}
is in force. 
\item~\label{item:2} Suppose $1 \le \alpha < 2$ and $\uN/N \gneqq 0$
(recall~\eqref{eq:26} in   Definition~\ref{def:not}).  Then $\{\xi^{n}\}$ is the
$\Omega$-$\delta_{0}$-Beta$(\gamma, 2-\alpha,\alpha)$-coalescent as
 in Definition~\ref{def:xi-kingman-beta}.~\breyta{Given  $n$ blocks the
 transition rate for a 
(simultaneous multiple) merger with group sizes $k_{1},\ldots,k_{r}$
for  $r \in [4]$ is}   (recall $m_{\infty}$ from~\eqref{eq:5})
\begin{equation}
\label{eq:lambdankBK}
\begin{split}
\lambda_{n;\boldsymbol{k};s} &  =  \one{r=1,k_{1}=2} \frac{C_{\kappa}}{C_{\alpha,\gamma}} \\
 &  +  \frac{  \alpha c f_{\alpha}^{(\infty)} }{ C_{\alpha,\gamma}  m_{\infty}^{\alpha}}  \sum_{\ell=0}^{s\wedge (4-r)} \binom{s}{\ell} \frac{(4)_{r+\ell} }{4^{k+\ell} }  \int_{0}^{1}\one{0 < x \le \gamma}x^{k+\ell - \alpha - 1} {(1-x)}^{n-k-\ell +\alpha - 1} dx
 \end{split}
\end{equation}
where we have, recalling $B(\gamma,2-\alpha,\alpha)$ from
Definition~\ref{def:xi-kingman-beta},  and
\breyta{$f_{a}^{(\infty)}$, $g_{\kappa}^{(\infty)}$,  $\overline{f_{\kappa}}(2)$}   from~\eqref{eq:isfg}
\begin{subequations}
\begin{align}
\label{eq:gamma}
\gamma &  =    \one{ \frac{\uN }{N} \to K   }  \frac{K   }{K + m_{\infty} } +    \one{ \frac{\uN }{N} \to \infty   }   \\
\label{eq:9}
C_{\kappa} & =   \one{\kappa = 2} \frac{2\overline{f_{\kappa}} (2) }{ 4m_{\infty}^{2}  } +   \one{\kappa > 2} \frac{2 \overline {f_{\kappa}} (2) }{4m_{\infty}^{2}}\frac{c_{\kappa} }{2^{\kappa}(\kappa - 2)(\kappa - 1) }  \\
\label{eq:10}
C_{\alpha,\gamma} & = C_{\kappa} +   \frac{\alpha c f_{\alpha}^{(\infty)} }{4m_{\infty}^{\alpha}}B(\gamma,2-\alpha,\alpha)\\
\label{eq:11}
& \breyta{g_{\kappa}^{(\infty)}}   \left( p_{1}  +      \frac{\pi^{2}}{6} - 1 \right) <
m_{\infty} < 2\breyta{f_{\kappa}^{(\infty)}}\left( p_{1} + \frac{\pi^{2} }{6}  - 1 \right) \quad \text{ when  $\kappa = 2$}  \\
\label{eq:16}
 &  \breyta{g_{\kappa}^{(\infty)}} \svigi{ p_{1} +  \frac{3^{1-\kappa}}{\kappa - 1 } - \frac{2^{-\kappa}}{\kappa} } 
 < m_{\infty} <  \breyta{f_{\kappa}^{(\infty)}} \svigi{ p_{1} +   \frac{\kappa}{\kappa - 1} } \quad  \text{ when  $\kappa > 2$} 
\end{align}
\end{subequations}
where in~\eqref{eq:11}  and~\eqref{eq:16} \linelabel{p1} $p_{1} = \prob{X = 1}$
with  $X \vartriangleright \mathds L(\kappa,\zeta(N))$  (recall
~\eqref{eq:28}).  Moreover,  it holds that  $\kappa + 2  < c_{\kappa} <
\kappa^{2}$~\eqref{eq:9}  when $\kappa > 2$,   and  $\lambda_{2;2;0} =
1$.~\linelabel{bassumption}\breyta{Recall that 
$\zeta(N)/N \gneqq 0$ means (by assumption) that  either  $\zeta(N)/N$
converges to a strictly  positive constant, or that $\zeta(N)/N$ diverges, as
reflected in~\eqref{eq:gamma}.}
\item~\label{item:8} Suppose \breyta{$1/2 < \alpha < 1$},   and
$\uN / N^{1/\alpha} \to \infty$.  Then $\{\xi^{n}\}$ is the
$\Omega$-$\delta_{0}$-Poisson-Dirichlet$(\alpha,0)$-coalescent as in
Definition~\ref{def:diploid-kingman-poisson-dirichlet-coalescent}. A 
transition  proceeds  by  sampling  group sizes using the $\delta_{0}$-Poisson-Dirichlet$(\alpha,0)$-coalescent, the blocks (ancestral
lineages) in each group are split among  4 subgroups  independently
and uniformly at random, and  the  blocks  assigned to the same
subgroup are merged.    When
\linelabel{total}\breyta{there  are  $n$ blocks,  the total rate
of observing a (simultaneous multiple)  merger,  with   group sizes
$k_{1}, \ldots, k_{r} \ge 2$  such that $\sum_{i}k_{i} \le n$,     is  }
\begin{equation}
\label{eq:omegadPDrates}
\begin{split}
\lambda_{n; k_{1},\ldots, k_{r}; s} &  =   \one{r=1, k_{1} = 2} \binom{n}{2} \frac{ \breyta{4} C_{\kappa}}{ \breyta{4}C_{\kappa} + c(1-\alpha) } \\
& +   \binom{n}{k_{1}\ldots k_{r} \, s } \frac{1}{\prod_{j=2}^{n}\svigi{ \sum_{i}\one{k_{i} = j}  }! } \frac{c }{\breyta{4} C_{\kappa} + c(1-\alpha)} p_{n;k_{1}, \ldots, k_{r}; s}.
\end{split}
\end{equation}
\breyta{In~\eqref{eq:omegadPDrates}, $C_{\kappa}$ is as
in~\eqref{eq:9}, $m_{\infty}$ as in~\eqref{eq:11} or~\eqref{eq:16}},  $2 \le k_{1}, \ldots, k_{r} \le n$,   $\sum_{i}k_{i} \le n$,
and $s = n - \sum_{i}k_{i}$.~\linelabel{c2}\breyta{The positive  constant $c$
in~\eqref{eq:omegadPDrates} is such that
$C_{\kappa}^{N}\varepsilon_{N} = c$ for all $N\in \N$.}    In~\eqref{eq:omegadPDrates} 
 $ p_{n;k_{1}, \ldots, k_{r}; s} $ is as in~\eqref{eq:24}.
\end{enumerate}
For all the cases above we have, as $N\to \infty$ (recall~\eqref{eq:simc}),
\begin{equation}
\label{eq:scalecN}
C_{\kappa}^{N}   c_{N} \overset c \sim    1
\end{equation}
with $c_{N}$ as  in Definition~\ref{def:cNhapl} (recall~\eqref{eq:6}); 
   $C_{\kappa}^{N} =  \one{\kappa > 2}N + \one{\kappa = 2}N/\log N$ as in~\eqref{eq:CNmap}.
\end{thm}
\breyta{\begin{remark}[The range of $\alpha$ in Case~\ref{item:8} of Theorem~\ref{thm:diploid-alpha-random-all}]\label{rm:range-alpha}
The proof of   Case~\ref{item:8} of
Theorem~\ref{thm:diploid-alpha-random-all} shows that we require
$\zeta(N)/N^{1/\alpha}\to \infty$.  Suppose  $\zeta(N) = N^{\delta +
1/\alpha}$ for some fixed $0 < \delta \ll 1$.  Assuming 
$N\varepsilon_{N} \overset c \sim 1$  forces us to take 
$1/2 < \alpha < 1$ to  guarantee that $m_{\infty} < \infty$
by Lemma~\ref{lma:finitemN}.  In Case~\ref{item:8} of
Theorem~\ref{thm:diploid-alpha-random-all},  the bound $\zeta(N)$ is 
 such that  $\zeta(N)/N^{1/\alpha} \to \infty$, and seen to be at most of
the form $N^{\delta + 1/\alpha}$.  
\end{remark}}
\begin{remark}
Taking $\uN = N^{x}$ for some fixed  $x > 0$ we see that the
condition $\uN^{3-\alpha}/N^{2}\to 0$ in Case~\ref{item:1} of
Theorem~\ref{thm:diploid-alpha-random-all} corresponds to
$x < 2/(3-\alpha) < 1$ when  $0<\alpha < 1$, which is a stronger
ask of   $\uN$ than the requirement  $\uN/N\to 0$ when 
$1 \le \alpha < 2$.
\end{remark}
\begin{remark}[The parameters of the $\Omega$-$\delta_{0}$-Beta$(\gamma,2-\alpha,\alpha)$-coalescent]
The transition rates of the
$\Omega$-$\delta_{0}$-Beta$(\gamma,2-\alpha,\alpha)$-coalescent  are
functions of  the parameters $\alpha$, $\gamma$, $c$, and $\kappa$, and
also of the mean  $m_{\infty}$. 
\end{remark}\breyta{\begin{remark}[Poisson-Dirichlet-coalescents and
stick-breaking constructions]
The Poisson-Dirichlet$(\alpha,0)$-coalescent can be sampled using a
stick-breaking construction~\citep{feng2010poisson}.   Similar
 constructions feature  in  $\Xi$-coalescents  generated  by
recurrent  selective sweeps~\cite[Section~3]{DS05}.    
\end{remark}}

As detailed in~\cite[\S~1]{BLS15}, and formalised in
~\cite[Corollary~1.2]{BLS15},  by tracking only completely dispersed
states   we   have weak convergence  on $D([0,\infty), \mathcal E_{2n})$ (the set of $\mathcal E_{2n}$ valued
càdlàg paths with Skorokhod's $J_{1}$ topology~\cite[Chapter~3.4]{ethier05:_markov}).      The proof of
Corollary~\ref{cor:thmarandallweakconv} is identical to the one of 
~\cite[Corollary~1.2]{BLS15}.  For $\xi^{n}\in\mathcal{S}_{n}$ (recall
~\eqref{eq:Sn}) define the map
$\text{\tt cd}: \mathcal{S}_{n}\to \mathcal{E}_{2n}$ (with {\tt cd}
standing for `complete dispersal') by
\begin{equation}
\label{eq:cd}
{\tt cd}(\xi^{n}) \equiv  \set{ \xi_{1}^{n}, \ldots, \xi_{b}^{n}  }, 
\end{equation}
where $b$ is the number of blocks in $\xi^{n}$ (recall~\eqref{eq:Sn}).
Then $\xi^{n} = {\tt cd}(\xi^{n})$ for all
$\xi^{n} \in \mathcal{E}_{2n}$.
\begin{corollary}[Weak convergence of \set{\xi^{n,N}} on $D([0,\infty), \mathcal E_{2n})$]\label{cor:thmarandallweakconv}
Let
$\widetilde \xi^{n,N}(m) := {\sf cd}\svigi{\xi^{n,N}(m) }\in \mathcal
E_{2n} $ \breyta{for all $m$}.   Under the conditions of
Theorem~\ref{thm:diploid-alpha-random-all}
$\set{\widetilde \xi^{n,N}\svigi{ \lfloor t/c_{N} \rfloor}; t \ge 0 }
\to \set{\xi^{n}} \equiv \set{\xi^{n}(t); t \ge 0} $ weakly on
$D\svigi{[0,\infty), \mathcal E_{2n}}$ as $N\to \infty$,  with
$\set{\xi^{n}} $ as given in
Theorem~~\ref{thm:diploid-alpha-random-all}.
\end{corollary}

We will also consider an environment where the 
$X_{1}, \ldots, X_{N}$ \breyta{for all $N$}  stay independent but  may not always be
identically distributed.
\begin{defn}[A random  environment]\label{df:alpha-random-one-environment}
Suppose a diploid population evolves as in Definition~\ref{dschwpop}.
Fix $0 < \alpha < 2 $, and $\kappa \ge 2$.   Recall $X_{1},\ldots, X_{N}$ from
Definition~\ref{def:not}.  Write $E_{1}$ for the event, when there
exists exactly one $i\in [N]$ where
$X_{i}\vartriangleright \mathds{L}(\alpha,\uN)$, and
$X_{j} \vartriangleright \mathds L(\kappa,\uN)$ for all
$j\in [N] \setminus \set{i}$, \breyta{and for each $N\in \IN$.}   When $E_{1}$ occurs the index $i$ is
picked uniformly at random.  Write $E_{1}^{\sf c}$ when $\kappa$
replaces $\alpha$ in $E_{1}$; when $E_{1}^{\sf c}$ occurs the
$X_{1},\ldots, X_{N}$ are i.i.d.\ copies of $X$ where
$X \vartriangleright \mathds L (\kappa,\uN)$, \breyta{for each $N\in
\IN$.}    Let
${(\overline \varepsilon_{N})}_{N\in \N}$ be a positive sequence with
$0<\overline \varepsilon_{N} < 1$ for all $N$.   It may hold that
$\overline \varepsilon_{N}\to 0$ as $N\to \infty$.  Suppose,
\breyta{for all $N$,}
\begin{displaymath}
\prob{ E_{1}  } = \overline\varepsilon_{N}, \quad \prob{E_{1}^{\sf c}} = 1 - \overline\varepsilon_{N}
\end{displaymath}
\end{defn}
Definition~\ref{df:alpha-random-one-environment} says,  that  in each generation there exists an $i\in [N]$
where
$X_{i} \vartriangleright \mathds L\left(\one{E_{1}}\alpha +
\one{E_{1}^{\sf c}}\kappa, \uN\right)$, and
$X_{j} \vartriangleright \mathds L\left(\kappa, \uN\right)$ for all
$j\in [N]\setminus \set{i}$ \breyta{and $N\in \N$}.  The $X_{1},\ldots, X_{N}$ are
exchangeable \breyta{for each $N\in \N$;} when $E_{1}$ occurs the index $i$ is picked
uniformly at random from $[N]$. The coalescents resulting from
Definition~\ref{df:alpha-random-one-environment} are described in
Theorem~\ref{thm:diploid-alpha-random-one-environment};
\S~\ref{proof:thmrandone} contains a proof of
Theorem~\ref{thm:diploid-alpha-random-one-environment}.

\begin{thm}[Coalescents under Definition~\ref{df:alpha-random-one-environment}]\label{thm:diploid-alpha-random-one-environment}
Suppose a diploid population evolves according to
Definitions~\ref{dschwpop} and~\ref{df:alpha-random-one-environment}
and that Assumption~\ref{rm:assumptionmN2+} holds.  Then
$\set{ \xi^{n,N}( \lfloor t/c_{N} \rfloor ); t \ge 0}$ converges, 
 in the sense of convergence of finite-dimensional
distributions, to $\{ \xi^{n}\} \equiv \{\xi^{n}(t); t \ge 0 \} $ as
specified in each case.
\begin{enumerate}
\item~\label{item:7} Suppose   $\uN / N\to 0$.  Then $\{
\xi^{n} \}$ is the Kingman-coalescent. 
\item~\label{item:9}  Suppose $0 < \alpha \le 1$ and $\uN /N \gneqq 0$ (recall
~\eqref{eq:26} in Definition~\ref{def:not}).  Then $\{ \xi^{n}\}$ is
the $\Omega$-$\delta_{0}$-Beta$(\gamma, 2 - \alpha,\alpha)$-coalescent
defined in Definition~\ref{def:xi-kingman-beta} with transition rates
as in~\eqref{eq:lambdankBK}.
\end{enumerate}
In both cases~\eqref{eq:scalecN} is in force 
($C_{\kappa}^{N} c_{N} \overset c \sim 1$)  with
$C_{\kappa}^{N}$ as in~\eqref{eq:CNmap}.
\end{thm}

As in Corollary~\ref{cor:thmarandallweakconv},  restricting to
completely dispersed states we have weak convergence on
$D([0,\infty), \mathcal E_{2n})$.   The proof of
Corollary~\ref{cor:thmarandoneweakconv}  is identical to the one of
~\cite[Corollary~1.2]{BLS15}.
\begin{corollary}[Weak convergence of \set{\xi^{n,N}} on $D([0,\infty), \mathcal E_{2n})$]\label{cor:thmarandoneweakconv}
As in Corollary~\ref{cor:thmarandallweakconv}, let
$\widetilde \xi^{n,N}(m) \equiv  {\sf cd}\svigi{\xi^{n,N}(m) }$
\breyta{for all $m$}.   Under the
conditions of Theorem~\ref{thm:diploid-alpha-random-one-environment}
$\set{\widetilde \xi^{n,N}\svigi{ \lfloor t/c_{N} \rfloor}; t \ge 0 }
\to \set{\xi^{n}} \equiv \set{\xi^{n}(t); t \ge 0} $ weakly on
$D\svigi{[0,\infty), \mathcal E_{2n}}$ as $N\to \infty$ with
$\set{\xi^{n}} $ as given in
Theorem~\ref{thm:diploid-alpha-random-one-environment}
\end{corollary}

\subsection{Varying population size}\label{sec:vary-popul-size}

Given the time-scaling~\eqref{eq:scalecN}, meaning that gene
genealogies span on average numbers of generations proportional to the
population size, it is plausible  that the population size 
varies enough  \breyta{during the history of a sample}  to affect
gene genealogies.~\cite{freund2020cannings} gives
conditions for deriving \breyta{time-changed}  $\Lambda$-coalescents
describing gene genealogies in
 haploid panmictic  populations with varying population size.
 We adapt the arguments for
~\cite[Theorem~3]{freund2020cannings} to diploid populations evolving
according to Definition~\ref{dschwpop} and
Definition~\ref{dfn:alpharandomall} (or
Definition~\ref{df:alpha-random-one-environment}).

Suppose $2N_{r-1}$ diploid potential offspring produced by $N_{r}$
arbitrarily formed parent pairs survive to maturity in generation $r$
\breyta{where, for all $r \in \lfloor t/c_{N} \rfloor$,   $t > 0$ fixed,  
  $N \equiv N_{0}$,  and with  $c_{N}$ being  the coalescence probability for the
fixed population size case,}
\begin{equation}
\label{eq:20}
\begin{split}
& 0 < h_{1}(t)N \le  N_{r} \le  h_{2}(t)N  < \infty  \\
& N_{\lfloor t/c_{N} \rfloor}/N \to v(t) \quad \text{ as  $N\to \infty$} 
\end{split}
\end{equation}
for some bounded positive functions  $h_{1},h_{2},v: [0,\infty) \to
(0,\infty)$~\cite[Equation~4]{freund2020cannings}.   Since $c_{N}$ is
regularly varying (recall~\eqref{eq:scalecN}) it holds that   $M_{1}(t) \le c_{N_{r}} /
c_{N} \le M_{2}(t)$ for all $0 \le r \le \lfloor t/c_{N} \rfloor$ for
some $M_{1}(t), M_{2}(t) \in (0,\infty)$
~\cite[Equation~13]{freund2020cannings}. Moreover, writing $X_{1}(r),
\ldots, X_{N_{r}}(r)$ for the number of potential offspring produced
in generation $r$ (from whom $2N_{r-1}$ will be sampled to survive to maturity),   it follows as in the
proof of~\cite[Theorem~3]{freund2020cannings} that  \breyta{(if 
 $(x_{1},x_{2},\ldots) \in o_{\Sigma}(c_{N})$ then $x_{1} + \cdots + x_{\lfloor t/c_{N}
 \rfloor } \to 0$)} 
\begin{displaymath}
 \frac{\svigi{N_{r}}_{\ell} }{\svigi{2N_{r-1}}_{a_{1} + \cdots + a_{\ell}} }\EE{ \svigi{X_{1}(r)}_{a_{1}} \cdots  \svigi{X_{\ell}(r)}_{a_{\ell}}   } =    \frac{\svigi{N_{r}}_{\ell} }{\svigi{2N_{r}}_{a_{1} + \cdots + a_{\ell}} }\EE{ \svigi{X_{1}(r)}_{a_{1}} \cdots  \svigi{X_{\ell}(r)}_{a_{\ell}}   } +  o_{\Sigma}(c_{N})
\end{displaymath}
for $a_{1} \ge a_{2} \ge \ldots \ge 1$~\cite[Equation~5]{freund2020cannings}.   Coupled with  $\lim_{N\to
\infty}c_{N} = 0$, and  Corollaries~\ref{cor:thmarandallweakconv}  and
~\ref{cor:thmarandoneweakconv}, 
it holds that 
$\set{\widetilde  \xi^{n,N}\svigi{\lfloor G_{N}^{-1}(t) \rfloor }; t \ge 0 }$
converges weakly to $\set{\xi^{n}}$ where $G_{N}^{-1}$ is as in
~\cite[Equation~1]{freund2020cannings}.   Then, by~\eqref{eq:20}   and
~\eqref{eq:scalecN} it follows that   $\set{\widetilde \xi^{n,N}\svigi{ \lfloor
t/c_{N} \rfloor }; t \ge 0}$ converges to  $\set{\xi^{n}(G(t)); t \ge 0
}$, where \linelabel{G}\breyta{$G(t) =  \int_{0}^{t}\svigi{v(s)}^{-1}{\rm d}s$ with
$v$ as in~\eqref{eq:20}}
~\cite[Lemma~4]{freund2020cannings}, and  $\set{\xi^{n}\svigi{G(t)}; t \ge 0
}$ as given in  Corollaries~\ref{cor:thmarandallweakconv} resp.\
~\ref{cor:thmarandoneweakconv}.  Moreover,  we have an \linelabel{analogue}\breyta{analogue} of 
~\cite[Lemma~1]{freund2020cannings}. 
\begin{lemma}[$\prob{S_{N,r} \le 2N_{r-1}}$ vanishes]\label{lm:Freundlemma1}
Fix $t > 0$.  Suppose $\svigi{N_{r-1} - N_{r}}/N \to 0$ as
$N\to \infty$ for all $0 \le r \le t/c_{N} $. Then there exists a
constant $0 < c  < 1$    such that
$\prob{\sum_{i=1}^{N_{r}} X_{i}(r) <  2N_{r-1} } \le c^{N}$ with  $N =
N_{0}$.    
\end{lemma}
\begin{proof}[Proof of Lemma~\ref{lm:Freundlemma1}]
Write $S_{N,r} \equiv  \sum_{i=1}^{N_{r}}X_{i}(r)$.    For $0 \le s
\le 1$ it holds that 
\begin{displaymath}
\EE{s^{S_{N,r}}} = \sum_{0\le k} s^{k}\prob{S_{N,r} = k} \ge  s^{2N_{r-1}} \sum_{k=0}^{2N_{r-1}}\prob{S_{N,r} = k } = s^{2N_{r-1}}\prob{S_{N,r} \le 2N_{r-1}}
\end{displaymath}
Write $\rho(s) = \EE{s^{X_{i}(r)}}$. Then  $\svigi{s^{2}}^{N_{r-1}}\prob{S_{N,r}
\le 2N_{r-1} } \le  \rho(s)^{N_{r}}$.  Define $d_{N,r} \equiv N_{r-1}
- N_{r}$  and $d_{N,r}^{\prime} \equiv d_{N,r}/N_{r} $   such that    $N_{r-1} = N_{r}\svigi{1
+ d_{N,r}^{\prime}}$.  It follows that  $\prob{S_{N,r} \le 2N_{r-1} }
\le   \svigi{ s^{-2(1 + d_{N,r}^{\prime} ) }\rho(s)  }^{N_{r}} $.
Since $\rho(1) = 1$,  and $\rho^{\prime}(1) =  \EE{X_{i}(r)} > 2$,  there
is an $0 < s_{0} < 1$ and an $\epsilon > 0$ such that
$s_{0}^{2(1+\epsilon)} > \rho(s_{0})$.  
\end{proof}

Thus, the arguments for
~\cite[Theorem~3]{freund2020cannings} can be extended to $\Xi$-coalescents.
\begin{thm}[Time-changed $\Xi$-coalescents]\label{thm:time-varyingXi}
Suppose a diploid population evolves as in Definition~\ref{dschwpop}
and either   Definition~\ref{dfn:alpharandomall} or
Definition~\ref{df:alpha-random-one-environment}.  For any $v:
[0,\infty) \to (0,\infty)$ there exist  deterministically varying
  sizes $\svigi{N_{r}}_{r\in \N_{0}}$ such that~\eqref{eq:20} holds
 for the given $v$,  and  that $\set{\widetilde \xi^{n,N}\svigi{\lfloor
 t/c_{N} \rfloor } }$ converge weakly to $\set{ \xi^{n}\svigi{G(t)}; t
 \ge 0}$ where $G(t) = \int_{0}^{t}\svigi{v(s)}^{-1}{\rm d}s$.~\breyta{Here,}   
 $\set{\xi^{n}\svigi{t}; t \ge 0 }$ is as given in
 Theorem~\ref{thm:diploid-alpha-random-all} or~\ref{thm:diploid-alpha-random-one-environment}.
\end{thm}

\section{Comparing  processes}\label{sec:incr-sample-size}

\breyta{Recall that $\set{\xi^{n,N}} \equiv  \set{\xi^{n,N}(r); r \in \N_{0}}$ denotes an
(annealed) ancestral process (tracking the random ancestral relations of sampled
gene copies in a finite population)  with time measured in generations, and
\breyta{$\set{\xi^{n}} \equiv \set{\xi^{n}(t); t \ge 0}$ denotes a
(annealed) 
continuous-time coalescent.}  In this section we use simulations to
investigate how well a given coalescent $\set{\xi^{n}}$ is
approximated by the ancestral process in the domain of attraction to
$\set{\xi^{n}}$.  We will also  compare  
  annealed and quenched (see below) gene genealogies (using
  simulations).   In both cases we will 
compare  empirical means of  functionals of processes.   First, we
compare annealed processes  $\set{\xi^{n}}$   
and   $\set{\xi^{n,N}}$, with $\set{\xi^{n,N}}$ in the domain of  
attraction of the given  $\set{\xi^{n}}$.
}

\subsection{Annealed ancestral processes and coalescents}\label{sec:comp-anne-proc}

\begin{figure}[htp]
\centering
\includegraphics[scale=1]{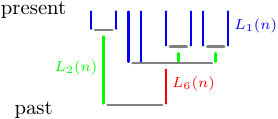}
\caption{\breyta{An example gene genealogy and the functionals
$L_{i}(n)$; 
 $L_{1}(n)$ is the sum of the length of the  blue branches, and
$L_{2}(n)$ the sum of the length of the {green} branches.} }\label{fig:illtree}
\end{figure}

  We write   $|A|$  for   the number of elements  in  a given
finite set $A$. For
$i=1,2,\ldots, 2n-1$ consider the functionals \breyta{($n$ represents
the number of sampled diploid individuals yielding $2n$ sampled  gene copies)}
\begin{displaymath}
\begin{split}
& L_{i}^{N}(n)  \equiv  \sum_{j=0}^{\tau^{N}(n)} |\set{ \xi \in  \xi^{n,N}(j) :  |\xi| = i }|,  \quad L^{N}(n) \equiv    \sum_{j=0}^{\tau^{N}(n)} | \xi^{n,N}(j)|, \\
& L_{i}(n) \equiv  \int_{0}^{\tau(n) }   |\set{ \xi \in \xi^{n}(t) :  |\xi| = i }| dt,   \quad  L(n) \equiv   \int_{0}^{\tau(n)}   |\xi^{n}(t)| dt
\end{split}
\end{displaymath}
where $\tau^{N}(n) \equiv \inf\set{ j \in \N : |\xi^{n,N}(j)| = 1}$,
 $\tau(n) \equiv \inf\set{ t \ge 0 : |\xi^{n}(t)| = 1}$,   $L^{N}(n) = \sum_{i=1}^{2n-1}L_{i}^{N}(n)$, and
$L(n) = \sum_{i=1}^{2n-1}L_{i}(n)$  
\citep[Section~1.1]{BLS15}. Interpreting $\set{\xi^{n,N}}$ and $\set{\xi^{n}}$ as
\breyta{trees}  one can interpret  $L_{i}^{N}(n)$ and
$L_{i}(n)$ as the random length of branches supporting $i$
leaves  \breyta{(Figure~\ref{fig:illtree})}.

We assume that
$ \xi^{n,N}(0) = \set{\set{1,2}, \ldots, \set{2n-1,2n}}$, i.e.\ we
sample $n$ diploid individuals and so $2n$ gene copies partitioned at
time 0 as just described. When sampling $\set{\xi^{n,N}}$  we track the pairing of
gene copies in diploid individuals;  gene copies residing  in the same
 diploid individual necessarily disperse (recall  Definition~\ref{dschwpop}
 and  Illustration~\ref{rm:illdschwpop}).

Define, for $i = 1,2,\ldots, 2n-1$,
\begin{equation}
\label{eq:33}
\begin{split}
& R_{i}^{N}(n)  \equiv  \frac{L_{i}^{N}(n)}{ L_{1}^{N}(n)  + \cdots +  L_{2n-1}^{N}(n) }, \quad  
R_{i}(n)    \equiv   \frac{L_{i}(n)}{ L_{1}(n)  + \cdots +  L_{2n-1}(n) }, 
\end{split}
\end{equation}
where $L^{N}(n) \ge 2n+2$ and $L(n) > 0$ both almost surely.
The   functionals   we will be concerned with   are 
\begin{subequations}
\begin{align}
&   \varrho_{i}^{N}(n) \equiv   \EE{R_{i}^{N}(n)}, \quad   \varrho_{i}(n) \equiv  \EE{R_{i}(n)} \label{eq:estimates}  \\
&   \overline\varrho_{i}^{N}(n) \approx  \EE{R_{i}^{N}(n)}, \quad   \overline\varrho_{i}(n) \approx   \EE{R_{i}(n)} \label{eq:approxestimates}  
\end{align}
\end{subequations}
\linelabel{varrho}\breyta{with $\overline \varrho_{i}^{N}(n)$ resp.\
$\overline\varrho_{i}(n)$ in~\eqref{eq:approxestimates}
denoting  the  corresponding empirical means computed from a  finite
number of independent realisations (using simulations) of  $R_{i}^{N}(n)$  and
$R_{i}(n)$, respectively.}

When \breyta{evaluating $\overline \varrho_{i}^{N}(n)$}  
we keep track of the configuration of blocks in individuals (recall Definition~\ref{dschwpop} and
Remark~\ref{rm:illdschwpop}).  Due to the instantaneous complete
dispersion of blocks of $\set{\xi^{n}}$, we assume
$\xi^{n}(0) = \set{\set{1},\ldots, \set{2n}}$ and
$\xi^{n}(t) \in \mathcal E_{2n}$ for all $t \ge 0$, i.e.\ the blocks
of any partition $\xi \in \set{\xi^{n}}$ are always assumed to be completely dispersed.  For
any given coalescent $\set{\xi^{n}} $ and an ancestral process
$\set{\xi^{n,N}\svigi{\lfloor t/c_{N} \rfloor }; t \ge 0 }$ in the
domain-of-attraction of $\set{\xi^{n}}$ it should hold that
$\EE{R_{i}^{N}(n)}$ predicted by $ \set{\xi^{n,N}} $ converges to
$\EE{R_{i}(n)}$ as predicted by $\set{\xi^{n}}$.  Comparing 
 $\overline \varrho_{i}(n)$ and $\overline \varrho_{i}^{N}(n)$ is then a  way to
see how well  $\set{\xi^{n}}$ and   $\set{\xi^{n,N}}$ agree.

\breyta{For evaluating   $\overline\varrho_{i}^{N}(n)$  we will consider a 
population of  constant size $2N$  evolving}   according to
Definition~\ref{dschwpop}   with the law
for the random number $X$ of potential offspring of an arbitrary
parent pair being 
\begin{equation}
\label{eq:32}
\prob{X = k } = \one{2 \le k \le \uN} h_{a}(N)\left( {k^{-a}} - {{(k+1)}^{-a}} \right)
\end{equation}
where $h_{a}(N)$ is  such that $\prob{X \in \{2,\ldots, \uN \} } = 1$.  It
then holds that $X_{1} + \cdots + X_{N} \ge 2N$ almost surely.
Moreover, $h_{a}(\zeta(N))\to 2^{a}$ as $\uN \to \infty$.    The
law in~\eqref{eq:33} is a special case  of the one in~\eqref{eq:PXiJ}.


Using the upper bound in~\eqref{eq:boundbiggera} in Lemma~\ref{lm:boundskernel} we see
\begin{displaymath}
\begin{split}
\limsup_{N\to\infty} m_{N} &  \le \limsup_{N\to\infty} h_{\kappa}(N) \kappa  \sum_{k=2}^{\zeta(N)} k^{-\kappa}  \\
\end{split}
\end{displaymath}
Similarly from the lower bound in~\eqref{eq:boundbiggera} we see
\begin{displaymath}
\liminf_{N\to\infty} m_{N} \ge  \liminf_{N\to \infty} h_{\kappa}(N) \sum_{k=2}^{\zeta(N)} \frac{k}{(k+1)^{1+\kappa}} =  \liminf_{N\to \infty} h_{\kappa}(N)\svigi{\sum_{j=3}^{  \uN + 1}j^{-\kappa}  -  \sum_{j=3}^{\uN + 1}j^{-1-\kappa}    }
\end{displaymath}
When $\kappa = 2$ it then holds that (recalling the 2-series
$\sum_{n=1}^{\infty}n^{-2} = \pi^{2}/6$)
\begin{displaymath}
\begin{split}
\limsup_{N\to \infty} m_{N} &  \le  8 \svigi{\pi^{2}/6  -  1 }   \\
\liminf_{N\to \infty}m_{N} & \ge  \frac{2}{3}\pi^{2} - \frac{11}{2}
\end{split}
\end{displaymath}
Recalling~\eqref{eq:32},  and noting that  $2 \pi^{2}/3 - 11/2 < 2$, 
we  approximate   $m_{\infty}$ with $2 < \mathbbm{m} <   8
\svigi{(\pi^{2}/6)  -  1 } $ when $\kappa = 2$ \breyta{(recall 
$X_{i} \ge 2$ for all $i\in [N]$  by~\eqref{eq:32}).}


When $\kappa > 2$   similar calculations show that
   \begin{displaymath}
     2^{\kappa}\svigi{ \frac{3^{1-\kappa}}{\kappa - 1 } - \frac{2^{-\kappa}}{\kappa} }  \le  1 \le   \liminf_{N\to \infty}m_{N} \le \limsup_{N\to \infty} m_{N} \le   2^{\kappa} \frac{\kappa}{\kappa - 1}
\end{displaymath}
Then, recalling again~\eqref{eq:32} 
we  approximate $m_{\infty}$, when $\kappa > 2$,  with
\begin{displaymath} 
  2     <  \mathbbm{m} < 2^{\kappa} \frac{\kappa}{\kappa - 1}
\end{displaymath}

It follows from Case~\ref{item:2} of Theorem~\ref{thm:diploid-alpha-random-all},
and Case~\ref{item:9} of
Theorem~\ref{thm:diploid-alpha-random-one-environment} that the
transition rate of a size-ordered  $(k_{1},\ldots, k_{r})$-merger  for all $r\in
[4]$ is  (recall~\eqref{eq:lambdankBK} and~\eqref{eq:gamma}--\eqref{eq:16})
\begin{align}
\notag
\lambda_{m;k_{1},\ldots, k_{r};s} &  = \binom{m}{k_{1}\ldots k_{r}s }\frac{1}{\prod_{j=2}^{m}\left( \sum_{i}\one{k_{i} = j} \right)!  }\lambda_{m;k_{1},\ldots, k_{r};s} ^{\prime} \\ \notag
\lambda_{m;k_{1},\ldots, k_{r};s} ^{\prime} & = \one{r=1,k_{1}=2}\frac{C_{\kappa} }{C_{\kappa,\alpha,\gamma}} \\ \notag
& +   \frac{\alpha c 2^{\alpha} }{C_{\kappa,\alpha,\gamma} {\mathbbm{m}^{\alpha} } } \sum_{\ell=0}^{s\wedge (4-r) }\binom{s}{\ell} \frac{(4)_{r+\ell} }{4^{k+\ell} }B(\gamma, k + \ell-\alpha, n-k-\ell +\alpha)  \\ \label{eq:17}
C_{\kappa} & =  \one{\kappa = 2} \frac{2 }{\mathbbm{m}^{2} } +  \one{\kappa > 2}\frac{2^{\kappa + 1}}{4\mathbbm{m}^{2}} \frac{c_{\kappa}}{2^{\kappa}(\kappa - 2)(\kappa - 1) } \\ \notag
C_{\kappa,\alpha,\gamma} & = C_{\kappa} +  \frac{c\alpha 2^{\alpha} }{4\mathbbm{m}^{\alpha} }B(\gamma,2-\alpha,\alpha) \\ \notag
\gamma & \equiv \one{ \frac{\zeta(N)}{N} \to K } \frac{K }{K +  \mathbbm{m} } +     \one{ \frac{\zeta(N)}{N} \to \infty } \\ \notag 
 \mathbbm{m} & =  \frac 12 \svigi{2 +   8(\pi^{2}/6 -1)} =   2\pi^{2}/3 - 3  \quad \text{when $\kappa = 2$} \\ \notag
 \mathbbm{m} & =  1 + 2^{\kappa - 1}\kappa/(\kappa - 1)   \quad \text{when $\kappa > 2$}
\end{align}
In~\eqref{eq:17} $0<\alpha<2$, $0 < \gamma \le 1$, $c > 0$, $\kappa
\ge 2$.

\begin{figure}[htp]
\centering
\captionsetup[subfloat]{labelfont={scriptsize,sf,md,up},textfont={scriptsize,sf}}
\subfloat[$c=1$, $\gamma = 1$]{\includegraphics[angle=0,scale=.6]{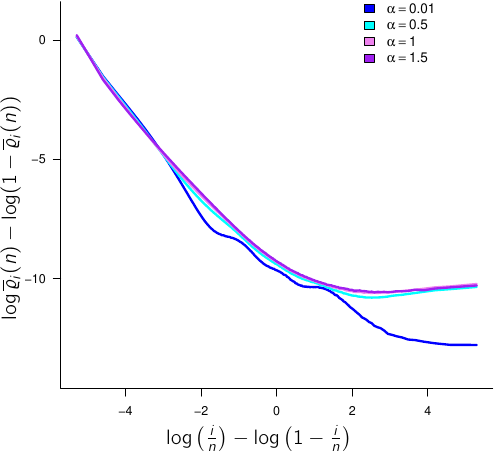}}
\subfloat[$c=10^{2}$, $\gamma = 1$]{\includegraphics[angle=0,scale=.6]{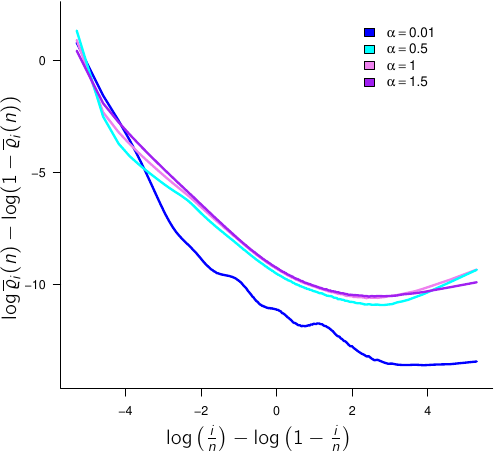}}
\subfloat[$c=10^{4}$, $\gamma = 1$]{\includegraphics[angle=0,scale=.6]{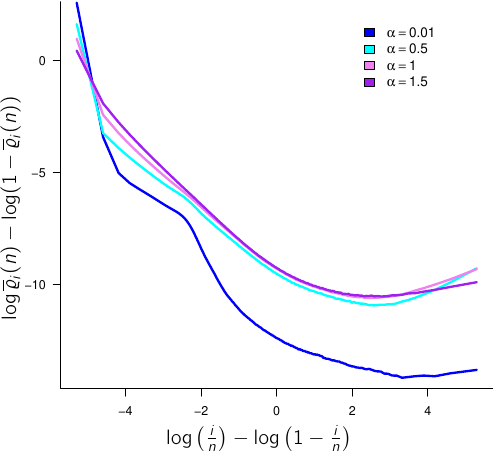}} \\
\subfloat[$c=1$, $\gamma = 0.1$]{\includegraphics[angle=0,scale=.6]{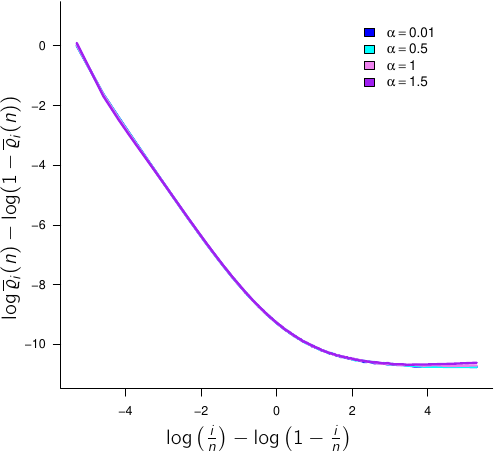}}
\subfloat[$c=10^{2}$, $\gamma = 0.1$]{\includegraphics[angle=0,scale=.6]{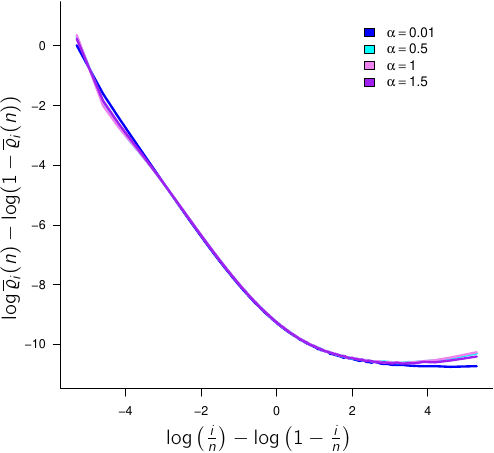}}
\subfloat[$c=10^{4}$, $\gamma = 0.1$]{\includegraphics[angle=0,scale=.6]{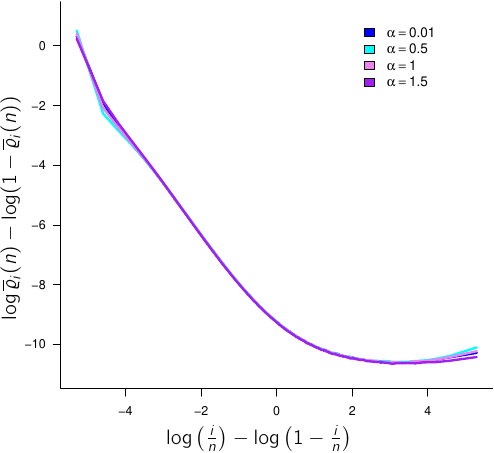}}
\caption{\breyta{The
$\Omega$-$\delta_{0}$-Beta$(\gamma,2-\alpha,\alpha)$-coalescent.}
\breyta{The empirical mean}    $\overline \varrho_{i}(n)$  \breyta{of $R_{i}(n)$
(recall~\eqref{eq:33},~\eqref{eq:estimates},~\eqref{eq:approxestimates})}  when $\set{\xi^{n}}$ is
the 
$\Omega$-$\delta_{0}$-Beta$(\gamma,2-\alpha,\alpha)$-coalescent;
when $n=100$, $\kappa = 2$, $c$, $\alpha$, and $\gamma$ as shown;
see Appendix~\ref{sec:estimatingERin} for a brief description of the
algorithm for sampling from the
$\Omega$-$\delta_{0}$-Beta$(\gamma,2-\alpha,\alpha)$ coalescent;
results from $10^{6}$ experiments }\label{fig:xindbetaoverparams}
\end{figure}

Examples of $\overline \varrho_{i}(n)$
\breyta{(recall~\eqref{eq:approxestimates})}  when $\set{\xi^{n}}$ is the
$\Omega$-$\delta_{0}$-Beta$(\gamma,2-\alpha,\alpha)$-coalescent
(transition rates as in~\eqref{eq:17}) are in
Figure~\ref{fig:xindbetaoverparams}.  The
$\Omega$-$\delta_{0}$-Beta$(\gamma,2-\alpha,\alpha)$-coalescent can
predict a range of different site-frequency spectra, including
$U$-shaped spectra similar to that observed in the highly fecund
diploid Atlantic cod \citep{Arnasonsweepstakes2022}.  Morever,
Figure~\ref{fig:xindbetaoverparams} clearly shows the effect of the
upper bound $\zeta(N)$ (through $\gamma$; recall~\eqref{eq:gamma}) on
the site-frequency spectrum.

\begin{figure}[htp]
\centering
\captionsetup[subfloat]{labelfont={scriptsize,sf,md,up},textfont={scriptsize,sf}}
\subfloat[$\alpha = 0.01$, Definition~\ref{df:alpha-random-one-environment}]{\label{fig:xindbetannealedM1}\includegraphics[angle=0,scale=.6]{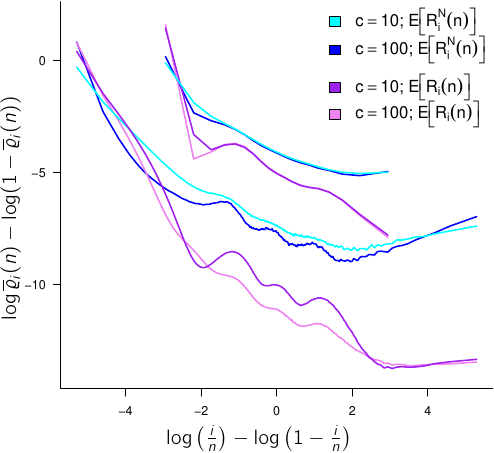}}
\subfloat[$\alpha = 1$, Definition~\ref{dfn:alpharandomall}]{\label{fig:xindbetannealedM0}\includegraphics[angle=0,scale=.6]{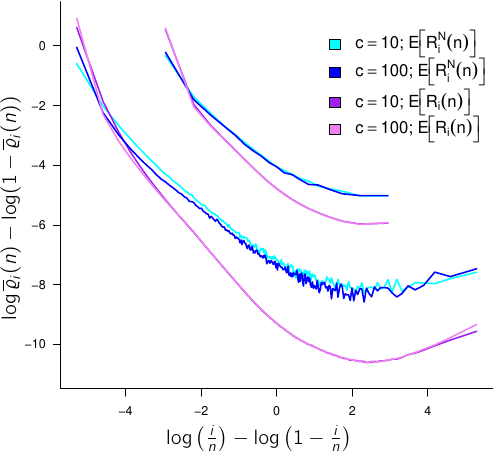}}
\caption{\breyta{The
$\Omega$-$\delta_{0}$-Beta$(\gamma,2-\alpha,\alpha)$-coalescent.}
Comparing  $\overline \varrho_{i}(n)$
\breyta{(recall~\eqref{eq:estimates} and~\eqref{eq:approxestimates})}  and
$\overline \varrho_{i}^{N}(n)$ 
 when    $\set{\xi^{n}} $ is    the  
$\Omega$-$\delta_{0}$-Beta$(\gamma,2-\alpha,\alpha)$-coalescent and
when the population evolves according to  Definitions~\ref{dschwpop}
and~\ref{dfn:alpharandomall} and~\ref{df:alpha-random-one-environment} and 
~\eqref{eq:32}  with    $N=2500$, $\zeta(N) = N\log N$,   $\alpha$ as
shown,     $\kappa = 2$, $\zeta(N)
= N\log N$, sample size $n=100$, $n=10$,  $\gamma = 1$; $c$ as shown;
with $\varepsilon_{N}$ as in~\eqref{eq:varepsilon} and 
 $\overline \varepsilon_{N}$ as in~\eqref{eq:varepsarandone}}\label{fig:compare_xindbeta_annealed}
\end{figure}

However, there is discrepancy between $\overline \varrho_{i}^{N}(n)$
and $\overline \varrho_{i}(n)$ \breyta{(recall~\eqref{eq:estimates}
and~\eqref{eq:approxestimates})}   when $\set{\xi^{n}}$ is 
 the $\Omega$-$\delta_{0}$-Beta$(\gamma,2-\alpha,\alpha)$-coalescent
(Figure~\ref{fig:compare_xindbeta_annealed}).  In
Figure~\ref{fig:compare_xindbeta_annealed} we compare
$\overline \varrho_{i}^{N}(n)$ (\breyta{recall~\eqref{eq:approxestimates};} blue and cyan lines) to
$\overline\varrho_{i}(n)$ (violet and purple lines) when the
population evolves as in Definition~\ref{dschwpop} and
Definition~\ref{df:alpha-random-one-environment}
(Figure~\ref{fig:xindbetannealedM1}) resp.\
Definition~\ref{dfn:alpharandomall}
(Figure~\ref{fig:xindbetannealedM0}) with numbers of potential
offspring distributed as in~\eqref{eq:32}.  One reason might be that
$\set{\xi^{n,N}}$ converges slowly to $\set{\xi^{n}}$, in particular
for small $\alpha$.  Should $\set{\xi^{n}}$ be  a good approximation of
$\set{\xi^{n,N}}$ only for population sizes exceeding
\breyta{realistic}  ones
(the population size used in
Figure~\ref{fig:compare_xindbeta_annealed} is likely nowhere near
that), the usefulness of coalescent-based inference methods will then
be left in doubt.  In contrast,  the agreement between  Kingman
 and  Wright-Fisher \breyta{trees}     is
in general quite good \citep{Fu2006}.
Figure~\ref{fig:compare_xindbeta_annealed}  also  reveals a
qualitative  difference  within $\overline \varrho_{i}^{N}(n)$
depending on the pre-limiting model
(Definitions~\ref{df:alpha-random-one-environment}  and
~\ref{dfn:alpharandomall}); the peaks in  the graph of   $\overline
\varrho_{i}^{N}(n)$ (blue lines in Figure~\ref{fig:xindbetannealedM1})
are absent  in Figure~\ref{fig:xindbetannealedM0}. Similarly,  the
peaks in the graphs for  $\overline\varrho_{i}(n)$ in
Figure~\ref{fig:xindbetannealedM1}  are absent in
Figure~\ref{fig:xindbetannealedM0}.

Recall from Case~\ref{item:8} of
Theorem~\ref{thm:diploid-alpha-random-all} that convergence to the
$\Omega$-$\delta_{0}$-Poisson-Dirichlet$(\alpha,0)$-coalescent as the
limit of $\set{\xi^{n,N}}$ determined by Definitions~\ref{dschwpop}
and~\ref{dfn:alpharandomall}, and~\eqref{eq:32}, requires the rather
strong (in the sense of being applicable to real populations; recall
that $2N$ is the population size)  assumption that
$\zeta(N)/N^{1/\alpha} \to \infty$.   Thus, we hesitate to
claim that the 
$\Omega$-$\delta_{0}$-Poisson-Dirichlet$(\alpha,0)$-coalescent is
relevant   for
explaining population genetic data, even for broadcast spawners.
Nevertheless, we record in
Figures~\ref{fig:time-changed_omega_delta_poissondiri} and~\ref{fig:diploiddpd}  examples of
\breyta{$\overline\varrho_{i}(n)$ (recall~\eqref{eq:estimates} and
~\eqref{eq:approxestimates})}   for the
$\Omega$-$\delta_{0}$-Poisson-Dirichlet$(\alpha,0)$-coalescent.

\subsection{Annealed and  quenched  ancestral  processes}\label{sec:comp-quench-anne}

\breyta{The annealed  empirical means}  $\overline \varrho_{i}^{N}(n)$   are obtained by
(implicitly) averaging over the \breyta{unknown}  ancestral relations of the sampled
gene copies.       An alternative way of estimating mean branch lengths is
to condition on  \breyta{a given}   population ancestry (the ancestral relations of
 gene copies in the population).   Given \breyta{a}  population
ancestry 
the gene genealogy of the  gene copies in a given sample is
complete and fixed, \breyta{it is quenched.}   We let $\EE{\widetilde R_{i}^{N}(n) }$ denote the
\breyta{quenched}   mean of relative branch lengths when averaging over population
ancestries.  Our approach is different from the one of
~\cite{Diamantidis2024}, who consider the  average over gene
genealogies within one fixed  population pedigree (the ancestral
relations of diploid individuals).
\begin{equation}
\label{eq:21}
\xymatrix  @C=1pt @R =2pt  {%
\text{\sf a} & \ar[rrr]^{\rm time} & && \rotatebox{-90}{\scalebox{.8}{past}} &&& && \text{\sf b} &&&&& \text{\sf c} &&&&&& \text{\sf d} \\
*+[o][F]{\phantom \bullet}& *+[o][F]{\phantom \bullet}& *+[o][F]{\phantom \bullet}&  *+[o][F]{\phantom \bullet}&    &&&&& *+[o][F]{\bullet} &&&&&   *+[o][F]{\bullet} &&  *+[o][F]{\bullet} &
*+[o][F]{\bullet}  &&&    *+[o][F]{\bullet}   &   *+[o][F]{\bullet}   \\
*+[o][F]{\phantom \bullet}& *+[o][F]{\phantom \bullet}& *+[o][F]{\phantom \bullet}&   *+[o][F]{\phantom \bullet}&    &&&&& *+[o][F]{\bullet} & *+[o][F]{\bullet} &&&&  *+[o][F]{\bullet} &
*+[o][F]{\bullet\bullet} &  *+[o][F]{\bullet}  &&&&
*+[o][F]{\bullet} &   *+[o][F]{\bullet}   &   *+[o][F]{\bullet}  \\\\
\text{\sf e} &&&&&   \text{\sf f} \\
*+[F]{\bullet} &   *+[F]{\bullet}   &   *+[F]{\bullet} &&&  *+[F]{\bullet} &  *+[F]{\bullet}  &  *+[F]{\bullet} & &  *+[F]{\bullet} \\
*+[F]{\bullet} &   *+[F]{\bullet}   &                     &&&   *+[F]{\bullet}&   *+[F]{\bullet} &  *+[F]{\bullet} & *+[F]{\bullet \bullet} &  *+[F]{\bullet} &  *+[F]{\bullet} \\
}
\end{equation}
In (\ref{eq:21}b--d) are 3 possible gene genealogies of 2 gene copies
($\bullet$) all going through the same pedigree (\ref{eq:21}a) (the
ancestry of diploid individuals (\scalebox{1.5}{$\circ$}); from left
to right is into the past).  In contrast to restricting to a single
pedigree, we average over population ancestries \breyta{(a record of
the ancestral relations of gene copies)}, recording the branch
lengths of one \breyta{complete sample} gene genealogy per population
ancestry (\ref{eq:21}e--f; the clusters of $\square$ represent 
population ancestries); any sample of gene copies shares one gene
genealogy within one and the same population ancestry.\linelabel{point2}\linelabel{Diaman}~\breyta{
 There are many  possible gene genealogies within a fixed 
pedigree.  Moreover, as~\cite{Diamantidis2024} show,  restricting to a
fixed pedigree  matters  for what the limiting  process (in a large
population) will look like~\citep[Theorem~1]{Diamantidis2024}.   The
view  here is that the natural approach   would be to
condition on a population ancestry,   read 1 complete sample tree
from each given ancestry, and average the sample trees over
independent ancestries.        It
remains to be seen  what the limiting quenched  coalescent  then looks
like, and if it is any different from the annealed coalescent. }

\linelabel{clarify}\breyta{
We  clarify the bookkeeping method for keeping track of the
ancestral relations of the gene copies in the population.  For each
gene copy in the population we record the level it occupies, the time
(generation)   it occupies the given level, and  the level occupied by
the gene copy's  immediate ancestor (the value). Using  a haploid panmictic
population for an example,  the offspring of the  copy on level $\ell$ at
time $g$ all have value $\ell$,  while the gene copy itself derives
from  the one occupying level $i$ at time $g-1$ (see~\eqref{eq:27})
\begin{equation}
\label{eq:27}
\xymatrix @C =2pt @R =0.5pt @M=0.5pt  {
& \text{level} \\
\text{time} & 1 & \ldots & \ell & \ldots \\
g& & \ldots & i & \ldots \\
g+1&   &  \ell & \ell & \ell &   \\
}
\end{equation}
A diploid
individual is seen as a pair of gene copies;  for some $i\in [2N]$
gene copies occupying  levels $2i-1$ and $2i$ at a given
time  form an individual. An example  ancestry of  a diploid
panmictic  population of $4N$ gene copies  is shown in~\eqref{eq:25}.
\begin{equation}
\label{eq:25}
\xymatrix @C =0.5pt @R =0.5pt  {
& \text{level} \\
\text{time} & 1 & \ldots  & i & i+1 & \ldots &j   & j+1   & \ldots  & 4N \\
\vdots \\
g & &  \ldots  & \ell_{1} & \ell_{2} & \ldots & \ell_{3} & \ell_{4} & \ldots  \\
g + 1 & &  \ldots & i & j & \ldots &  j+1 & i+1 & \ldots \\
\vdots \\
}
\end{equation}
In~\eqref{eq:25} an   individual whose  gene copies  occupy levels
$i$ and $i+1$, and another individual whose gene copies occupy levels
$j$ and $j+1$  at time  $g$  form a parent pair; the gene copies of
the    surviving
offspring of this parent pair  $(\set{i,i+1},\set{j,j+1})$  will then  have values taken from
$\set{i,i+1,j,j+1}$. We exclude selfing such that one gene copy of an
offspring of     $(\set{i,i+1},\set{j,j+1})$   will
sample its value  from $\set{i,i+1}$.   The other gene copy of the
given  offspring  will
sample its value from $\set{j,j+1}$.  The parent gene copies
themselves  have values (immediate ancestors)
$\ell_{1},\ldots,\ell_{4}$ as shown in~\eqref{eq:25};  the gene copy
on level $i$ at time $g$  derives from  the one occupying level
$\ell_{1}$ at time $g-1$, and so on.  In this way we  record the
population ancestry.   Given a population ancestry the  complete
sample tree of  a given sample of gene copies is fixed, as illustrated
in~\eqref{eq:30}.  
\begin{equation}
\label{eq:30}
\xymatrix @C=1pt @R=1pt @M=0.2pt{
&\text{\sf A} &&&&&&  \text{\sf B} &&&&&&&& \text{\sf C}  \\
\text{past}&&  \text{level} &&&&&&&& \text{level} \\\\
& 1 & \ldots &  4N &&&& 1 &&& \cdots && 4N \\
&&&&&&&&&&&&&&&& \bullet \\
 & &\vdots &   &&&&  &&  & \vdots  &&&&&  \vdots  \\
\phantom{\text{present}} & \circ   & \cdots   &  \circ  &&&& \circ &   && \cdots && \circ &&& \bullet & \bullet   \\
\text{present} &  \circ  & \cdots & \circ &&&& \bullet &  \cdots &  \bullet & \circ & \cdots & \circ &&& \bullet & \bullet &\cdots  &\bullet  \\
}
\end{equation}
In~\eqref{eq:30} we broadly  illustrate the process for generating a quenched
gene genealogy.  
In~(\ref{eq:30}A) we record  a single  population ancestry (the
ancestral relations of the gene copies in the entire population)   with $\circ$
denoting the gene copies (recall~\eqref{eq:25});    once the identity of the sampled gene
copies (shown as $\bullet$ in~(\ref{eq:30}B)) is known    all that remains
is to track the fixed complete  sample tree   (\ref{eq:30}C).  The
  \breyta{empirical means}   $\overline \rho_{i}^{N}(n)$ of quenched relative
branch lengths  are then obtained by
repeating the method    shown   in~\eqref{eq:30},  each time starting with a new
population ancestry and reading 1 complete  sample tree from each
ancestry (see also Appendix~\ref{sec:estimatingERiNA}). 
}

\breyta{The annealed  ancestral process is sampled by  sampling
offspring numbers  and merging blocks assigned to the same  parent
chromosome  (see Appendix~\ref{sec:estimatingERiN}); the complete
sample tree in the sense of the quenched method is never generated.     We   track  block sizes of
sampled blocks (and of blocks ancestral to the sampled blocks) 
and  the partitioning of blocks in individuals as we 
transition  backwards in time.    The examples
in~\eqref{eq:29} show 2 possible transitions for  4 blocks labelled
$a$, $b$, $c$, and $d$  all in separate  diploid offspring of a given
parent pair.  
\begin{equation}
\label{eq:29}
\xymatrix @C=2pt @R=0.5pt {
 & \text{\sf A} & &&&&& \text{\sf B} \\
\text{parents} & &  *+[o][F]{\set{c+d} }  &   *+[o][F]{\set{a+b} }     &   &&  &&  *+[o][F]{\set{b} \set d }  &   *+[o][F]{\set{a}  \set c }   \\
\text{offspring}  & *+[F]{\set a} &  *+[F]{\set b } &  *+[F]{\set c } &  *+[F]{\set d}  && &   *+[F]{\set a} &  *+[F]{\set b} &  *+[F]{\set c} &    *+[F]{\set d}  \\  
}
\end{equation}
In~(\ref{eq:29}A)  blocks $c$ and  $d$ merge,  and blocks  $a$
and $b$ merge in an example of a simultaneous merger;  
(\ref{eq:29}B)  is an example of an instantaneous state in which
blocks end up in the same diploid individual without merging.
In~(\ref{eq:29}A)  we would continue with 
$\xymatrix@1@C=2pt{ *+[F]{\set{c+d}}& *+[F]{\set{a+b}}}$,   and with 
$\xymatrix@1@C=2pt{ *+[F]{\set{a}\set{c}} & *+[F]{\set{b}\set{d}}}$
in~(\ref{eq:29}B).
%
}

\breyta{The  means  $\overline\rho_{i}^{N}(n)$  are evaluated
}
for a diploid panmictic population of constant size $2N$ diploid
individuals ($4N$ gene copies) by evolving the population forward in
time as in Definition~\ref{dschwpop} and~\eqref{eq:32} and recording
the ancestry of the gene copies \breyta{(recall~\eqref{eq:25})}.
Every now and then  \breyta{(meaning that we increment the population
ancestry for a fixed number of generations in between drawing a sample)} we randomly
sample $n$ diploid individuals (and so $2n$ gene copies), each time
tracing the ancestry of the sampled gene copies, until a sample is
obtained whose gene copies have a common ancestor (sampled gene copies
without a common ancestor are discarded). The realised gene genealogy
(gene tree) tracing the ancestry of the sampled gene copies is then
said to be \emph{complete}. The gene tree of the sampled gene copies
is then fixed (for the given population ancestry and sample), and all
that remains to do is to read the branch lengths off the fixed tree
\breyta{(recall~\eqref{eq:30})}.
Repeating this a given number of times, each time starting from
scratch with a new population, gives us \breyta{the empirical mean 
$\overline\rho_{i}^{N}(n)$, an  estimate of   $\EE{\widetilde
R_{i}^{N}(n)}$ (see Appendix~\ref{sec:estimatingERiNA}).}

In Figure~\ref{fig:annealedA} we compare 
$\overline \varrho _{i}^{N}(n) $ (\breyta{an estimate  of 
$\EE{R_{i}^{N}(n)}$}; blue lines) and $\overline \rho _{i}^{N}(n) $
(\breyta{an estimate of    $\EE{\widetilde R_{i}^{N}(n)}$}; red lines)
when the population evolves as in Definitions~\ref{dschwpop}
and~\ref{dfn:alpharandomall}, and~\eqref{eq:32}.  When $\zeta(N) = 2N$
(Figure~\ref{fig:qacompareB}) then $\overline \varrho _{i}^{N}(n) $
and $\overline \rho _{i}^{N}(n) $ broadly agree; however when
$\zeta(N) = 4N^{2}$ (Figure~\ref{fig:qacompareA})
$\overline\rho_{i}^{N}(n)$ and $\overline\varrho_{i}^{N}(n)$ are
qualitatively different (Figure~\ref{fig:qaC} in
Appendix~\ref{sec:estimatingERiNA} contains  further examples).
\begin{figure}[htp]
\centering
\captionsetup[subfloat]{labelfont={scriptsize,sf,md,up},textfont={scriptsize,sf}}
\subfloat[$\zeta(N)=2N$]{\label{fig:qacompareB}\includegraphics[angle=0,scale=.6]{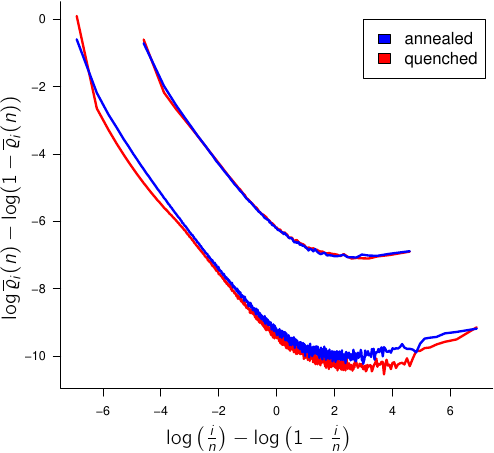}}
\subfloat[$\zeta(N)=4N^{2}$]{\label{fig:qacompareA}\includegraphics[angle=0,scale=.6]{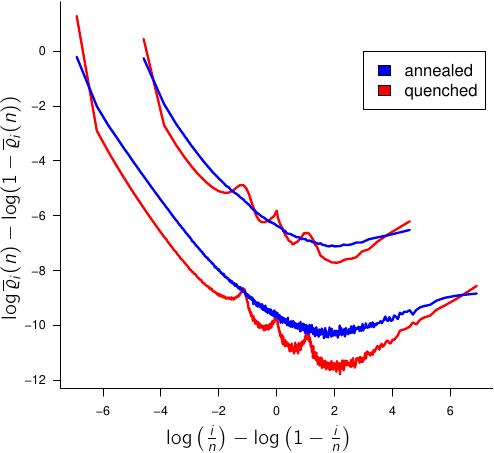}}
\caption{\breyta{Annealed and quenched  gene genealogies.} Comparing
$\overline \varrho_{i}^{N}(n)$ (\breyta{an estimate of   $\EE{R_{i}^{N}(n)}$}; blue
lines) and $\overline \rho_{i}^{N}(n)$ (\breyta{an estimate  of 
$\EE{\widetilde R_{i}^{N}(n)}$}; red lines) when
the population evolves according to Definitions~\ref{dschwpop} and
~\ref{dfn:alpharandomall}, and~\eqref{eq:32},   for $N=250$ (population size $2N$ of
diploid individuals), upper bound $\zeta(N)$ as shown,
$\varepsilon_{N} = 0.1$, $\alpha = 1$, $\kappa = 2$; sample size
$n=50$ and $n=500$ diploid individuals; results from $10^{5}$
experiments. Appendices~\ref{sec:estimatingERiN} and
~\ref{sec:estimatingERiNA} contain brief descriptions of the sampling
algorithms }\label{fig:annealedA}
\end{figure}


\subsection{Annealed time-varying  coalescents}\label{sec:time-vary-proc-1}

\begin{figure}[htp]
\centering
\captionsetup[subfloat]{labelfont={scriptsize,sf,md,up},textfont={scriptsize,sf}}
\subfloat[$c=1$, $\alpha = 0.01$]{\includegraphics[angle=0,scale=.6]{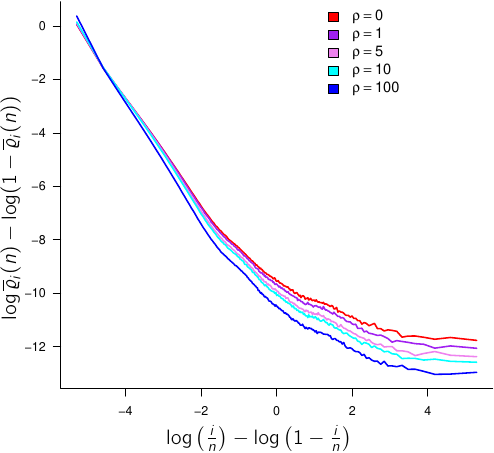}}
\subfloat[$c=1$, $\alpha = 0.5$]{\includegraphics[angle=0,scale=.6]{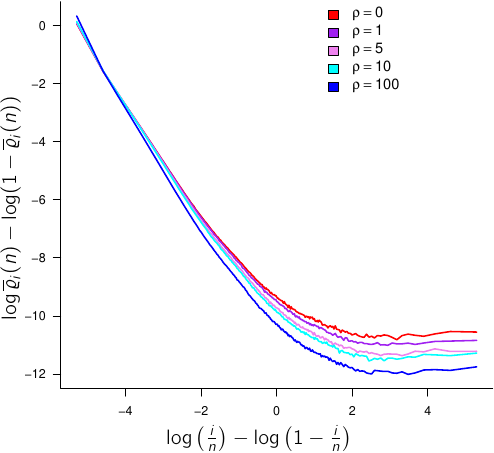}}
\subfloat[$c=1$, $\alpha = 1$]{\includegraphics[angle=0,scale=.6]{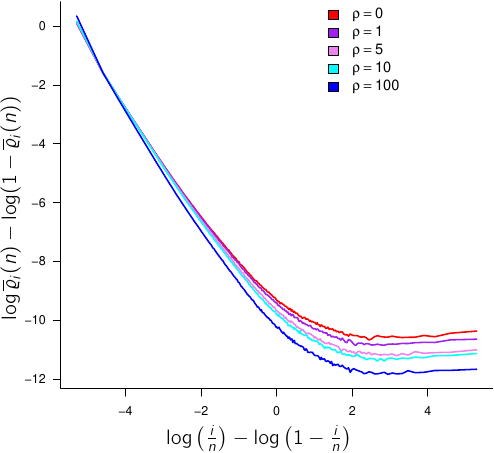}} \\
\subfloat[$c=10^{2}$, $\alpha = 0.01$]{\includegraphics[angle=0,scale=.6]{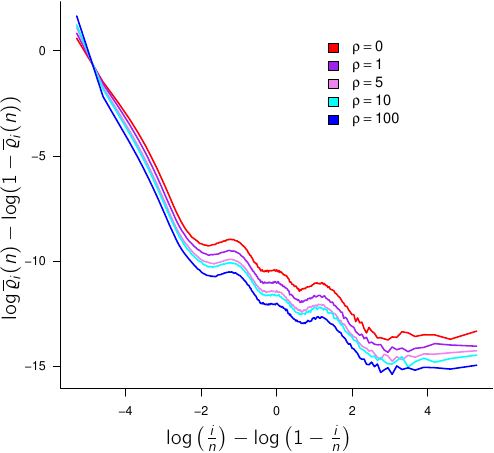}}
\subfloat[$c=10^{2}$, $\alpha = 0.5$]{\includegraphics[angle=0,scale=.6]{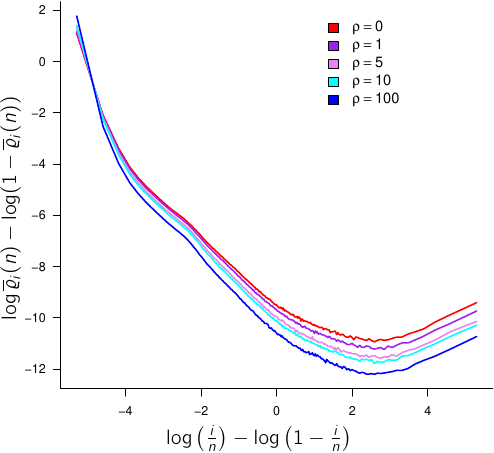}}
\subfloat[$c=10^{2}$, $\alpha = 1$]{\includegraphics[angle=0,scale=.6]{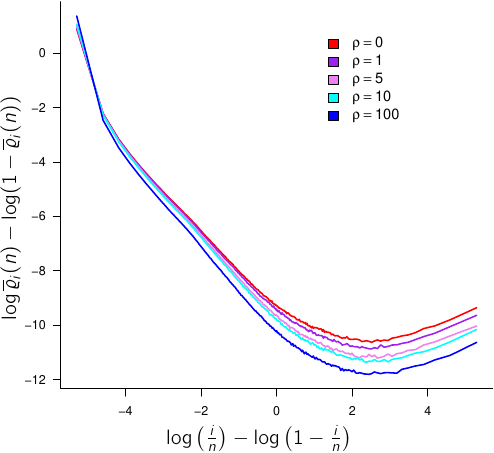}} \\
\subfloat[$c=10^{4}$, $\alpha = 0.01$]{\includegraphics[angle=0,scale=.6]{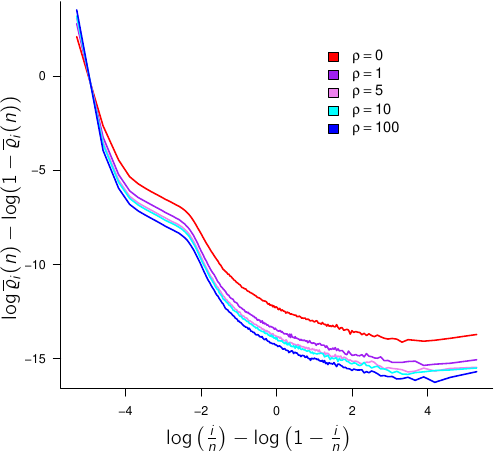}}
\subfloat[$c=10^{4}$, $\alpha = 0.5$]{\includegraphics[angle=0,scale=.6]{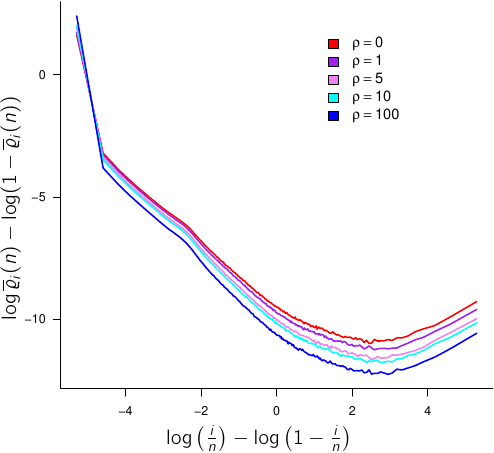}}
\subfloat[$c=10^{4}$, $\alpha = 1$]{\includegraphics[angle=0,scale=.6]{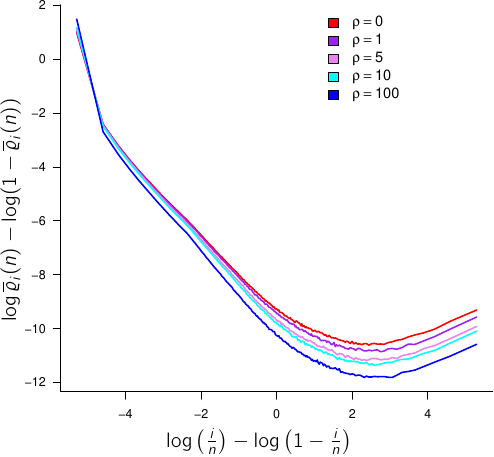}} \\
\caption{\breyta{A (time-changed)
$\Omega$-$\delta_{0}$-Beta$(\gamma,2-\alpha,\alpha)$-coalescent.}
\breyta{Empirical means}   $\overline\varrho_{i}(n)$  \breyta{of $\EE{R_{i}(n)}$}    (recall
~\eqref{eq:estimates})  when $\set{\xi^{n}}$ is the   time-changed
$\Omega$-$\delta_{0}$-Beta$(\gamma,2-\alpha,\alpha)$-coalescent with
time-change $G(t) = \int_{0}^{t}e^{\rho s}{\rm d}s = \one{\rho > 0}(1/\rho)(e^{\rho
t} - 1) + \one{\rho = 0}t$; $\gamma = 1$,
$\alpha, c, \rho$ as shown  for
$n=100$;  $10^{5}$ experiments }\label{fig:deltanullbetaweg}
\end{figure}

\begin{figure}[htp]
\centering
\captionsetup[subfloat]{labelfont={scriptsize,sf,md,up},textfont={scriptsize,sf}}
\subfloat[$\alpha = 0.5$, $c=1$]{\includegraphics[angle=0,scale=.6]{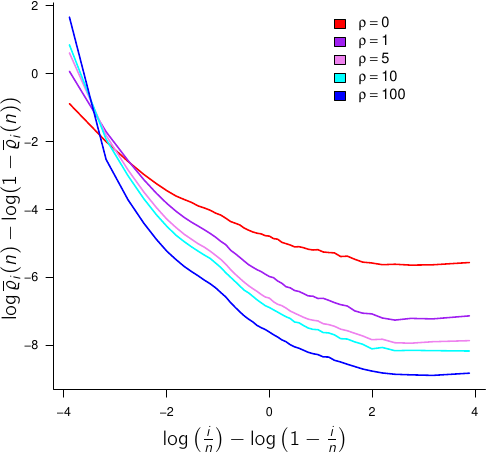}}
\subfloat[$\alpha = 0.99$, $c=1$]{\includegraphics[angle=0,scale=.6]{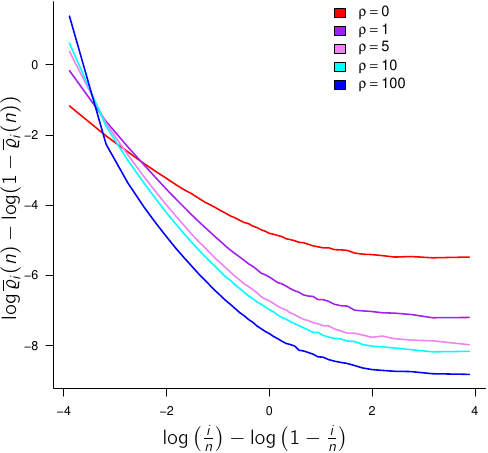}} \\
\subfloat[$\alpha = 0.5$, $c=100$]{\includegraphics[angle=0,scale=.6]{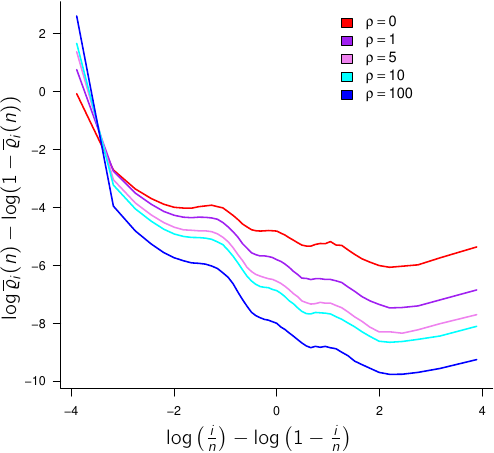}}
\subfloat[$\alpha = 0.99$, $c=100$]{\includegraphics[angle=0,scale=.6]{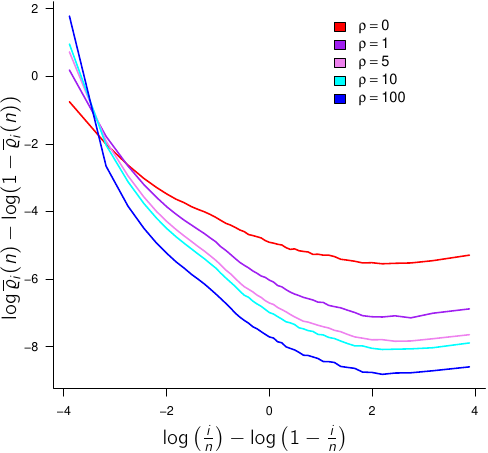}} 
\caption{\breyta{A (time-changed) 
$\Omega$-$\delta_{0}$-Poisson-Dirichlet$(\alpha,0)$-coalescent.}
\breyta{Empirical means}    $\overline \varrho_{i}(n)$ (recall
~\eqref{eq:estimates})  when $\set{\xi^{n}}$ is the 
time-changed
$\Omega$-$\delta_{0}$-Poisson-Dirichlet$(\alpha,0)$-coalescent with
time-change   $G(t) = \int_{0}^{t}e^{\rho s}{\rm d}s = \one{\rho > 0}(1/\rho)(e^{\rho
t} - 1) + \one{\rho = 0}t$ with $\rho,\alpha,c$  as shown, $\kappa =
2$; results from $10^{5}$ experiments }\label{fig:time-changed_omega_delta_poissondiri}
\end{figure}

In Theorem~\ref{thm:time-varyingXi} we extend
~\cite[Theorem~3]{freund2020cannings} to $\Xi$-coalescents where the
time-change is independent of $\alpha$.  In 
Figure~\ref{fig:deltanullbetaweg}  we give examples of the effect of
exponential population growth      (time-change function $v(t) = e^{-\rho
t}$) \citep{Donnelly1995}   when  $\set{\xi^{n}}$ is the 
$\Omega$-$\delta_{0}$-Beta$(\gamma,2-\alpha,\alpha)$-coalescent, and
in  Figure~\ref{fig:time-changed_omega_delta_poissondiri} when
$\set{\xi^{n}}$ is the 
$\Omega$-$\delta_{0}$-Poisson-Dirichlet$(\alpha,0)$-coalescent. Population
growth extends the relative length of external branches  of the
 examples of $\Xi$-coalescents studied here, 
similar to the effect  on Kingman trees
\citep{Donnelly1995}.

\begin{figure}[htp]
\centering
\captionsetup[subfloat]{labelfont={scriptsize,sf,md,up},textfont={scriptsize,sf}}
\subfloat[$\alpha = 0.01$, $c=10^{2}$]{\includegraphics[angle=0,scale=.6]{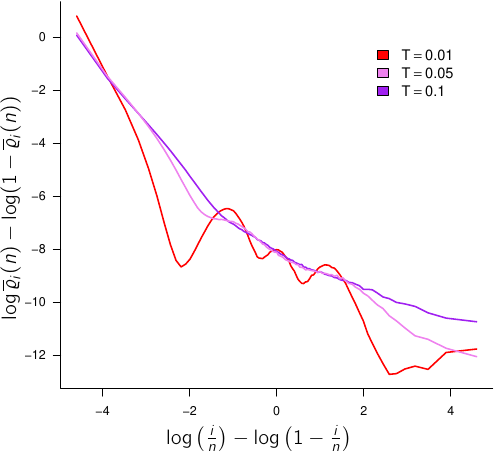}}
\subfloat[$\alpha = 0.5$, $c=10^{2}$]{\includegraphics[angle=0,scale=.6]{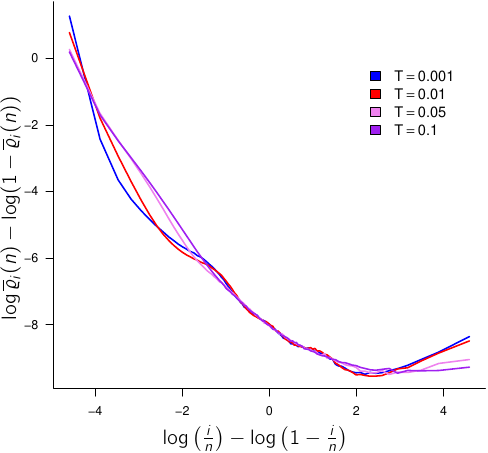}}
\subfloat[$\alpha = 1$, $c=10^{2}$]{\includegraphics[angle=0,scale=.6]{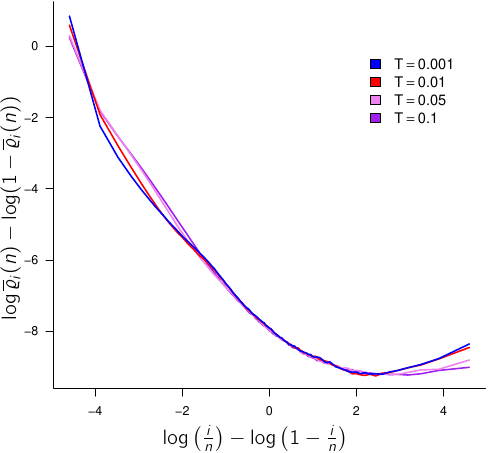}} \\
\subfloat[$\alpha = 0.01$, $c=10^{4}$]{\includegraphics[angle=0,scale=.6]{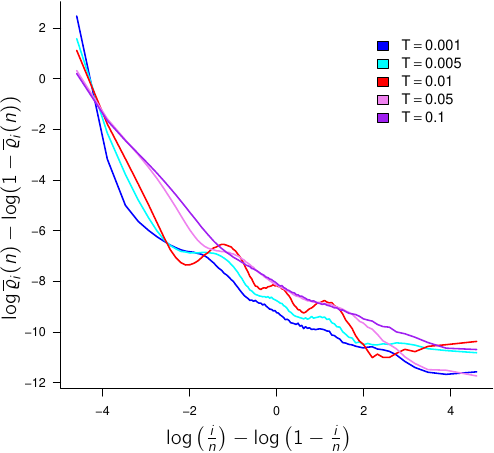}}
\subfloat[$\alpha = 0.5$, $c=10^{4}$]{\includegraphics[angle=0,scale=.6]{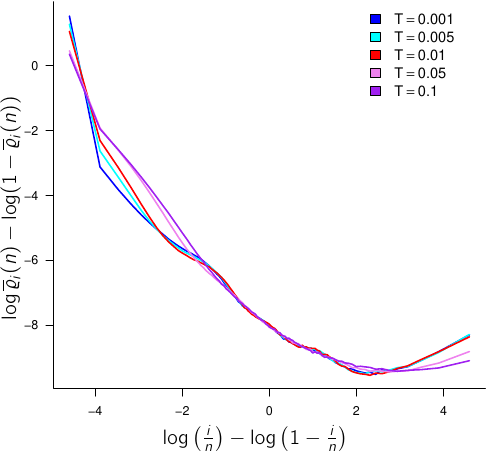}}
\subfloat[$\alpha = 1$, $c=10^{4}$]{\includegraphics[angle=0,scale=.6]{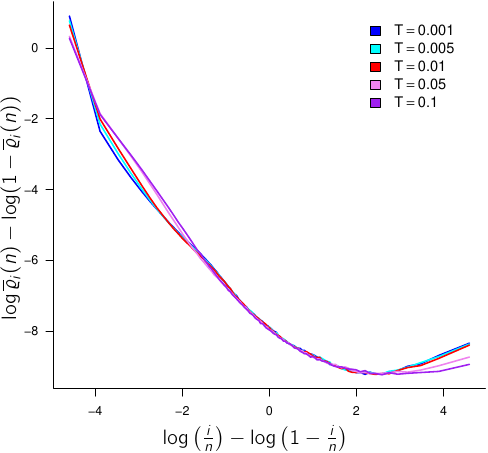}} 
\caption{\breyta{A time-changed 
$\Omega$-$\delta_{0}$-Beta$(\gamma_{t},2-\alpha,\alpha)$-coalescent.}
\breyta{Empirical means}  $\overline \varrho_{i}(n)$
\breyta{estimating  $\EE{R_{i}(n)}$}
(recall~\eqref{eq:approxestimates})  when $\set{\xi^{n}}$ is the 
time-varying 
$\Omega$-$\delta_{0}$-Beta$(\gamma_{t},2-\alpha,\alpha)$-coalescent
with $\gamma_{t}$ a step function $\gamma_{t} = \one{0 \le t \le T}
0.1 + \one{t > T}$  with $T$ as shown,  $\kappa = 2$;
results from    $10^{5}$ experiments  }\label{fig:timevaryingxibeta}
\end{figure}

\breyta{Since 
~\eqref{eq:scalecN} is in force, such that gene genealogies (on
average) span lengths of time  proportional to $N$ generation,}      it is plausible  that some components of
~\eqref{eq:PXiJ} may vary  over time.   In
Figure~\ref{fig:timevaryingxibeta} we give examples taking
$\gamma_{t}$ to be a simple  step-function. Here, we take 
  $\svigi{\zeta_{r}(N)}_{r\in \N_{0}}$ to be a sequence of cutoffs
  (with $\zeta_{r}(N)$ being the cutoff in~\eqref{eq:PXiJ} at
  generation $r$ into the past)  such that $\zeta_{\lfloor t/c_{N}
  \rfloor}(N)/N$ converges uniformly to some bounded positive
  function.  Assuming a constant population size,   our calculations
  for Case~\ref{item:2} of Theorem~\ref{thm:diploid-alpha-random-all},
  and Case~\ref{item:9} of
  Theorem~\ref{thm:diploid-alpha-random-one-environment}  then show
  that   the ancestral process $\set{\xi^{n,N}}$   converges (in
  finite-dimensional distributions) 
  to  a time-varying
  $\Omega$-$\delta_{0}$-Beta$(\gamma_{t},2-\alpha,\alpha)$-coalescent.

\section{Conclusion}\label{sec:conclusion}

Our main results are \textit{(i)} continuous-time coalescents that are
either the Kingman coalescent or simultaneous multiple-merger
coalescents as specific families of Beta- or
Poisson-Dirichlet-coalescents \breyta{including}  an atom at zero; \textit{
(ii)} in arbitrarily large populations time is measured in units
proportional to either  $N/\log N$ or $N$  generations; \textit{(iii)}
it follows that  population size
changes (satisfying specific assumptions)   lead to time-changed coalescents where the  time-change  is independent of
the skewness parameter $\alpha$;  \textit{(iv)}
in
scenarios with increased effect of sweepstakes  (e.g.\
$\zeta(N)/N \to \infty$)  \breyta{the  annealed and quenched   empirical means}   $\overline \varrho
_{i}^{N}(n)$   and
$\overline \rho _{i}^{N}(n)$  are
quite  different.

Based on our construction of a population model where the population
evolves as in Definition~\ref{dschwpop} and incorporates sweepstakes 
 as in Definitions~\ref{dfn:alpharandomall} or
~\ref{df:alpha-random-one-environment} with an upper bound on the
number of potential offspring (recall~\eqref{eq:PXiJ}) the resulting
coalescent is driven by a measure of the form (recall~\eqref{eq:26} in  Definition~\ref{def:not})
\begin{displaymath}
\Xi = \delta_{0} +    \one{\frac{ \uN }{N} \gneqq  0} \Xi_{+}
\end{displaymath}
where  $\Xi_{+}$ is a finite measure on $\Delta_{+}$ (recall
~\eqref{eq:simplex}).  In   our formulation $\uN$ determines
if the limiting coalescent admits multiple mergers  or not.
For comparison, when the
$X_{1}, \ldots, X_{N}$ are allowed to be arbitrarily large and with a
regularly varying tail (recall~\eqref{eq:4}) as in~\cite[Equation~26]{BLS15}
then
$\Xi = \one{\alpha \ge 2}\delta_{0} + \one{1 < \alpha < 2}\Xi_{+}$
~\cite[Proposition~2.5]{BLS15}.    When 
$\Xi_{+}(\Delta_{+}) < 1$ one can assign the mass
$1 - \Xi_{+}(\Delta_{+})$ to $(0,\ldots)$~\citep[Section~1]{BLS15}.  We elect to
explicitly model scenarios as in Definitions~\ref{dfn:alpharandomall}
and~\ref{df:alpha-random-one-environment}  where most of the time small families are
generated (corresponding to most of the time sampling the zero atom
$(0, \ldots)$), but occasionally there is an increased chance of  large families
(corresponding to sampling elements from $\Delta_{+}$).~\breyta{Our
formulation enables us to compare  $\set{\xi^{n,N}}$ and
$\set{\xi^{n}}$, and  the annealed and quenched processes,  using the
empirical means   $\overline\varrho_{i}^{N}$, $\overline\varrho_{i}$,
and $\overline\rho_{i}^{N}$.}  In addition, we show that 
explicitly   incorporating  a random environment in the sense of
Definitions~\ref{dfn:alpharandomall} and
~\ref{df:alpha-random-one-environment}  alters the timescaling
(recall~\eqref{eq:scalecN} and~\eqref{eq:CNmap}).~\breyta{The unit of
time in these models has relevance for  explaining observed genetic
variation; one should be able to recover the  mutations used to  obtain
parameter  estimates for a given model without  unrealistic assumptions about  the
population size or the mutation rate.  
\begin{equation}
\label{eq:31}
\xymatrix @R=1pt @C=1pt @M=1pt {
& \text{1: use mutations to estimate $\alpha$}\\
*+[F-:<3pt>]{\text{mutations}} \ar@/^2pc/[rrr]|*+[F]{\text{model$(\alpha)$}}  &&  &  *+[F-:<3pt>]{\text{estimate }\widehat \alpha}  \ar@/^2pc/[lll]|*+[F]{\text{model$(\widehat\alpha)$}} \\
& \text{2: use $\widehat \alpha$ to recover mutations } \\
}
\end{equation}
}
\breyta{In~\eqref{eq:31} ``model''  refers to coalescent transition
rates,  the time-scaling $(1/c_{N})$  used for passing to the coalescent limit, and
assumptions about the  mutation mechanism.}
  Mutation rates (per site per
generation)  can be estimated using molecular
methods~\citep{Milholland2017,Xue2009,Wooldridge2025,Seplyarskiy2023,Wang2026}.

The models considered here (recall Definitions~\ref{dschwpop},
~\ref{dfn:alpharandomall}, and~\ref{df:alpha-random-one-environment})
can be a basis for constructing ancestral influence graphs for diploid
highly fecund populations~\citep{Koskela2019}, and so
for investigating the effects of elements such as population
structure, range expansion, recurrent bottlenecks, and natural
selection on genetic variation.  Moreover, the time-scaling
~\eqref{eq:scalecN} implies,  with gene genealogies now spanning 
(on average) time intervals proportional to  (at least) $N/\log N$
generations,  that  some components of~\eqref{eq:PXiJ} may vary over
time.~\breyta{ Our numerical results should encourage 
comparison of   processes as in 
\S~\ref{sec:incr-sample-size}, and mathematical studies of quenched (conditional) gene genealogies.}

\section{Proofs}\label{sec:proofs}

In this section we give proofs of
Theorems~\ref{thm:diploid-alpha-random-all} and
~\ref{thm:diploid-alpha-random-one-environment}.   First,  we record a
useful approximation of the bounds in~\eqref{eq:PXiJ}.
\begin{lemma}[\cite{Eldonhaploid2026},   bounds on $k^{-a} - {(1+k)}^{-a}$]\label{lm:boundskernel}
Suppose $0 < a \le 1$  and   $k\in\N$. Then
\begin{equation}
\label{eq:boundsmallera}
a {(1+k)}^{-1-a} \le k^{-a} - {(1 + k)}^{-a} \le  k^{-1-a}
\end{equation}
If  $a \ge 1$,  and $k\ge 2$,  it holds that
\begin{equation}
\label{eq:boundbiggera}
k^{-1-a} \le {(k-1)}^{-a} - k^{-a} \le  a {(k-1)}^{-1-a}
\end{equation}
\end{lemma}

The following lemma is a straightforward extension of~\cite[Lemma~6;
Equation~16]{schweinsberg03}; recall that, \breyta{if the  population size is
constant at $2N$ diploid individuals, it holds 
$\nu_{1} + \cdots + \nu_{N} = 2N$.}
\begin{lemma}[Relation between transition probabilities]\label{lm:dSchwlm6}
With $\nu_{1},\ldots, \nu_{N}$ and $X_{1},\ldots, X_{N}$ as in
Definition~\ref{def:not} we have, with $k_{1}, \ldots, k_{r} \ge 2$
and $k = k_{1} + \cdots + k_{r}$ for any $r\in \N$ (recall $S_{N}$
from~\eqref{SN})
\begin{equation}
\label{eq:dSchwlm6}
\lim_{N\to \infty}    \frac{{N}^{r}}{  {(2N)}^{k}   c_{N}}   \EE{ (\nu_{1})_{k_{1}}\cdots  (\nu_{r})_{k_{r}}   } =     \lim_{N\to \infty} \frac{ N^{r}  }{c_{N}} \EE{ \frac{ (X_{1})_{k_{1}}\cdots (X_{r})_{k_{r}}}{ S_{N}^{k_{1} + \cdots + k_{r}}  }\one{S_{N} \ge 2N} } 
\end{equation}
in the sense that when one of the limits exists so does the other, and
when they do exist they are equal.  Moreover, as $N\to \infty$,
recalling $c_{N}$ from Definition~\ref{def:cNhapl},
\begin{displaymath}
c_{N} =  \frac{N \EE{ \nu_{1}(\nu_{1} - 1) }}{2N(2N - 1) } \frac{1}{4}  \sim  \frac{1}{4} N\EE{ \frac{ X_{1}(X_{1} - 1)}{S_{N}^{2} }\one{S_{N} \ge 2N} }
\end{displaymath}
\end{lemma}
\begin{remark}
The existence of
$\lim_{N\to \infty}\svigi{N^{r-k}/c_{N}} \EE{ (\nu_{1})_{k_{1}}\cdots
(\nu_{r})_{k_{r}} } $ implies the existence of
\begin{displaymath}
\lim_{N\to \infty}\frac{ N^{r-k}}{ c_{N} } \EE{ (\nu_{1})_{k_{1}}\cdots
(\nu_{r})_{k_{r}} \nu_{r+1} \cdots \nu_{r+s} } 
\end{displaymath}
for any $s \in \N$~\cite[Lemma~3.5]{MS01}.
\end{remark}

\begin{propn}[\cite{Eldon2026}; approximating $\sum_{k} H(k)(G(k) - G(k+1))$]\label{pr:approxsum}
Suppose $G$ and $H$ are positive functions on $[1,\infty)$,  $H$ 
monotone increasing, $G$  monotone decreasing,  and $\int HG^{\prime}$
exists.  For $\ell,m\in \N$, \breyta{$\ell \le m$}, 
\begin{displaymath}
-\int_{\ell}^{m+1}H(x-1)G^{\prime}(x)dx \le \sum_{k=\ell}^{m}  H(k)(G(k) - G(k+1)) \le -\int_{\ell}^{m+1}H(x)G^{\prime}(x)dx
\end{displaymath}
\end{propn}


\begin{propn}[Bounding $\sum_{k} k(k-1){(k+M)}^{-2}\svigi{k^{-a} - (k+1)^{-a}}$]\label{prop:boundsumkM}
Write $s(a) \equiv \sum_{k=\ell}^{m} k(k-1)(k+M)^{-2}\svigi{ k^{-a} -
(k+1)^{-a}}$.~\breyta{Suppose $M>1$ is fixed.}
\begin{enumerate}
\item When $0 < a < 1$ or $1 < a < 2$ it holds that
\begin{equation}
\label{eq:casealone}
\frac{a }{(M-1)^{a}}\int_{ \frac{\ell }{\ell + M - 1 } }^{\frac{m+1 }{M+m } }u^{1-a}(1-u)^{a-1}du + O\svigi{\tfrac 1{M^{2}}}  \le s(a) \le \frac{a}{M^{a}}\int_{\frac{\ell}{\ell + M} }^{\frac{m+1 }{M+m+1 } }u^{1-a}(1-u)^{a-1}du +  O\svigi{\tfrac 1{M^{2}}} 
\end{equation}
\item When $a=1$ we have
\begin{equation}
\label{eq:caseaone}
\frac{m-\ell + 1 }{(M+\ell -1)(M+m) } + O\svigi{\frac{\log M }{M^{2}}  } \le s(a) \le \frac{m-\ell +1 }{(M+\ell)(M + m + 1)  } +   O\svigi{\frac{\log M }{M^{2}}  }
\end{equation}
\item When $a = 2$ we have
\begin{equation}
\label{eq:caseatwo}
s(a) = 2M^{-2}\log M +  O\svigi{M^{-2}}
\end{equation}
\item When $a > 2$ it holds that
\begin{equation}
\label{eq:caseabtwo}
\begin{split}
& \frac{a}{a-2}\frac{\ell^{2-a} }{(\ell + M - 1)^{2} } - \frac{3a}{a-1 }\frac{\ell^{1-a} }{(\ell + M - 1)^{2} } +  \frac{2\ell^{-a} }{(\ell + M-1)^{2}}    + O(m^{1-a}(m + M)^{-2}) + O(M^{-3}) \\
& \le s(a) \le \frac{a}{a - 2}\frac{ \ell^{2-a}}{ (\ell + M)^{2} }  -  \frac{a}{a-1}\frac{\ell^{1-a}}{(\ell + M)^{2} }   + O(m^{2-a}(m+ 1 + M)^{-2})   + O(M^{-a})
\end{split}
\end{equation}
\end{enumerate}

\end{propn}
\begin{proof}[Proof of Proposition~\ref{prop:boundsumkM}]
 Using Proposition~\ref{pr:approxsum}  with  $G(x) = x^{-a}$, and $H(x) = x(x-1)/{(x + M)}^{2}$, we see
 \begin{equation}
 \label{eq:boundssum}
 \int_{\ell}^{m+1}\frac{(x-1)(x-2) }{(x-1+M)^{2} }\frac{a}{x^{a+1}} dx  \le \sum_{k=\ell}^{m} \frac{k(k-1)}{(k+M)^{2} } \svigi{ \frac{1}{k^{a}} - \frac{1}{(1+k)^{a}}  } \le  \int_{\ell}^{m+1}\frac{x(x-1)}{(x+M)^{2}} \frac{a}{x^{a+1}}dx 
\end{equation}
Evaluating the integrals in~\eqref{eq:boundssum}   using standard
integral calculus techniques   finishes the proof.
\end{proof}

\begin{lemma}[$\prob{X_{1} + \cdots + X_{N} <  2N}$ vanishes]\label{lm:PSNltwoN}
Suppose $X_{1}, \ldots, X_{N}$ are independent non-negative
integer-valued random variables such that  $2 < \EE{X_{i}} < \infty $ for
all $i\in [N]$.   A positive constant $c < 1$ then exists such that
$\prob{X_{1} + \cdots + X_{N} < 2N} \le c^{N}$ for all $N \in \N$.
\end{lemma}
\begin{proof}[Proof of Lemma~\ref{lm:PSNltwoN}]
The proof follows the one  of~\cite[Lemma~5]{schweinsberg03}.  Write
$S_{N}\equiv X_{1}+\cdots + X_{N}$.  Let
$\rho: [0,1]\to [0,1]$ be given by $\rho(s) \equiv \EE{s^{X_{1}}}$ so
that $\rho^{\prime}(1) = \EE{X_{1}} > 2$ and $\rho(1) = 1$.  It then
holds that $\rho(r) < r^{2}$ for some number $r\in (0,1)$.   Then
$\EE{s^{S_{N} }} = \sum_{k}s^{k}\prob {S_{N} = k} \ge \sum_{0\le k \le
2N}s^{2N}\prob{S_{N} = k} = s^{2N}\prob{S_{N} \le 2N}$ so that
$\prob{S_{N}\le 2N} \le (s^{-2}\rho(s))^{N}$ for all $0 < s \le 1$.
Take $c= \rho(r)/r^{2}$.
\end{proof}

\subsection{Proof of  Theorem~\ref{thm:diploid-alpha-random-all}}\label{sec:proofthmall}

In this section we prove 
Theorem~\ref{thm:diploid-alpha-random-all}.  Recall
Definition~\ref{dfn:alpharandomall}, and the notation in
Definitions~\ref{def:notation} and~\ref{def:not}.  We first give
conditions on $\varepsilon_{N}$ for $m_{\infty}$ from~\eqref{eq:5} to
be finite.
\begin{lemma}[Finite $m_{\infty}$]\label{lma:finitemN}
Under the conditions of Theorem~\ref{thm:diploid-alpha-random-all} with
\begin{equation}
\label{eq:eNfinitemN}
\varepsilon_{N} \in \begin{cases} O\svigi{ \uN^{\alpha - 1}} & \text{when $0 < \alpha < 1$ } \\
O(1/\log \zeta(N)) & \text{when $\alpha = 1$ } \\
O(1) & \text{when $\alpha > 1$} 
\end{cases}
\end{equation}
as $N\to \infty$   it holds that     $\limsup_{N\to \infty}\EE{X_{1}} < \infty$.
\end{lemma}
\begin{proof}[Proof of Lemma~\ref{lma:finitemN}]
 When $0 < \alpha < 1$ the upper bound in
~\eqref{eq:boundsmallera}  gives, recalling~\breyta{${f_{\alpha}^{(\infty)}}$} from~\eqref{eq:isfg},
\begin{displaymath}
\EE{X_{1}| E} \le  {f_{\alpha}^{(\infty)}} \sum_{k=1}^{\uN} k^{-\alpha}  \le  {f_{\alpha}^{(\infty)}}  +   {f_{\alpha}^{(\infty)}} \int_{1}^{\uN} x^{-\alpha}dx =   {f_{\alpha}^{(\infty)}} + \frac{ {f_{\alpha}^{(\infty)}}}{1-\alpha } \svigi{ \uN^{1-\alpha} -1} 
\end{displaymath}
The boundedness of $\EE{X_{1}|E^{\sf c}}$ follows by similar
arguments.  By Definition~\ref{dfn:alpharandomall}
$\EE{X_{1}} = \EE{X_{1}| E} \varepsilon_{N} + \EE{X_{1}|E^{\sf
c}}(1-\varepsilon_{N})$, hence choosing $\varepsilon_{N}$ as in the
lemma the result follows.~\breyta{The other cases in
~\eqref{eq:eNfinitemN} are obtained using similar calculations.} 
\end{proof}
\breyta{\begin{lemma}[$S_{N}/(Nm_{N}) \to 1$ almost surely]\label{lm:almostsureconvSN}
Suppose $X_{1},\ldots, X_{N}$ for each $N\in \IN$  are i.i.d.\ and  distributed as in
Definition~\ref{dfn:alpharandomall} with $\varepsilon_{N}$ as in
Lemma~\ref{lma:finitemN} such that
$\limsup_{N\to\infty} m_{N} < \infty$ for each $N\in \IN$ (then $m_{N}$ converges to $m_{\infty} \in
(2,\infty)$  by~\eqref{eq:5} and Remark~\ref{rm:assumptionmN2+}).  It
holds  $S_{N}/(Nm_{N})\to 1$
almost surely.
\end{lemma}}
\begin{proof}[Proof of Lemma~\ref{lm:almostsureconvSN}]
\breyta{Recall $m_{N}$ from~\eqref{eq:5}, $S_{N}$ from~\eqref{SN}.  Write $\overline X_{i} := (X_{i} -
m_{N})/(Nm_{N})$ for all $i\in [N]$; for  $i>N$  choose $\overline
 X_{i}$ such that $\EE{\overline X_{i}} = 0$,   the  $\overline X_{i}$ be  independent, and
 $\EE{|\overline X_{i}|^{a}} \le  i^{-2}$ for some $1 < a < 2$.
It follows that
  $\lim_{N\to \infty}  \sum_{i=1}^{N} \EE{|\overline X_{i}|^{a}}/N^{a}
  $ is finite, all the $\overline X_{i}$ are independent,   and   so  $\sum_{i}\overline X_{i}\to 0$ almost surely
  \citep[Theorem~8.4.6]{athreya06:_measur}.   Since
$S_{N}/(Nm_{N}) = 1 + \sum_{i=1}^{N}\overline X_{i}$,  the lemma
follows.}
\end{proof}\breyta{\begin{remark}[Behaviour of $\zeta(N)$]
Lemma~\ref{lm:almostsureconvSN} holds   with 
$m_{N}$  finite for each $N$. However, we are interested in cases
where $m_{\infty}$ is finite. Therefore,  we  restrict the lemma to
cases where    $\limsup_{N}m_{N} < \infty$.   For $m_{\infty}$ to be finite we require 
$\zeta(N)$  not increase too quickly with $N$.   
Whenever  $\zeta(N)/N \to \infty$  in Case~\ref{item:2}  of   Theorem~\ref{thm:diploid-alpha-random-all}   with $1\le \alpha < 2$
we expect    $\zeta(N)$ to be at most of the form  $N^{x}$ for
some fixed  $x \ge  1$.   For example,   $\alpha = 1$  requires that 
 $x < 2a/(a-1)$,  $1 < a < 2$,     for   Lemma~\ref{lm:almostsureconvSN} to hold. 
\end{remark}}
We verify~\eqref{eq:scalecN} when $1 \le \alpha < 2$ under the
conditions of Theorem~\ref{thm:diploid-alpha-random-all}.
\begin{lemma}[Verifying~\eqref{eq:scalecN} when $1 \le \alpha < 2$]\label{lm:cNrandall}
Suppose the conditions of Theorem~\ref{thm:diploid-alpha-random-all}
hold,  and that $1\le \alpha < 2$.  Let $L \equiv L(N)$ be a positive  function of
$N$ with $L/N\to 0$ as $N\to \infty$.   Suppose
\breyta{$f_{a}^{(\infty)} = g_{a}^{(\infty)}$, and that 
$\underline {g_{a}}(2) = \overline {f_{a}}(2)$
(recall~\eqref{eq:isfg})}. Take, \breyta{for some $c > 0$ fixed
(recall $\one{\cdot}$ from~\eqref{eq:one}, $S_{N}$ from~\eqref{SN})}
\begin{equation}
\label{eq:varepsilon}
\varepsilon_{N}  =  cN^{\alpha - 2}\left( \one{\kappa > 2} +  \one{\kappa = 2}\log N\right)
\end{equation}
     Then~\eqref{eq:scalecN} is in force, i.e.\  $\lim_{N\to
     \infty}C_{\kappa}^{N}c_{N} = C_{\alpha,\gamma}$ with
     $C_{\alpha,\gamma}$ as in~\eqref{eq:10}.
\end{lemma}
\begin{proof}[Proof of Lemma~\ref{lm:cNrandall}]
 Fix   $0<\delta < 1$ such that
$(1-\delta)m_{N} > 2$ (recall~\eqref{eq:5} and
Remark~\ref{rm:assumptionmN2+}). Define, for each $N\in \IN$, 
\begin{equation}
\label{eq:YMR}
\begin{split}
M_{+} &  :=  (1+\delta)Nm_{N}, \quad  M_{-} := (1-\delta)Nm_{N}, \\
Y_{+} & :=  \frac{X_{1}(X_{1}-1)}{ {(X_{1} + M_{+})}^{2} }, \quad  Y_{-}  :=  \frac{X_{1}(X_{1}-1)}{ {(X_{1} + M_{-})}^{2} }, \\
R_{N} & :=  \one{S_{N}\ge 2N}    X_{1}(X_{1}-1) / S_{N}^{2}.
\end{split}
\end{equation}
Fix $\epsilon > 0$. Choosing $\varepsilon_{N}$ (recall
Definition~\ref{dfn:alpharandomall}) as in~\eqref{eq:eNfinitemN} gives
$m_{\infty} < \infty$ by Lemma~\ref{lma:finitemN}.  Using
Lemma~\ref{lm:almostsureconvSN} we can adapt the arguments in the
proof of~\cite[Lemma~13]{schweinsberg03} to obtain
\begin{equation}
\label{eq:boundERN}
(1-\epsilon)\EE{Y_{+}}   \le  \EE{R_{N}} \le  \epsilon \EE{\frac{(X_{1})_{2} }{ \max\set{X_{1}^{2}, 4N^{2} }  } } +  \EE{ Y_{-} }
\end{equation}
We will use Proposition~\ref{pr:approxsum} with $H(x) = x(x-1)(x+M)^{-2}$ and
$G(x) = x^{-a}$ with $M$ either $M_{+}$ or $M_{-}$ from
~\eqref{eq:YMR} and $a$ as given each time to approximate $\EE{Y_{-}}$
and $\EE{Y_{+}}$. First,  we check  that
\begin{equation}
\label{eq:claimboundedEXmax}
\limsup_{N\to\infty} C_{\kappa}^{N} N  \EE{ \frac{ (X_{1})_{2} }{ \max\set{X_{1}^{2}, 4N^{2}} } }  <  \infty
\end{equation}
with  $\alpha \ge 1$.    Since $(X_{1} + 2N)^{2} \le 4\max\set{X_{1}^{2}, 4N^{2}}$ we 
have
\begin{displaymath}
\EE{\frac{ (X_{1})_{2}}{ \max\set{X_{1}^{2}, 4N^{2}}} } \le 4 \EE{ \frac{  (X_{1})_{2}}{ (X_{1} + 2N)^{2}} }
\end{displaymath}
Using the upper bound in~\eqref{eq:boundbiggera} in
Lemma~\ref{lm:boundskernel} we see, recalling event $E$ from 
Definition~\ref{dfn:alpharandomall},
\begin{displaymath}
\EE{ \frac{(X_{1})_{2} }{(X_{1} + 2N)^{2} } |E } \le \alpha  {f_{\alpha}^{(\infty)}} \sum_{k=2}^{\uN} \frac{k(k-1)}{(k+2N)^{2} }\frac{1}{k^{1+\alpha}} \le  {f_{\alpha}^{(\infty)}} \sum_{k=2}^{\uN} \frac{1}{(k+2N)^{2} } \le  {f_{\alpha}^{(\infty)}}  \int_{1}^{\uN} \frac{1}{(x + 2N)^{2}} dx 
\end{displaymath} 
when $\alpha = 1$;~\eqref{eq:claimboundedEXmax}   follows with $\varepsilon_{N}$ as in~\eqref{eq:varepsilon}.

We approximate $\EE{Y_{-}}$ with $Y_{-}$ as in~\eqref{eq:YMR}.  We
see, using Proposition~\ref{pr:approxsum} with
$G^{\prime}(x) = -\alpha x^{-\alpha -1}$, and $H(x) = x(x-1)(x + M_{-})^{-2}$,
\begin{displaymath}
\begin{split}
\EE{Y_{-} | E} & \le  \overline {f_{\alpha}}(2) \sum_{k=2}^{\uN} \frac{k(k-1)}{(k+M_{-})^{2}} \svigi{ \frac{1 }{k^{\alpha}} - \frac{1}{(1+k)^{\alpha}} }    \le \alpha \overline {f_{\alpha}}(2) \int_{2}^{\uN + 1} \frac{x(x-1)}{(x+M_{-})^{2}}\frac{1}{x^{1+\alpha}}dx \\ 
& \le   \frac{\alpha \overline {f_{\alpha}} (2) }{M_{-}^{\alpha}} \int_{ \tfrac{2}{2 + M_{-}} }^{ \tfrac{\uN + 1}{\uN + 1 + M_{-} }    } u^{1-\alpha}(1-u)^{\alpha - 1}du    +  O(N^{-2})
 \end{split}
\end{displaymath}
using the substitution $y = M_{-}/(x + M_{-})$ (so that
$x= M_{-}y^{-1} - M_{-}$ and $dx = -M_{-}y^{-2}dy$) on
$\int x^{1-\alpha}(x + M_{-})^{-2}dx $ and noting that  $\int (x +
M_{-})^{-2}x^{-2} dx \le \int (x + M_{-})^{-2}x^{-\alpha} dx $  over
$[1,\infty)$  when $1 \le \alpha < 2$.  Write (recall $K > 0$ is fixed) 
\begin{equation}
\label{eq:gammahat}
\begin{split}
\overline\gamma &  :=   \one{ \tfrac \uN N \to K }\frac{K }{K +  (1-\delta)m_{\infty}} +  \one{\tfrac \uN N \to \infty } \\
\widehat \gamma &  :=     \one{ \tfrac \uN N \to K }\frac{K }{K +  (1+\delta)m_{\infty}} +  \one{\tfrac \uN N \to \infty }
\end{split}
\end{equation}
Using the  assumption $L/N\to 0$ with $\varepsilon_{N}$ as in~\eqref{eq:varepsilon} and $C_{\kappa}^{N}$ as in~\eqref{eq:CNmap} we obtain 
\begin{displaymath}
\limsup_{N\to \infty} C_{\kappa}^{N}  N\EE{Y_{-} | E}\varepsilon_{N} \le  \frac{\alpha c  f_{\alpha}^{(\infty)} }{ ((1-\delta)m_{\infty})^{\alpha} } \int_{0}^{1} \one{ 0 < u \le \overline \gamma}  u^{1-\alpha}(1-u)^{\alpha - 1}du 
\end{displaymath}
  Analogous calculations give (recall $1 \le \alpha < 2$ by assumption) 
\begin{displaymath}
\liminf_{N\to \infty} C_{\kappa}^{N} N\EE{Y_{+} | E}\varepsilon_{N}  \ge \frac{ \alpha c g_{\alpha}^{(\infty)} }{ ((1+\delta)m_{\infty})^{\alpha}  } \int_{0}^{1} \one{0 < u \le \widehat \gamma} u^{1-\alpha}(1-u)^{\alpha - 1}du
\end{displaymath}

When $\kappa = 2$ we see, using Proposition~\ref{pr:approxsum} with $H(x)= x(x-1)(x + M_{-})^{-2}$ and $G(x) = x^{-2}$,  
\begin{displaymath}
\begin{split}
\EE{Y_{-} | E^{\sf c}} &   \le  \overline {f_{\kappa}}(2) \sum_{k=2}^{\uN} \frac{k(k-1)}{(k + M_{-})^{2}}\svigi{\frac{1}{k^{2}} - \frac{1}{(1+k)^{2}} }   \le  2 \overline {f_{\kappa}}(2) \int_{2}^{\uN + 1} \frac{x(x-1) }{(x+M_{-})^{2}} \frac{1}{x^{3}}dx    \\
\end{split}
\end{displaymath}
Integration by partial fractions gives
\begin{displaymath}
\begin{split}
\int_{2}^{\uN + 1} \frac{1 }{x(x+ M_{-})^{2}}dx &  =  \frac{1}{M_{-}^{2}} \int_{2}^{\uN + 1}\left( \frac{1}{x} - \frac{1}{x + M_{-}} - \frac{M_{-}}{(x+ M_{-})^{2}} \right)dx  \\
& = \frac{1}{M_{-}^{2}}\left( \log \frac{\uN + 1 }{M_{-}+ \uN + 1 } + \log \frac{2 + M_{-}}{2} +  \frac{ (1 - \uN)M_{-} }{(\uN + 1 +M_{-})(M_{-} +2 ) } \right)
\end{split}
\end{displaymath}
and we conclude, recalling $C_{\kappa}^{N}$ from~\eqref{eq:CNmap} and that  $\kappa = 2$, 
\begin{displaymath}
\limsup_{N\to \infty} C_{\kappa}^{N} N \EE{Y_{-}| E^{\sf c} }(1 - \varepsilon_{N})  \le  2\overline {f_{\kappa}}(2) ((1-\delta)m_{\infty})^{-2}
\end{displaymath}
after checking that   $\limsup_{N\to \infty} C_{\kappa}^N N \int_{2}^{\uN}x^{-2}(x+M_{-})^{-2}dx  = 0$  using integration by parts  (where   $\int x^{-2}(x+ M_{-})^{-2}dx  =    2M_{-} ^{-3}\log( 1 + M_{-}/x)    -  (M_{-} + 2x)/(M_{-}^{3}x + M_{-}^{2}x^{2}) + c$).
Again using  Proposition~\ref{pr:approxsum} with $H(x)= x(x-1)(x + M_{+})^{-2}$ and  $G(x) = x^{-2}$ we see
\begin{displaymath}
\begin{split}
\EE{Y_{+} | E^{\sf c}} & \ge  \underline {g_{\kappa}}(2) \sum_{k=2}^{\uN} \frac{k(k-1) }{(k + M_{+})^{2}} \left( \frac{1}{k^{2}} - \frac{1}{(1+k)^{2}}\right)  \ge  2\underline {g_{\kappa}}(2) \int_{2}^{\uN + 1} \frac{(x-1)(x-2)}{(x+ M_{+} - 1)^{2}}\frac{1}{x^{3}}dx  \\
& \ge 2\underline {g_{\kappa}}(2) \int_{2}^{\uN + 1} \frac{x(x-1) }{(x+M_{+} - 1)^{2}}\frac{1}{x^{3}}dx -   4\underline {g_{\kappa}}(2) \int_{2}^{\uN + 1} \frac{x-1 }{(x + M_{+} - 1)^{2}}\frac{1}{x^{3}}dx 
\end{split}
\end{displaymath}
and we can conclude that, when $\kappa = 2$,  
\begin{displaymath}
\liminf_{N\to \infty} C_{\kappa}^{N} N \EE{Y_{+}|E^{\sf c}}(1-\varepsilon_{N}) \ge  2\underline {g_{\kappa}}(2)((1-\delta)m_{\infty})^{-2}
\end{displaymath}

When $\kappa > 2$ we see, using Proposition~\ref{pr:approxsum}, 
\begin{displaymath}
\EE{Y_{-} | E^{\sf c}}  \le \kappa  \overline {f_{\kappa}}(2) \int_{2}^{\uN + 1}\frac{x(x-1) }{(x + M_{-})^{2}}\frac{1}{x^{1+\kappa}} dx 
\end{displaymath}
Suppose  $2 < \kappa <3$.  Integration by parts and the substitution $y = M_{-}/(x + M_{-})$ then give 
\begin{displaymath}
\begin{split}
\int_{2}^{\uN + 1}\frac{x^{1 - \kappa}}{(x+ M_{-})^{2}}dx & =  \frac{1}{2-\kappa}\left[ \frac{x^{2-\kappa}}{(x + M_{-})^{2}} \right]_{2}^{\uN + 1} - \frac{2}{\kappa - 2} \int_{2}^{\uN + 1} \frac{ x^{2-\kappa}}{(x + M_{-})^{3} }dx  \\
& =  \frac{2^{2-\kappa}}{\kappa -2 }\frac{1}{(2 + M_{-})^{2}}  + O\left( \frac{ \uN^{2-\kappa}}{(\uN + M_{-})^{2} } \right) + O\left( M_{-}^{-\kappa} \right)
\end{split}
\end{displaymath}
Iterating the calculation  for  $\int x^{-\kappa}(x+ M_{-})^{-2}dx$ we conclude 
\begin{displaymath}
\limsup_{N\to \infty} C_{\kappa}^{N} N\EE{Y_{-} | E^{\sf c}}(1-\varepsilon_{N}) \le   \frac{\kappa \overline {f_{\kappa}}(2)}{((1-\delta)m_{\infty})^{2}}\svigi{ \frac{ 2^{2-\kappa}  }{\kappa - 2 } -  \frac{2^{1-\kappa} }{\kappa - 1 } } =  \frac{2^{1-\kappa}\kappa^{2} \overline {f_{\kappa}}(2) }{ ((1-\delta)m_{\infty})^{2} (\kappa - 2)(\kappa -1) }
\end{displaymath}
Similarly, when $2 <\kappa < 3$, again using  Prop~\ref{pr:approxsum}, 
\begin{displaymath}
\begin{split}
\EE{Y_{+}| E^{\sf c}} &  \ge \kappa \underline {g_{\kappa}}(2) \int_{2}^{\uN + 1} \frac{(x-1)(x-2) }{(x + M_{+} -1)^{2}}\frac{1}{x^{1+\kappa}}dx  =   \kappa \underline {g_{\kappa}}(2) \int_{2}^{\uN + 1} \frac{x^{2} -3x + 2 }{(x + M_{+} -1)^{2}}\frac{1}{x^{1+\kappa}}dx      \\
\end{split}
\end{displaymath}
and we obtain
\begin{displaymath}
\liminf_{N\to \infty}C_{\kappa}^{N} N\EE{Y_{+} | E^{\sf c}}(1-\varepsilon_{N}) \ge  \frac{ 2^{1-\kappa} \underline {g_{\kappa}}(2)  }{((1+\delta)m_{\infty})^{2} }\frac{ \kappa + 2  }{(\kappa -2)(\kappa -1) }
\end{displaymath}
One checks in the same way that the same leading terms hold for $\kappa \ge 3$.  
The lemma now follows from~\eqref{eq:6} and Lemma~\ref{lm:dSchwlm6}   after  taking $\epsilon$ and $\delta$ to 0.
\end{proof}

\begin{lemma}[Convergence to the Kingman coalescent]\label{lm:proofconvkingmantypeA}
Under the conditions of Case~\ref{item:1} of
Theorem~\ref{thm:diploid-alpha-random-all} it holds that
\begin{displaymath}
\limsup_{N\to \infty}  \frac{N}{c_{N}}\EE{ \frac{ {(X_{1})}_{3} }{S_{N}^{3}  } \one{S_{N} \ge 2N} } = 0
\end{displaymath}
\end{lemma}
\begin{proof}[Proof of Lemma~\ref{lm:proofconvkingmantypeA}]
On $S_{N} \ge 2N$ we have
$S_{N}^{3} \ge \max\set{X_{1}^{3}, {(2N)}^{3} }$ so that, on $E$
~\breyta{(recall Definition~\ref{dfn:alpharandomall})} and
with $0 < \alpha < 1$, using the upper bound in
~\eqref{eq:boundsmallera} in Lemma~\ref{lm:boundskernel}, and that
$|x+y|^{p} \le 2^{p-1}\svigi{|x|^{p} + |y|^{p}}$ for reals $x,y$ and $p \ge 1$~\cite[Proposition~3.1.10{\it (iii)}, Equation~1.12]{athreya06:_measur} 
\begin{displaymath}
\begin{split}
\frac{N}{c_{N}} \EE{  \frac{ (X_{1})_{3} }{S_{N}^{3}  } \one{S_{N} \ge 2N}  |E   } \varepsilon_{N} \le  \frac{ {f_{\alpha}^{ (\infty)}} \varepsilon_{N} }{8N^{2}c_{N}} \sum_{k=3}^{\uN} \frac{k^{3}}{ (k-1)^{1+\alpha}  }  \le    \frac{ {f_{\alpha}^{ (\infty)}} \varepsilon_{N} }{2N^{2}c_{N}} \sum_{k=3}^{\uN} \frac{ (k - 1)^{3} + 1}{ (k-1)^{1+\alpha}  } 
\end{split}
\end{displaymath}
and the result follows when $0<\alpha < 1$ assuming
$\limsup_{N\to\infty}\varepsilon_{N}/c_{N} < \infty$  as $N\to \infty$. The result when
$1 \le \alpha < 2$ follows from a similar calculation.~\linelabel{K}\breyta{Convergence to
the Kingman coalescent follows from  Lemma~\ref{lm:dSchwlm6} 
and   \citep[Proposition~2]{schweinsberg03} (see~\cite[Section~4]{mohle2000total}).}
\end{proof}

We check that the probability of two or more large families vanishes
in a large population when $1 \le \alpha < 2$.  However, as we will
see, diploidy ensures that we retain the possibility of (up to 4-fold)
simultaneous mergers.
\begin{lemma}[The probability of two or more large families vanishes when $1\le \alpha < 2$]\label{proof:twopluslargefams}
Under the conditions of Case~\ref{item:2} of Theorem~\ref{thm:diploid-alpha-random-all} it holds that 
\begin{displaymath}
\lim_{N\to \infty} \frac{ N^{2}}{c_{N}} \EE{ \frac{(X_{1})_{2} (X_{2})_{2}} {S_{N}^{4}}\one{S_{N}\ge 2N} } = 0
\end{displaymath}
\end{lemma}
\begin{proof}[Proof of Lemma~\ref{proof:twopluslargefams}]
We see, recalling that the $X_{1}, \ldots, X_{N}$ \breyta{for each
$N\in \IN$}  are i.i.d.,   
\begin{displaymath}
\begin{split}
 &  \frac{N^{2}}{c_{N}} \EE{\frac{(X_{1})_{2}(X_{2})_{2} }{ S_{N}^{4}  } \one{S_{N} \ge 2N} }\le  \frac{N^{2}}{c_{N}}\EE{\frac{(X_{1})_{2}(X_{2})_{2} }{ \max\set{X_{1}^{2},4N^{2}}\max\set{X_{2}^{2}, 4N^{2} } }\one{S_{N} \ge 2N} } \\
& \le  \frac{N^{2}}{c_{N}}\left( \EE{\frac{(X_{1})_{2}}{\max\set{X_{1}^{2}, 4N^{2}} } } \right)^{2} \le  16 \frac{N^{2}}{c_{N}}\left( \EE{\frac{(X_{1})_{2}}{(X_{1} + 2N)^{2}} }  \right)^{2}
\end{split}
\end{displaymath}
Recalling event $E$ from Definition~\ref{dfn:alpharandomall} we see, when $1 \le \alpha < 2$,  
\begin{displaymath}
\EE{\frac{(X_{1})_{2} }{(X_{1} + 2N)^{2}} | E } \le  {f_{\alpha}^{ (\infty)}} \sum_{k=2}^{\uN} \frac{k(k-1)}{(k + 2N)^{2}}\svigi{\frac{1}{k^{\alpha}} - \frac{1}{(1+k)^{\alpha}} }  \le \alpha     {f_{\alpha}^{ (\infty)}} \int_{1}^{\uN} \frac{k(k-1)}{(x+ 2N)^{2}k^{1+\alpha} }dx \overset c \sim \frac 1 {N^{\alpha}}
\end{displaymath}
as $N\to \infty$ (recall $\uN/N \gneqq 0$), using
~\eqref{eq:boundbiggera} in Lemma~\ref{lm:boundskernel} and the
substitution $y = 2N/(x + 2N)$ when $1 < \alpha < 2$.  The result now
follows from Lemma~\ref{lm:cNrandall} with $\varepsilon_{N}$ as in
~\eqref{eq:varepsilon}.
 \end{proof}
\begin{lemma}[Identifying the  measure $\Xi_{+}$]\label{lm:Fj}
Under the conditions of Case~\ref{item:2} of Theorem~\ref{thm:diploid-alpha-random-all} we have
\begin{displaymath}
\lim_{N\to \infty} \frac{N}{c_{N}} \prob{\frac{X_{1}}{S_{N}}\one{S_{N} \ge 2N} \ge x } =  \frac{\alpha c f_{\alpha}^{(\infty)} }{ C_{\kappa,\alpha,\gamma}m_{\infty}^{\alpha}  } \int_{x}^{1}\one{0 < u \le \gamma }u^{-1-\alpha} {(1-u)}^{\alpha - 1}du
\end{displaymath}
for $0<x<1$ and with  $C_{\kappa,\alpha, \gamma}$  as in~\eqref{eq:10} and $\gamma$ as in~\eqref{eq:gamma} and  $c$ from~\eqref{eq:varepsilon} and $f_{\infty}$ as in~\eqref{eq:isfg}. 
\end{lemma}
\begin{proof}[Proof of Lemma~\ref{lm:Fj}]
Let $\epsilon, \delta > 0$ be fixed with $(1-\delta)m_{N} > 2$ for all
$N$.  Then, by Lemma~\ref{lm:almostsureconvSN} we have for all  $N$ large enough 
that, with $\widetilde S_{N}$ as in~\eqref{SN2} in Definition~\ref{def:not}, 
\begin{displaymath}
\prob{M_{-} \le  \widetilde S _{N} \le  M_{+}} > 1-\epsilon
\end{displaymath}
is in force where $M_{-}$ and $M_{+}$ are as in~\eqref{eq:YMR}.    Then, for
any $0 < x < 1$,
\begin{displaymath}
(1-\epsilon)\prob{ \frac{X_{1}}{X_{1} + M_{+}} \ge x  } \le \prob{\frac{X_{1}}{S_{N}}\one{S_{N}\ge 2N } \ge x } \le  \epsilon\prob{X_{1} \ge 2Nx } + \prob{ \frac{ X_{1}  }{X_{1} + M_{-}} \ge x  }
\end{displaymath}
using the arguments for~\cite[Equations~44 and 45 in
Lemma~14]{schweinsberg03}.  Recalling event $E$ from
Definition~\ref{dfn:alpharandomall} we see
\begin{displaymath}
\begin{split}
& \prob{\frac{X_{1}}{X_{1} + M_{+}} \ge x | E }   =  \prob{X_{1} \ge  \frac{x}{1-x}M_{+} | E }  \ge   \underline{g_{\alpha}}\svigi{ \frac{x}{1-x}M_{+}}  \svigi{ \svigi{\frac{1-x}{x}}^{\alpha} M_{+}^{-\alpha} - (1+\uN)^{-\alpha} } \\
& =  \frac{  \underline{g_{\alpha}}\svigi{ \frac{x}{1-x}M_{+}} }{ M_{+}^{\alpha} }\svigi{ \svigi{\frac{1-x }{x} }^{\alpha} -  \svigi{ \frac{ M_{+}   }{1+ \uN }  }^{\alpha} }    \\
\end{split}
\end{displaymath}
For $a > 0 $ and $0 < x \le  1/(1+c)$ for any constant $c \ge 0$  we
have (cf.\ \citep{Eldon2026})
\begin{equation}
\label{eq:Fintegral}
\svigi{\frac{1-x}{x}}^{a} - c^{a} =  a\int_{x}^{\frac {1}{1+c}} u^{-1-a}(1-u)^{a-1}du
\end{equation}
seen using the substitution $y = (1-u)/u$.   Write $K_{+}^{N} :=  M_{+}/(\uN + 1)$ so that
\begin{displaymath}
\lim_{N\to \infty} \frac 1 {1+K_{+}^{N}} =   \widehat \gamma
\end{displaymath}
where $\widehat \gamma$ is as in~\eqref{eq:gammahat}.  By
Lemma~\ref{lm:cNrandall} with $\varepsilon_{N}$ as in
~\eqref{eq:varepsilon} we have (recall $C_{\kappa,\alpha,\gamma}$ from~\eqref{eq:10})
\begin{displaymath}
\begin{split}
& \liminf_{N\to \infty} \frac{N}{c_{N}}  \prob{ \frac{ X_{1} }{X_{1} + M_{+} } \ge x | E } \varepsilon_{N} \ge \liminf_{N\to \infty} \frac{\alpha \underline {g_{\alpha}}\svigi{ \frac x{1-x}M_{+} } }{ ((1+\delta)m_{N})^{\alpha} }         \frac{N^{1-\alpha}\varepsilon_{N} }{c_{N}}\int_{x}^{ \frac{1}{1+K_{+}^{N}} } u^{-1-\alpha}(1-u)^{\alpha - 1}du  \\
& =   \frac{\alpha c g_{\alpha}^{(\infty)}}{C_{\kappa,\alpha,\gamma}((1+\delta)m_{\infty})^{\alpha} } \int_{x}^{1}\one{0<x \le \widehat \gamma }u^{-1-\alpha}(1-u)^{\alpha - 1}du =     \frac{\alpha c g_{\alpha}^{(\infty)}B(\widehat \gamma,2-\alpha,\alpha) }{C_{\kappa,\alpha,\gamma}((1+\delta)m_{\infty})^{\alpha} }  \int_{x}^{ \widehat \gamma }u^{-2} \text{Beta}(\widehat \gamma,2-\alpha,\alpha) du
\end{split}
\end{displaymath}
where Beta$(p, 2-\alpha,\alpha) $ is the  law on $(0,1]$ with density as in \eqref{eq:3}. 
  Analogous calculations give
\begin{displaymath}
\begin{split}
&  \limsup_{N\to \infty}\frac{N}{c_{N}}\prob{\frac{X_{1} }{X_{1} + M_{-} } \ge x | E } \le \limsup_{N\to \infty} \frac{\alpha \overline {f_{\alpha}}\svigi{\frac{x}{1-x}M_{-} } }{((1-\delta)m_{N})^{\alpha} } \frac{N^{1-\alpha}\varepsilon_{N} }{c_{N}} \int_{x}^{ \frac{1}{1+K_{-}^{N}} } u^{-1-\alpha}(1-u)^{\alpha - 1}du  \\
& =   \frac{\alpha c f_{\alpha}^{(\infty)} }{C_{\kappa,\alpha,\gamma}((1-\delta)m_{\infty})^{\alpha} } \int_{x}^{1}\one{0<x \le \overline \gamma }u^{-1-\alpha}(1-u)^{\alpha - 1}du =     \frac{\alpha c f_{\alpha}^{(\infty)}B(\overline \gamma,2-\alpha,\alpha) }{C_{\kappa,\alpha,\gamma}((1+\delta)m_{\infty})^{\alpha} }  \int_{x}^{ \overline \gamma }u^{-2} \text{Beta}(\overline \gamma,2-\alpha,\alpha) du
\end{split}
\end{displaymath}
where $K_{-}^{N} =  M_{-}/(\uN + 1)$ so that $\lim_{N\to \infty} 1/(1 + K_{-}^{N}) = \overline \gamma$ with  $\overline \gamma$ as in  \eqref{eq:gammahat}. The lemma follows after taking $\epsilon$ and $\delta$ to 0.   
\end{proof}

\linelabel{ProofThmall}\begin{proof}[Proof of Theorem~\ref{thm:diploid-alpha-random-all}]
Case~\ref{item:1} follows from Lemmas~\ref{lm:proofconvkingmantypeA} and 
~\ref{lm:dSchwlm6}, and from~\cite[Theorem~5.4; Equation~2]{MS03}.
The proof of Case~\ref{item:2} follows as in the proof of
~\cite[Theorem~1.1]{BLS15} (in turn based on the results of
~\cite{M98}); we only note that (recall $|A|$ is the number of elements
in a given finite  set $A$)
\begin{displaymath}
\prob{\xi^{n,N}(m) \to  {\sf cd}\svigi{\xi^{n,N}(m) } } \ge 1 - \binom{  | \xi^{n,N}(m) |   }{2}c_{N}
\end{displaymath}
where   $c_{N}$ is the coalescence
probability as in Defition~\ref{def:cNhapl} (recall
~\eqref{eq:6}). Since $c_{N}\to 0$ by Lemmas~\ref{lm:cNrandall} and
~\eqref{eq:dSchwlm6}, and $ | \xi^{n,N}(m) | \le 2n < \infty$,   so that
$\prob{\xi^{n,N}(m) \to {\sf cd}\svigi{\xi^{n,N}(m) } } \to 1 $ as
$N\to \infty$,  complete dispersion of ancestral blocks paired in
the same diploid individual occurs instantaneously in the limit, and
the limiting process is $\mathcal E_{2n}$-valued \breyta{(the set of
partitions of $[2n]$)}.

By Lemma~\ref{lm:cNrandall} and Lemma~\ref{lm:dSchwlm6} 
\begin{displaymath}
\lim_{N\to \infty} \frac{1}{ 4 Nc_{N}   }\EE{ (\nu_{1})_{2}  } = 1
\end{displaymath}
and the limits in~\eqref{eq:14} are zero for all
$r \in \N \setminus \set{1}$ by Lemma~\ref{proof:twopluslargefams},
the limits in~\eqref{eq:14} hold for all $j\in \N$. 
\breyta{Fix $0 < x < 1$, 
and $0 < \epsilon < x$. Then, by the arguments for~\citep[Theorem~4c]{schweinsberg03}, 
\begin{displaymath}
\begin{split}
\limsup_{N\to \infty} \frac{N}{c_{N}}\prob{\nu_{1} > 2Nx} &  \le \limsup_{N\to \infty}\frac{N}{c_{N}}\prob{\frac{X_{1}}{S_{N}}\one{S_{N}\ge 2N} \ge x - \epsilon}, \\
\liminf_{N\to \infty} \frac{N}{c_{N}}\prob{\nu_{1} > 2Nx} & \ge \liminf_{N\to \infty}\frac{N}{c_{N}}\prob{\frac{X_{1}}{S_{N}}\one{S_{N}\ge 2N} \ge x + \epsilon}.
\end{split}
\end{displaymath}}
\breyta{Then, after taking $\epsilon \to 0$, it holds  $\lim_{N\to \infty} (N/c_{N})\prob{ \nu_{1} > 2Nx} =
\int_{x}^{1}u^{-2}F_{1}(du) $,  where $F_{1}$ is the
Beta$(\gamma, 2-\alpha,\alpha)$-measure by Lemma~\ref{lm:Fj}.   Furthermore,}
\begin{displaymath}
\lim_{N\to \infty} \frac{N^{j}}{c_{N}} \prob{ \nu_{1} > 2Nx_{1}, \ldots , \nu_{j} > 2N x_{j}   } = 0
\end{displaymath}
for all $j\in \set{2,3,\ldots}$ by Lemma~\ref{proof:twopluslargefams}
and~\cite[Equation~18]{MS01}.  Hence, the limits in~\eqref{eq:15} hold
for all points of continuity for the measures $F_{r}$ for all
$r \in \N$.  Thus, 
$\mengi{ \xi^{n,N}\svigi{ \lfloor t/c_{N} \rfloor}; t \ge 0 } \to
\set{ \xi^{n}(t); t \ge 0 } $ in the sense of convergence of  finite dimensional
distributions.

We turn to Case~\ref{item:8} of
Theorem~\ref{thm:diploid-alpha-random-all}.    The proof follows
the one of~\cite[Theorem~4d]{schweinsberg03} (see also  the proof of
~\cite[Theorem~3.5, Case~3]{Eldonhaploid2026}).    Suppose    
$0 < \alpha < 1$,    and $\uN/N^{1/\alpha} \to \infty$.   Let
$Y_{(1)} \ge Y_{(2)} \ge \ldots \ge Y_{(N)}$ be the ranked values of
$N^{-1/\alpha}X_{1},\ldots, N^{-1/\alpha}X_{N}$.  Write
$p_{i} \equiv \prob{x_{i} \le Y_{1} < x_{i-1} | E } = \prob{X_{1} \ge
N^{1/\alpha}x_{i} | E } - \prob{X_{1} \ge N^{1/\alpha}x_{i-1} | E }
$.  Using~\eqref{eq:12}  we have, for $1 \le i \le  j$, recalling~\eqref{eq:isfg}, 
\begin{displaymath}
\begin{split}
& \frac{\underline {g_{\alpha}} \svigi{ N^{1/\alpha}x_{i}} }{ N  }\svigi{x_{i}^{-\alpha} - \svigi{ \frac{N^{1/\alpha} }{\uN + 1 }  }^{\alpha} } -  \frac{\overline {f_{\alpha}}\svigi{ N^{1/\alpha}x_{i-1}} }{ N  }\svigi{x_{i-1}^{-\alpha}  -  \svigi{ \frac{N^{1/\alpha} }{\uN + 1 }  }^{\alpha} }  \\
& =    \frac 1N  \svigi{ \underline {g_{\alpha}} \svigi{N^{1/\alpha} x_{i} }x_{i}^{-\alpha} -  \overline {f_{\alpha}} \svigi{N^{1/\alpha}x_{i-1} }x_{i-1}^{-\alpha}  } + O\svigi{ \uN^{-\alpha} }   \\
& \le  p_{i} \le  \frac{ \overline {f_{\alpha}}\svigi{N^{1/\alpha}x_{i} } }{N}\svigi{ x_{i}^{-\alpha} -  \svigi{ \frac{N^{1/\alpha}  }{\uN + 1 }  }^{\alpha}  } -    \frac{ \underline {g_{\alpha}}\svigi{N^{1/\alpha}x_{i-1} } }{N}\svigi{ x_{i-1}^{-\alpha} -  \svigi{ \frac{N^{1/\alpha}  }{\uN +1 }  }^{\alpha}  } \\
& =    \frac 1N  \svigi{ \overline {f_{\alpha}} \svigi{N^{1/\alpha} x_{i} }x_{i}^{-\alpha} -  \underline {g_{\alpha}} \svigi{N^{1/\alpha}x_{i-1} }x_{i-1}^{-\alpha}  } + O\svigi{ \uN^{-\alpha} }   \\
\end{split}
\end{displaymath}
Moreover,
$p \equiv 1 - p_{1} - \cdots - p_{j} = 1 - \prob{Y_{1} \ge x_{j} | E
}$. For positive integers $n_{1},\ldots, n_{j}$ we then have  
\begin{displaymath}
p^{N - n_{1} - \cdots - n_{j}} =  \svigi{ 1 - \prob{X_{1} \ge N^{1/\alpha}x_{j} | E  }  }^{N - n_{1} - \cdots - n_{j}}
\end{displaymath}
By~\eqref{eq:12} it then holds that
\begin{displaymath}
\begin{split}
  \svigi{1 -  \frac{\overline {f_{\alpha}} \svigi{ N^{1/\alpha} x_{j} } }{Nx_{j}^{\alpha} }\svigi{1 -    \svigi{ \frac{ x_{j} N^{1/\alpha} }{ \uN + 1  }}^{\alpha}  }   }^{N - \sum_{i}x_{i}} 
 &  \le p^{N - \sum_{i} x_{i}} \le  \\
 & \svigi{1 -  \frac{\underline {g_{\alpha}} \svigi{ N^{1/\alpha} x_{j} } }{Nx_{j}^{\alpha} }\svigi{1 -    \svigi{ \frac{ x_{j} N^{1/\alpha} }{ \uN + 1  }}^{\alpha}  }   }^{N - \sum_{i}x_{i}}
\end{split}
\end{displaymath}
Hence,  recalling that $\uN/N^{1/\alpha}\to \infty$ by assumption,  
\begin{displaymath}
\exp \svigi{-f_{\alpha}^{(\infty)} x_{j}^{-\alpha} } \le \lim_{N\to \infty} p^{N - \sum_{i}x_{i}} \le  \exp \svigi{-g_{\alpha}^{(\infty)} x_{j}^{-\alpha} }
\end{displaymath}
It follows that
\begin{displaymath}
\begin{split}
& \prod_{i=1}^{j} \exp\svigi{-f_{\alpha}^{(\infty)}\svigi{x_{i}^{-\alpha} - x_{i-1}^{-\alpha} } }\svigi{g_{\alpha}^{(\infty)}x_{i}^{-\alpha} - f_{\alpha}^{(\infty)}x_{i-1}^{-\alpha} }^{n_{i}} \\
& \le \lim_{N\to\infty} \frac{ (N)_{n_{1} + \cdots + n_{j} } }{n_{1}! \cdots n_{j}! } p_{1}^{n_{1}} \cdots p_{j}^{n_{j}} p^{N - \sum_{i}n_{i}} \\
& \le   \prod_{i=1}^{j} \exp\svigi{-g_{\alpha}^{(\infty)}\svigi{x_{i}^{-\alpha} - x_{i-1}^{-\alpha} } }\svigi{f_{\alpha}^{(\infty)}x_{i}^{-\alpha} - g_{\alpha}^{(\infty)}x_{i-1}^{-\alpha} }^{n_{i}}
\end{split}
\end{displaymath}
Taking $g_{\alpha}^{(\infty)} = f_{\alpha}^{(\infty)}$ we then have that
$\prob{\cap_{i=1}^{j} \set{L_{i}^{N} = n_{i} } } =
\prob{\cap_{i=1}^{j} \set{L_{i} = n_{i} } } $ where
$\svigi{L_{1}^{N},\ldots, L_{j}^{N} }$ and
$\svigi{L_{1},\ldots, L_{j} }$ are as in the proof of
~\cite[Lemma~20]{schweinsberg03}.  The
Poisson-Dirichlet$(\alpha,0)$-distribution can be constructed from the
ranked points $Z_{1} \ge Z_{2} \ge \ldots$ of a Poisson point process
on $(0,\infty)$ with characteristic measure
$\nu_{\alpha}((x,\infty)) = \one{x>0}Cx^{-\alpha}$, where the law of
$\svigi{Z_{j}/\sum_{i}Z_{i} }_{j\in \N}$ is the Poisson-Dirichlet
distribution with parameter $(\alpha, 0)$.  Let $\epsilon > 0$ and
choose $\delta > 0$ as in the proof of~\cite[Lemma~21]{schweinsberg03}
so that $\delta \to 0$ as $\epsilon \to 0$.  We can use the upper
bound in~\eqref{eq:boundsmallera} to show that
\begin{displaymath}
\begin{split}
\EE{\sum_{i=1}^{N}Y_{i} \one{Y_{i} \le \delta } | E }  &  =   N^{1 - 1/\alpha} \EE{ X_{1} \one{ X_{1} \le N^{1/\alpha} \delta } | E} \le  {f_{\alpha}^{ (\infty)}} N^{1 - 1/\alpha} \sum_{k=1}^{ \lfloor N^{1/\alpha} \delta \rfloor  } k^{-\alpha} \le  {f_{\alpha}^{ (\infty)}} \frac{\delta^{1-\alpha}}{1 - \alpha}
\end{split}
\end{displaymath}
By the arguments in the proof of~\cite[Lemma~21]{schweinsberg03} we
see  that $\svigi{Y_{1}, \ldots, Y_{j}, \sum_{i=j+1}^{N}Y_{i} }$
converges weakly to
$\svigi{Z_{1}, \ldots, Z_{j}, \sum_{i=j+1}^{\infty}Z_{i} }$ as
$N\to \infty$.  Write
$\svigi{ W_{j} }_{j\in \N} \equiv \svigi{Z_{j}/\sum_{i}Z_{i} }_{j\in
\N} $.  Suppose $C_{\kappa}^{N}c_{N} \overset c \sim 1 $, and that
$\varepsilon_{N}C_{\kappa}^{N} = c$ for some $c > 0$ fixed, and for
each $N\in \IN$.  Then for
all $k_{1},\ldots, k_{r} \ge 2$,  $r\in \N$, and with $\Xi_{\alpha}$
as in Definition~\ref{def:poisson-dirichlet} (recall $C_{\kappa}$ from
~\eqref{eq:9}),  
\begin{displaymath}
\begin{split}
& \lim_{N\to \infty}\frac{N^{r}}{c_{N}}\EE{ \frac{(X_{1})_{k_{1}} \cdots (X_{r})_{k_{r}}  }{S_{N}^{k_{1} + \cdots + k_{r}} } \one{S_{N} \ge 2N}  } \\
& =   \lim_{N\to \infty}\frac{N^{r}}{c_{N}}\EE{ \frac{(X_{1})_{k_{1}} \cdots (X_{r})_{k_{r}}  }{S_{N}^{k_{1} + \cdots + k_{r}} } \one{S_{N} \ge 2N} | E }\varepsilon_{N} +   \lim_{N\to \infty}\frac{N^{r}}{c_{N}}\EE{ \frac{(X_{1})_{k_{1}} \cdots (X_{r})_{k_{r}}  }{S_{N}^{k_{1} + \cdots + k_{r}} } \one{S_{N} \ge 2N} |E^{\sf c} }(1-\varepsilon_{N}) \\
& = \frac{c}{c(1-\alpha) + 4C_{\kappa} }\sum_{\substack{ i_{1},\ldots, i_{r} = 1 \\  \text{all distinct} }} \EE{W_{i_{1}}^{k_{1}} \cdots  W_{i_{r}}^{k_{r}} } + \one{r=1,k_{1} = 2 } \frac{4C_{\kappa} }{c(1-\alpha) +  4C_{\kappa} }  \\
& = \frac{c}{c(1-\alpha) + 4C_{\kappa} }\int_{\Delta_{+}} \sum_{\substack{ i_{1},\ldots, i_{r} = 1 \\  \text{all distinct} }  }^{\infty}  x_{i_{1}}^{k_{1}} \cdots x_{i_{r}}^{k_{r}} \frac{1}{\sum_{j=1}^{\infty}x_{j}^{2} }   \Xi_{\alpha}(dx)  +  \one{r=1, k_{1}=2} \frac{ 4C_{\kappa} }{ c(1-\alpha) + 4C_{\kappa}   }
\end{split}
\end{displaymath}
It remains to verify that $C_{\kappa}^{N} c_{N} \sim 1$.  The same
arguments as for~\cite[Equation~77]{schweinsberg03} give that
\begin{displaymath}
\lim_{N\to \infty}N\EE{(X_{1})_{2} S_{N}^{-2}\one{S_{N} \ge 2N} | E }
= 1 - \alpha.
\end{displaymath}
We also have that
$C_{\kappa}^{N} \overset c \sim N\EE{(X_{1})_{2}S_{N}^{-2}\one{S_{N}
\ge 2N} | E^{\sf c} }$; it follows that
$C_{\kappa}^{N} c_{N} \overset c \sim  c(1-\alpha) + C_{\kappa}$ as $N\to \infty$.   Given that
$C_{\kappa}^{N} \varepsilon_{N} = c$, it follows from
~\cite[Lemma~6, Equation~16]{schweinsberg03} that  
\begin{displaymath}
\begin{split}
\lim_{N\to \infty} \frac{\EE{ (\nu_{1})_{k_{1}} \cdots (\nu_{r})_{k_{r}}} }{ 2^{k} N^{ k - r} c_{N} }  & = \one{r = 1, k_{1} = 2} \frac{ 4C_{\kappa}  }{ c(1-\alpha) + 4C_{\kappa}} \\
& +   \frac{c}{c(1-\alpha) + 4C_{\kappa} } \int_{\Delta_{+}}  \sum_{\substack{ i_{1},\ldots, i_{r} = 1 \\  \text{all distinct} }  }^{\infty} x_{i_{1}}^{k_{1}} \cdots   x_{i_{r}}^{k_{r}} \frac{1}{\sum_{j=1}^{\infty}x_{j}^{2} } \Xi_{\alpha}(dx)
\end{split}
\end{displaymath}
To get the full rate of a particular combination of  up to $4r$
simultaneous mergers  from  $k_{1},\ldots, k_{r}$ group sizes  it
remains  to  sum over all the ways in which  such mergers can occur.
We omit this here; an example when  $r=1$ can be seen in connection
with the
$\Omega$-$\delta_{0}$-Beta$(\gamma,2-\alpha,\alpha)$-coalescent.  
Hence, the limits in~\eqref{xmoments}  exist  for all $r\in \N$  and
$k_{1},\ldots,k_{r}\ge 2$, and    the proof of
Case~\ref{item:8} of
Theorem~\ref{thm:diploid-alpha-random-all}  is complete.
\end{proof}

\subsection{Proof of Theorem~\ref{thm:diploid-alpha-random-one-environment}}\label{proof:thmrandone}

In this section we give a proof of
Theorem~\ref{thm:diploid-alpha-random-one-environment}.

First, we identify conditions on $\overline \varepsilon_{N}$ for
$m_{\infty} < \infty$ (recall \eqref{eq:5}) to hold.
\begin{lemma}[Finite $m_{\infty}$]
\label{lm:finitemarandone}
Suppose the  conditions of Theorem~\ref{thm:diploid-alpha-random-one-environment} are in force,  and that 
\begin{equation}
\label{eq:ovarepsiNfinitem}
\overline \varepsilon_{N} \in
\begin{cases}
O(1) & \text{when $\alpha = 1$} \\
O\svigi {N\uN ^{ \alpha - 1} } & \text{when $0<\alpha < 1$} \\
\end{cases}
\end{equation}
It then holds that   $\limsup_{N\to \infty}m_{N} < \infty$. 
\end{lemma}
\begin{proof}[Proof of Lemma~\ref{lm:finitemarandone}]
Recall the events $E_{1}$ and $E_{1}^{\sf c}$ from
Definition~\ref{df:alpha-random-one-environment}.  When $0<\alpha < 1$
we see, using the upper bound in~\eqref{eq:boundsmallera} and
~\eqref{eq:boundbiggera},
\begin{displaymath}
\EE{X | E_{1}} \le  \prob{X=1} +  \frac { {f_{\alpha}^{ (\infty)}} } N\sum_{k=2}^{\uN}\frac{k}{(k-1)^{1+\alpha}} +  \kappa {f_{\kappa}^{ (\infty)}} \frac{N-1}{N} \sum_{k=2}^{\uN}  {k^{-\kappa} }
\end{displaymath}
Since $\sum_{k \ge 1}k^{-\kappa}$ converges for $\kappa \ge 2$ it suffices to check that
\begin{displaymath}
\sum_{k=2}^{\uN}\frac{(k-1) + 1}{(k-1)^{1+\alpha}} =  \sum_{j=1}^{\uN}j^{-\alpha} + \sum_{j=1}^{\uN}j^{-1-\alpha} \le 2 + O\svigi{\uN^{1-\alpha}}
\end{displaymath}
With $X$ denoting the random number of potential offspring of an
arbitrary parent pair, the result for $0 < \alpha < 1$ now follows
from the relation
$\EE{X} = \EE{X|E_{1}}\overline \varepsilon_{N} + \EE{X|E_{1}^{\sf
c}}(1 - \overline \varepsilon_{N})$.  The result for $\alpha = 1$ is
obtained in the same way (here we omit the details).
\end{proof}

We now check that $S_{N}/(Nm_{N})\to 1$ almost surely as $N\to \infty$ (recall Lemma~\ref{lm:almostsureconvSN}).  
\begin{lemma}[$S_{N}/(Nm_{N})\to 1$ almost surely]
Under the conditions of
Theorem~\ref{thm:diploid-alpha-random-one-environment}, with  $1 < r < 2$ fixed, and  
$\overline \varepsilon_{N}$ fulfilling~\eqref{eq:ovarepsiNfinitem} in
Lemma~\ref{lm:finitemarandone} where
\begin{equation}
\label{eq:convasaone}
\overline \varepsilon_{N} \in
\begin{cases}
O(1) & \text{when } \uN/N \to K \\
o\svigi{ 1 \wedge \tfrac{N^{r}}{ \uN^{r - \alpha} } } & \text{when } \uN/N \to \infty
\end{cases}
\end{equation}
It then holds that $S_{N}/(N m_{N}) \to 1$ almost surely as
$N\to \infty$. 
\end{lemma}
\begin{proof}[Proof of Lemma~\ref{eq:convasaone}]
We will use the approach of~\cite[\breyta{Proof of Theorem~2.2i}]{PANOV2017379}.  Choosing
$\overline \varepsilon _{N}$ as in~\eqref{eq:convasaone} ensures that
$\limsup_{N\to \infty} m_{N} < \infty$ by
Lemma~\ref{lm:finitemarandone}.  Define,  \breyta{for $i \in [N]$,}
$\overline X_{i} \equiv (X_{i} - m_{N})/(N m_{N})$.  Then
\begin{displaymath}
\overline S_{N} \equiv   \sum_{i=1}^{N} \overline X_{i} =  \frac{S_{N}}{N m_{N}} - 1
\end{displaymath}
Suppose $\limsup_{N\to \infty} \EE{|\overline X_{i} |^{r} } <
\infty$ \breyta{for some $1 < r < 2$.}   It then holds by~\cite[Theorem~2]{10.1214/aoms/1177700291}
\begin{displaymath}
\EE{|\overline S_{N}|^{r} } \le  \frac{2}{N^{r}m_{N}^{r}}\sum_{i=1}^{N}\EE{|X_{i} - m_{N}|^{r} } =    \frac{2}{m_{N}^{r}}N^{1-r}\EE{|X_{1} - m_{N}|^{r} }
\end{displaymath}
We will now show that
$\limsup_{N\to \infty} \EE{|\overline S_{N}|^{r} } = 0$.  The upper
bound in~\eqref{eq:boundsmallera} and~\eqref{eq:boundbiggera}, and the inequality
$|x+y|^{s} \le 2^{s-1}\svigi{|x|^{s} + |y|^{s}}$ for any
$x,y\in \mathds R$ and $s \ge 1$~\cite[Proposition~3.1.10{\it (iii)},
Equation~1.12]{athreya06:_measur} combine to give
\begin{displaymath}
\begin{split}
\EE{|X_{1} - m_{N}|^{r} | E_{1}} &  \le |m_{N}|^{r}\prob{X_{1} = 0} +  |1 - m_{N}|^{r}\prob{X_{1} = 1}  +   \frac{ {f_{\alpha}^{ (\infty)}} }{N}\sum_{k=2}^{\uN} \frac{ |k - m_{N}|^{r}}{k^{1+\alpha}}  + \kappa {f_{\kappa}^{ (\infty)}} \sum_{k=2}^{\uN}  \frac{ |k - m_{N}|^{r}}{k^{1+\kappa}} \\
&  \le  2|m_{N}|^{r} +  \frac{2^{r-1}  {f_{\alpha}^{ (\infty)}} } {N} \sum_{k=2}^{\uN} \frac{k^{r} + m_{N}^{r} }{k^{1+\alpha} } + 2^{r-1}\kappa f_{\kappa}^{ (\infty)} \sum_{k=2}^{\uN} \frac{k^{r} + m_{N}^{r} }{k^{1+\kappa} }
\end{split}
\end{displaymath}
Since $0 < \alpha \le 1$,  $1< r < 2$, and  $2  \le \kappa$,  we see
\begin{displaymath}
\frac 1N \sum_{k=2}^{\uN} k^{r-\alpha -1} \le \frac 1N +  \one{r \le \alpha + 1}\frac 1N \int_{1}^{\uN} x^{r-\alpha-1}dx +  \one{r > \alpha + 1} \frac 1N \int_{1}^{\uN + 1}x^{r-\alpha-1}dx = O\svigi{\frac{\uN^{r-\alpha}}{N} }
\end{displaymath}
Then
$\EE{|X_{1} - m_{N}|^{r} | E_{1} } =  O\svigi{1+ N^{-1}\uN^{r-\alpha}
}$ so that with $\overline \varepsilon _{N}$ as in
~\eqref{eq:convasaone}  it holds that, for $1 < r < 2$, 
\begin{displaymath}
\limsup_{N\to \infty} N^{1-r}\EE{|X_{1} - m_{N}|^{r} | E_{1} } \overline \varepsilon_{N}  = 0
\end{displaymath}
One checks in the same way that
$\limsup_{N\to \infty} N^{1-r} \EE{|X_{1} - m_{N}|^{r} | E_{1}^{\sf c}
}(1-\overline \varepsilon _{N}) = 0$ (here we omit the details).  We
have shown that there is an $r > 1$ such that
$\limsup_{N\to \infty}\EE{|\overline S_{N}|^{r}} = 0$.~\breyta{For $i > N$ take
$\overline X_{i}$ to be  independent   with mean 0.   It
follows that 
$\overline S_{N}$ converges in probability to 0.   The 
 $\overline X_{i}$ are independent for all $i\in \N$, 
 and so  $\overline S_{N} $ converges almost
surely to 0 (e.g.\ \cite[Theorem 8.3.3]{athreya06:_measur}). Hence, 
$S_{N}/(Nm_{N}) = 1 + \overline S_{N}$} converges almost surely to 1.
\end{proof}

We now verify~\eqref{eq:scalecN} (i.e.\
$C_{\kappa}^{N}c_{N} \overset c \sim 1$ where $C_{\kappa}^{N}$ is as
in~\eqref{eq:CNmap}).
\begin{lemma}[Verifying~\eqref{eq:scalecN}]\label{lm:cNarandone}
Equation~\eqref{eq:scalecN} is in force under the conditions of
Theorem~\ref{thm:diploid-alpha-random-one-environment} when taking
\begin{equation}
\label{eq:varepsarandone}
\overline \varepsilon_{N} =
\begin{cases}
cN^{\alpha - 1} & \text{when $0 < \alpha < 1$ and $\kappa > 2$} \\
cN^{\alpha - 1}\log N & \text{when $0 < \alpha < 1$ and $\kappa = 2$} \\
\varepsilon & \text{when $\alpha = 1$ and  $\kappa > 2$} 
\end{cases}
\end{equation}
where $0<\varepsilon < 1$ and $c > 0$  are  fixed.
\end{lemma}
\begin{proof}[Proof of Lemma~\ref{lm:cNarandone}]
The proof follows by similar arguments as the proof of
Lemma~\ref{lm:cNrandall}. Given $\epsilon > 0$ fixed we use
Lemma~\ref{eq:convasaone} to get~\eqref{eq:boundERN} with $Y_{-}$ and
$Y_{+}$ as in~\eqref{eq:YMR}. Then, by
Definition~\ref{df:alpha-random-one-environment},
\begin{displaymath}
\begin{split}
\EE{Y_{-}} & \le  \svigi{ \frac{\overline {f_{\alpha}}(2)}{N}\sum_{k=2}^{\uN} \frac{k(k-1)}{(k+M_{-})^{2}}\svigi{ \frac{1}{k^{\alpha}} - \frac{1}{(1+k)^{\alpha}} } + \overline {f_{\kappa}}(2)\frac{N-1}{N}\sum_{k=2}^{\uN} \frac{k(k-1)}{(k+M_{-})^{2}}\svigi{ \frac{1}{k^{\kappa}} - \frac{1}{(1+k)^{\kappa}} } }\overline \varepsilon_{N} \\ 
& +   \overline {f_{\kappa}}(2)\sum_{k=2}^{\uN} \frac{k(k-1)}{(k+M_{-})^{2}}\svigi{ \frac{1}{k^{\kappa}} - \frac{1}{(1+k)^{\kappa}} }\svigi{1- \overline \varepsilon_{N}} \\
& \sim   \frac{\overline {f_{\alpha}}(2) }{N}\sum_{k=2}^{\uN} \frac{k(k-1)}{(k+M_{-})^{2}}\svigi{ \frac{1}{k^{\alpha}} - \frac{1}{(1+k)^{\alpha}} }\overline \varepsilon_{N} +  \overline {f_{\kappa}}(2) \sum_{k=2}^{\uN}  \frac{k(k-1)}{(k+M_{-})^{2}}\svigi{ \frac{1}{k^{\kappa}} - \frac{1}{(1+k)^{\kappa}} } \\
\end{split}
\end{displaymath}
The lower bound on $\EE{Y_{+}}$ is of the same form as the upper bound
on $\EE{Y_{-}}$ except with $\underline {g_{a}}(2)$ replacing
$\overline {f_{a}}(2)$.  The lemma then follows from
Propositions~\ref{pr:approxsum} and~\ref{prop:boundsumkM} as in the
proof of Lemma~\ref{lm:cNrandall}, and after checking that
~\eqref{eq:claimboundedEXmax} holds (here we omit the straightforward
calculations).
\end{proof}

The following lemma checks that    a  necessary and sufficient
condition for convergence to the \breyta{Kingman} coalescent holds under
the conditions of Case~\ref{item:7} of
Theorem~\ref{thm:diploid-alpha-random-one-environment}; the proof
follows by similar arguments as the proof of
Lemma~\ref{lm:proofconvkingmantypeA}, and is  omitted.
\begin{lemma}[Convergence to the Kingman coalescent]\label{lm:convdeltaarandone}
Under the conditions of Case~\ref{item:7} of
Theorem~\ref{thm:diploid-alpha-random-one-environment}  and with
$\overline \varepsilon_{N}$ as in~\eqref{eq:varepsarandone}   it holds that
$\lim_{N\to \infty}N\EE{ {(X_{1})}_{3}S_{N}^{-3}\one{S_{N} \ge 2N} }/c_{N} = 0$ 
\end{lemma}

\breyta{Under}  the conditions of Case~\ref{item:9} of
Theorem~\ref{thm:diploid-alpha-random-one-environment} simultaneous
mergers will not be generated by the occurrence of two or more large
families in any given generation in a large population (as
$N \to \infty$).  The proof follows similar arguments as the proof of
Lemma~\ref{proof:twopluslargefams} and is omitted.
\begin{lemma}[The probability of two or more large families vanishes as $N\to \infty$]
Under the conditions of Case~\ref{item:9} of
Theorem~\ref{thm:diploid-alpha-random-one-environment} and with
$\overline \varepsilon_{N}$ as in~\eqref{eq:varepsarandone} it holds
that
\begin{displaymath}
\lim_{N\to \infty}\frac{N^{2}}{c_{N}}\EE{ \frac{ {(X_{1})}_{2} {(X_{2})}_{2}  }{S_{N}^{4}}\one{S_{N} \ge 2N}} = 0
\end{displaymath}
\end{lemma}

\linelabel{proofThemone}\begin{proof}[Proof of Theorem~\ref{thm:diploid-alpha-random-one-environment}]
Case~\ref{item:7} of Theorem
~\ref{thm:diploid-alpha-random-one-environment} is
Lemma~\ref{lm:convdeltaarandone} and Lemma~\ref{lm:dSchwlm6}.~\breyta{
Convergence to the Kingman coalescent follows as in the proof of Lemma~\ref{lm:proofconvkingmantypeA}.}

Case~\ref{item:9} of
Theorem~\ref{thm:diploid-alpha-random-one-environment} follows by
identical  arguments as for  Case~\ref{item:2} of
Theorem~\ref{thm:diploid-alpha-random-all}.~\breyta{The proof of Theorem~\ref{thm:diploid-alpha-random-one-environment} is complete.}
\end{proof}

\appendix

\counterwithin*{equation}{section}
\counterwithin*{figure}{section}
\renewcommand\theequation{\thesection\arabic{equation}}
 \renewcommand\thefigure{\thesection\arabic{figure}}

\section{Computing $\overline \varrho_{i}(n)$   for the   $\Omega$-Beta$(2-\beta,\beta)$-Poisson-Dirichlet$(\alpha,0)$-coalescent}\label{sec:appr-eer_-omega}

\breyta{In this section we briefly present an algorithm for computing
the empirical mean $\overline\varrho_{i}(n)$ 
 of  annealed relative branch lengths $R_{i}(n)$
(recall~\eqref{eq:33},~\eqref{eq:estimates},
and~\eqref{eq:approxestimates})   of gene
genealogies  when  governed by the
$\Omega$-Beta$(2-\beta,\beta)$-Poisson-Dirichlet$(\alpha,0)$-coalescent.}

Fix \breyta{$1 < \beta < 2$, and   $1/\beta < \alpha < 1$}.  The 
Beta$(2-\beta,\beta)$-Poisson-Dirichlet$(\alpha,0)$-coalescent on the
partitions of $[n]$ is a continuous-time  $\Xi$-coalescent with measure $\Xi$ of the
form (recall $\Delta_{+}$ from~\eqref{eq:simplex} and $\Xi_{\alpha}$
from Definition~\ref{def:poisson-dirichlet})
\begin{displaymath}
\Xi(dx) =  \int_{0}^{1} \delta_{\svigi{x, 0, 0, \ldots}} \text{Beta}(2-\beta,\beta)(dx) +  \int_{\Delta_{+}}\delta_{\svigi{x_{1},x_{2}, \ldots }}\Xi_{\alpha}(dx)
\end{displaymath}
The  transition rates are  
\begin{displaymath}
\begin{split}
\lambda_{n;k_{1},\ldots,k_{r};s} &  = \frac{1}{C_{\beta} + c(1-\alpha)}\svigi{ \one{r=1}{\beta}{m_{\infty}^{-\beta}}B\svigi{k_{1}-\beta, n-k_{1} + \beta} +  {cp_{n;k_{1},\ldots,k_{r};s}}}
\end{split}
\end{displaymath}
where $C_{\beta} =  \beta m_{\infty}^{-\beta}B(2-\beta,\beta)$,  $1 +
2^{-\beta}/(\beta - 1) < m_{\infty} <  1 +  1/(\beta - 1)
$, and $p_{n;k_{1},\ldots, k_{r};s}$ is as in~\eqref{eq:24}.

Consider a diploid panmictic  population evolving according to
Definition~\ref{dschwpop} and Definition~\ref{dfn:alpharandomall}. 
The calculations  leading to     Cases~\ref{item:2} and~\ref{item:8} of
Theorem~\ref{thm:diploid-alpha-random-all} taking $\uN
/N^{1/\alpha}\to \infty$ and $\varepsilon_{N} \overset c \sim c_{N}$
where $N^{\beta - 1} c_{N} \overset c \sim 1$ all as $N\to \infty$  then lead to the
continuous-time 
$\Omega$-Beta$(2-\beta,\beta)$-Poisson-Dirichlet$(\alpha,0)$-coalescent. 
\begin{defn}[The $\Omega$-Beta$(2-\beta,\beta)$-Poisson-Dirichlet$(\alpha,0)$-coalescent]\label{df:omegabetapoidiri}
The
$\Omega$-Beta$(2-\beta,\beta)$-Poisson-Dirichlet$(\alpha,0)$-coalescent, 
\breyta{for $1 < \beta < 2$ and $1/\beta < \alpha < 1$,} 
is the Beta$(2-\beta,\beta)$-Poisson-Dirichlet$(\alpha,0)$-coalescent
where the blocks in each group  split among four subgroups
independently and uniformly at random, and the blocks assigned to the
same subgroup are merged.   
\end{defn}
\breyta{\begin{remark}[The range of $\alpha$ in Definition~\ref{df:omegabetapoidiri}]
The range of  $\alpha$ in Definition~\ref{df:omegabetapoidiri}
guarantees that the mean $m_{\infty} < \infty$, as one can easily
check using the upper bound in~\eqref{eq:boundsmallera} in
Lemma~\ref{lm:boundskernel} (Recall also Remark~\ref{rm:range-alpha}).  
\end{remark}}
The measure driving the 
$\Omega$-Beta$(2-\beta,\beta)$-Poisson-Dirichlet$(\alpha,0)$-coalescent
is of the form 
\begin{displaymath}
\Xi(dx) =  \int_{0}^{1} \delta_{\svigi{\frac x4, \frac x4, \frac x4, \frac x4, 0, 0, \ldots}} \text{Beta}(2-\beta,\beta)(dx) +  \int_{\Delta_{+}}\delta_{\svigi{\frac{x_{1}}{4}, \frac{x_{1}}{4}, \frac{x_{1}}{4}, \frac{x_{1}}{4}, \frac{x_{2}}{4}, \frac{x_{2}}{4}, \frac{x_{2}}{4}, \frac{x_{2}}{4}, \ldots }}\Xi_{\alpha}(dx)
\end{displaymath}

\begin{figure}[htp]
\centering
\captionsetup[subfloat]{labelfont={scriptsize,sf,md,up},textfont={scriptsize,sf}}
\subfloat[$\alpha=0.9$, $\beta=1.125$]{\label{fig:betapoidiria}\includegraphics[angle=0,scale=.7]{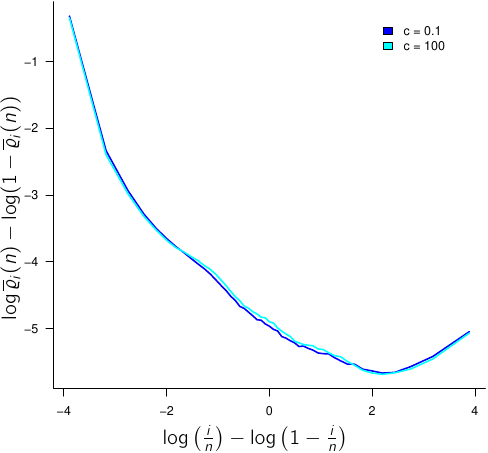}}
\subfloat[$\beta=1.5$, $c=0.1$]{\label{fig:betapoidirib}\includegraphics[angle=0,scale=.7]{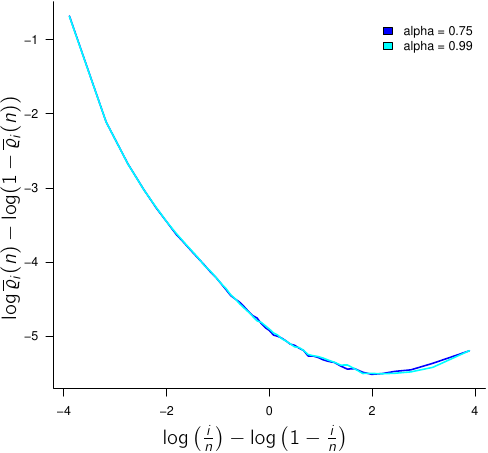}}
\subfloat[$\beta=1.5$, $c=100$]{\label{fig:betapoidiric}\includegraphics[angle=0,scale=.7]{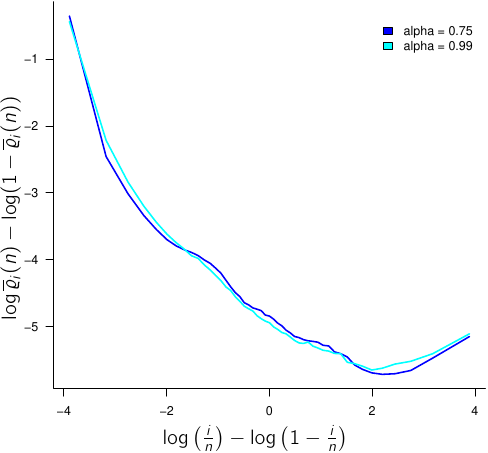}} \\
\caption{\breyta{The
$\Omega$-Beta$(2-\beta,\beta)$-Poisson-Dirichlet$(\alpha,0)$-coalescent.
Examples of  $\overline \varrho_{i}(n)$~(recall~\eqref{eq:33},~\eqref{eq:estimates}
and~\eqref{eq:approxestimates})}
when $\mengi{\xi^{n}}$ is the
$\Omega$-Beta$(2-\beta,\beta)$-Poisson-Dirichlet$(\alpha,0)$-coalescent
as in Definition~\ref{df:omegabetapoidiri}
for $n=50$, $c,\alpha,\beta$ as shown, and approximating $m_{\infty}$
with  $\svigi{2 + \svigi{1 + 2^{1-\beta}}/(\beta - 1)}/2$.  The scale of the ordinate
(y-axis) may vary between graphs; results from $10^{5}$ experiments}\label{fig:betapoissondiriI}
\end{figure}

Figure~\ref{fig:betapoissondiriI}  contains examples of
$\overline\varrho_{i}(n)$~\eqref{eq:estimates} predicted by the
$\Omega$-Beta$(2-\beta,\beta)$-Poisson-Dirichlet$(\alpha,0)$-coalescent.
The graphs in  Figure~\ref{fig:betapoidiric}  indicate that the
$\Omega$-Beta$(2-\beta,\beta)$-Poisson-Dirichlet$(\alpha,0)$-coalescent
can  predict a  non-monotone (U-shaped) site-frequency spectrum with
 peaks around the middle of the spectrum.  However,
 Figures~\ref{fig:betapoidiria} and~\ref{fig:betapoidirib} indicate
there are   cases  where the site-frequency spectrum is U-shaped
without the peaks,  and resembling the site-frequency spectrum
observed in population genomic data from the  diploid Atlantic cod
~\citep{Arnasonsweepstakes2022}.

\section{Computing $\overline \varrho_{i}(n)$    for the
$\Omega$-$\delta_{0}$-Beta$(\gamma,2-\alpha,\alpha)$-coalescent }\label{sec:estimatingERin}

In this section we briefly describe the algorithm for sampling from
the $\Omega$-$\delta_{0}$-Beta$(\gamma,2-\alpha,\alpha)$ coalescent
and computing  \breyta{the empirical mean  $\overline \varrho _{i}(n)$
of   $R_{i}(n)$
(recall~\eqref{eq:33},~\eqref{eq:estimates}, and~\eqref{eq:approxestimates}).}
Figure~\ref{fig:xindbetaoverparams}  records  examples of
$\overline \varrho_{i}(n)$.  

Let $\lambda_{m} := \sum_{m\ge k_{1}\ge \ldots \ge k_{r} \ge 2; \sum_{j}k_{j}\le m  } \lambda_{m;k_{1}, \ldots, k_{r};s}$ denote
the total jump rate out of $m$ blocks   
and $p_{m}(k) = \lambda_{m;k_{1}, \ldots, k_{r};s}/\lambda_{m} $ the
probability of seeing (ordered) merger sizes
$k = (k_{1}, \ldots, k_{r})$ where
$2 \le k_{1} + \cdots + k_{r} \le m$.  Let 
\begin{displaymath}
\mathcal K_{m} := \mengi{ (k_{1}, \ldots, k_{r}) : m \ge k_{1} \ge \ldots \ge k_{r} \ge 2, \quad 2  \le k_{1} + \cdots + k_{r} 
  \le m, \quad  r \in [4]}
\end{displaymath}
denote the set of all possible mergers when there are $m$ blocks and
$j_{m} : \mathcal K_{m} \to [\# \mathcal K_{m}]$ a map assigning
unique indexes to the mergers.  Let
$F_{m}: [\#\mathcal K_{m}] \to [0,1]$ denote the cumulative
probability mass function for the merger sizes where
$F_{m}(j) = \sum_{\ell=1}^{j} p_{m}(j_{m}(\ell))$.

Let $\svigi{\ell_{i}(n), \ldots, \ell_{n-1}(n) }$ denote the realised
branch lengths  and $\svigi{b_{1},\ldots, b_{m}}$ the
current block sizes where $b_{j}\in [2n]$ and $\sum_{j}b_{j} = 2n$.   

\begin{enumerate}
\item $\svigi{ \mathfrak r _{1}(n), \ldots,  \mathfrak r_{2n-1}(n)} \leftarrow (0,\ldots, 0)$
\item for each of $M$ experiments
\begin{enumerate}
\item $\svigi{\ell_{1}(n), \ldots, \ell_{2n-1}(n)} \leftarrow (0,\ldots, 0) $
\item set the current number of blocks  $m \leftarrow 2n$ 
\item $\svigi{b_{1}, \ldots, b_{m} } \leftarrow \svigi{1,\ldots, 1} $
\item {\bf while} $m > 1$:
\begin{enumerate}
\item sample a random exponential $t$ with rate $\lambda_{m}$ 
\item  $\ell_{b}(n) \leftarrow t +  \ell_{b}(n)  $ for $b =  b_{1}, \ldots, b_{m}$
\item sample merger sizes $\inf \mengi{j \in [ |\mathcal K_{m}| ] : U \le F_{m}(j) }$  where  $U$ a standard  random uniform 
\item merge blocks according to sampled merger sizes $k_{1},\ldots, k_{r}$ 
\item $m\leftarrow m - k_{1} - \cdots - k_{r} + r$
\end{enumerate}
\item  $ \mathfrak r_{i}(n) \leftarrow \mathfrak r_{i}(n) +  \ell_{i}(n)/\sum_{j}\ell_{j}(n)$  for $i \in [2n-1]$
\end{enumerate}
\item return an approximation   $\overline  \varrho_{i}(n) =  (1/M)\mathfrak r_{i}(n) $   of $\EE{R_{i}(n)}$ for $i = 1,2, \ldots, 2n-1$
\end{enumerate}


\section{Computing $\overline \varrho_{i}(n)$ for the 
$\Omega$-$\delta_{0}$-Poisson-Dirichlet$(\alpha,0)$-coalescent  }\label{sec:estimatingforodpd}

\begin{figure}[htp]
\centering
\captionsetup[subfloat]{labelfont={scriptsize,sf,md,up},textfont={scriptsize,sf}}
\subfloat[$c=1$]{\includegraphics[angle=0,scale=.7]{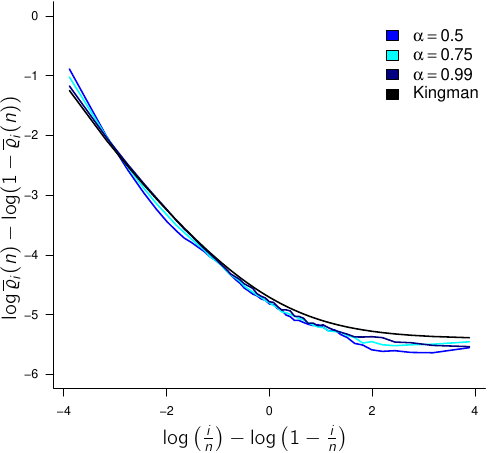}}
\subfloat[$c=10^{2}$]{\includegraphics[angle=0,scale=.7]{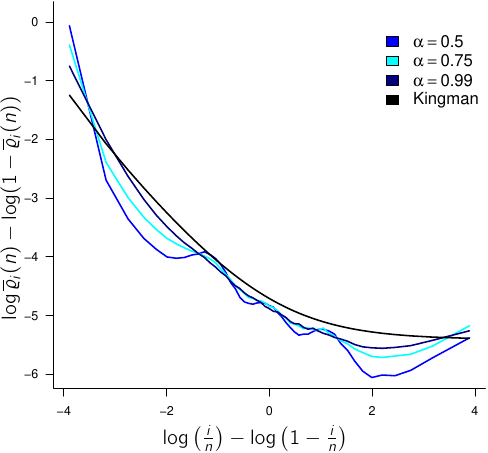}}\\
\caption{\breyta{The 
$\Omega$-$\delta_{0}$-Poisson-Dirichlet$(\alpha,0)$ coalescent.}
\breyta{Empirical mean}  $\overline \varrho_{i}(n)$
\breyta{of $R_{i}(n)$}   (recall~\eqref{eq:33},~\eqref{eq:estimates})
 when $\set{\xi^{n}}$ is the
$\Omega$-$\delta_{0}$-Poisson-Dirichlet$(\alpha,0)$ coalescent
(Definition~\ref{def:diploid-kingman-poisson-dirichlet-coalescent});
for sample size $n=50$, $\kappa = 2$, $c$ and $\alpha$ as shown;
results from $10^{5}$ experiments. The scale of the ordinate (y-axis)
may vary between graphs }\label{fig:diploiddpd}
\end{figure}

In this section we briefly describe the algorithm for computing
\breyta{the empirical mean}  $\overline \varrho_{i}(n)$ of
$R_{i}(n)$  (recall~\eqref{eq:33},
~\eqref{eq:estimates}, and~\eqref{eq:approxestimates})  as 
predicted by the 
$\Omega$-$\delta_{0}$-Poisson-Dirichlet$(\alpha,0)$ coalescent (recall
Definition~\ref{def:diploid-kingman-poisson-dirichlet-coalescent}). In 
  Figure~\ref{fig:diploiddpd} we record examples \breyta{of
  $\overline\varrho_{i}(n)$}   for sample size
  $n=50$.

\begin{enumerate}
\item $\svigi{\mathfrak r _{1}(n), \ldots, \mathfrak  r _{2n-1}(n) } \leftarrow \svigi{0,\ldots,0 } $
\item for each of  $M$ experiments: 
\begin{enumerate}
\item $\svigi{ \ell_{1}(n), \ldots, \ell_{2n-1}(n)  } \leftarrow \svigi{0,\ldots, 0}$
\item $m\leftarrow 2n$
\item $\svigi{b_{1}, \ldots, b_{2n}}\leftarrow \svigi{1,\ldots, 1}$
\item {\bf while} $m > 1$:
\begin{enumerate}
\item $t \leftarrow 0$
\item until  a merger occurs (at least one subgroup with at least 2 blocks):
\begin{enumerate}
\item sample group sizes $k_{1},\ldots, k_{r}$  with rate~\eqref{eq:omegadPDrates}
\item assign the blocks in each group to one of 4 subgroups independently and
uniformly at random 
\item $t \leftarrow t +  \text{Exp}(\lambda_{m})$ where $\lambda_{m}$
is the sum of the group sizes rates~\eqref{eq:omegadPDrates} for $m$ blocks
\end{enumerate}
\item $\ell_{b}(n) \leftarrow t +   \ell_{b}(n)  $ for $b  = b_{1},\ldots, b_{m}$
\item merge blocks in the same subgroup  and  update $m$ 
\end{enumerate}
\item update  $\mathfrak r _{i}(n)$ given one realisation $\svigi{\ell_{1}(n),\ldots, \ell_{2n-1}(n) } $ of branch lengths:
\begin{displaymath}
 \mathfrak r_{j}(n)  \leftarrow     \mathfrak r_{j}(n) +    \frac{
 \ell_{j}(n) }{  \sum_{i=1}^{2n-1}\ell_{i}(n) } 
\end{displaymath}
\end{enumerate}
\item return an approximation
 $ \overline \varrho_{i}(n) = (1/M) \mathfrak r_{i}(n) $ for
 of $\EE{R_{i}(n)}$ for $i=1,2,\ldots,2n-1$ 
\end{enumerate}

\section{Computing $\overline\varrho_{i}^{N}(n)$ }\label{sec:estimatingERiN}

\breyta{
In this section we briefly describe the algorithm for computing 
$\overline \varrho_{i}^{N}(n)$~\eqref{eq:approxestimates}, the
empirical mean of $R_{i}^{N}(n)$ \breyta{(recall~\eqref{eq:33})}
from a finite number of annealed   gene genealogies. }
 Suppose our sample comes from a population
evolving according to Def~\ref{dschwpop}. Since we are only estimating
branch lengths it suffices to keep track of the current block sizes.
However, our population is diploid and so we need to keep track of the
pairing of blocks in diploid individuals. Thus, we record the current
sample configuration in the form of pairs of block sizes. At time 0 we
have $n$ pairs of $\set{1,1}$, $\set{\set{1,1}, \ldots, \set{1,1}} $;
a common ancestor is reached upon entering the configuration
$\set{\set{2n,0}}$.  Thus, the sample configuration can be seen as a
record of the current marked individuals, where a `marked' individual
carries at least one ancestral block. Single-marked individuals
carrying one marked block are recorded as $\set{b,0}$. Let $n$ denote
the number of sampled diploid individuals so that $2n$ is the number
of sampled gene copies.

\begin{enumerate}
\item  $\left( \mathfrak r_{1}^{N}(n),\ldots,  \mathfrak  r_{2n-1}^{N}(n) \right) \leftarrow (0,\ldots, 0) $ realised relative branch lengths
\item for each of $M$ experiments
\begin{enumerate}
\item $ \left(\ell_{1}^{N}(n), \ldots, \ell_{2n-1}^{N}(n) \right) \leftarrow   \left(0,\ldots, 0 \right)$
\item initialise block sizes $\set{\set{b_{1}, b_{2}}, \ldots, \set{b_{2n-1}, b_{2n}} }\leftarrow \set{\set{1,1}, \ldots, \set{1,1}} $
\item {\bf while} all block sizes are smaller than  $2n$:
\begin{enumerate}
\item $\ell_{b}^{N}(n) \leftarrow  1 +  \ell_{b}^{N}(n)  $ for current block sizes $b = b_{1}, \ldots, b_{m}$
\item sample numbers of potential offspring $X_{1},\ldots, X_{N}$
\item given $X_{1},\ldots, X_{N}$ assign marked diploid  individuals to  families 
\item merge blocks and record the  new block sizes and configuration (pairing in individuals) 
\end{enumerate}
\item given branch lengths $\ell_{i}^{N}(n)$ update  $r_{i}^{N}(n)$ for $i = 1,2,\ldots, 2n-1$
\begin{displaymath}
\mathfrak r_{i}^{N}(n) \leftarrow  \mathfrak r_{i}^{N}(n) +   \frac{ \ell_{i}^{N}(n) }{ \ell_{1}^{N}(n) + \cdots +  \ell_{2n-1}^{N}(n)  }
\end{displaymath}
\end{enumerate}
\item return an approximation   $ \overline \varrho_{i}^{N}(n)  =  (1/M) \mathfrak r _{i}^{N}(n)$  of  $\EE{R_{i}^{N}(n)} $   
\end{enumerate}

\section{Computing $\overline \rho_{i}^{N}(n)$}\label{sec:estimatingERiNA}

\breyta{
In this section we briefly describe the algorithm for computing
$\overline\rho_{i}^{N}(n)$, the  empirical mean of  
quenched relative branch lengths $\widetilde R_{i}^{N}(n)$ (see also
\S~\ref{sec:comp-quench-anne}).}  Let
$\mathds{A}^{(N,n)} = {(A_{i}(g))}_{i\in [4N], g\in \N\cup \{0\}}$
denote a realised ancestry, recording the ancestral relations of the
$4N$ gene copies (chromosomes) in the population.  In each generation
there are $4N$ gene copies, and each chromosome occupies a level.  If
the chromosome on level $\ell$ at time $g$ produces $k$ surviving
copies then $A_{\ell_{1}}(g+1) = \cdots = A_{\ell_{k}}(g+1) = \ell$,
and if $A_{i}(g) = A_{j}(g)$ then the chromosomes on level $i$ resp.\
$j$ at time $g$ share an immediate ancestor. In this way the ancestry
records the ancestral relations of $4N$ chromosomes, where $N$ is the
number of parent pairs producing potential offspring in each
generation.

\begin{figure}[htp]
\centering
\captionsetup[subfloat]{labelfont={scriptsize,sf,md,up},textfont={scriptsize,sf}}
\subfloat[$\zeta(N)=2N, \alpha = 0.5$]{\includegraphics[angle=0,scale=.5]{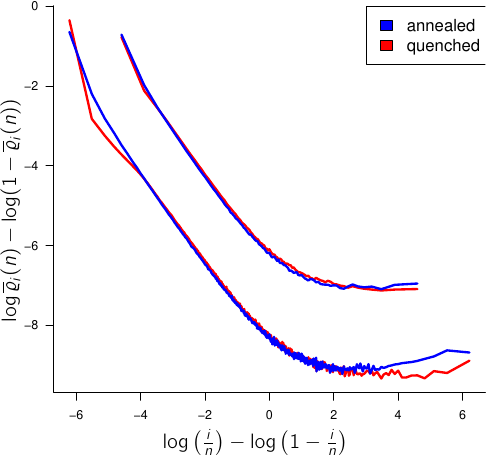}}
\subfloat[$\zeta(N)=4N^{2}, \alpha = 0.5$]{\includegraphics[angle=0,scale=.5]{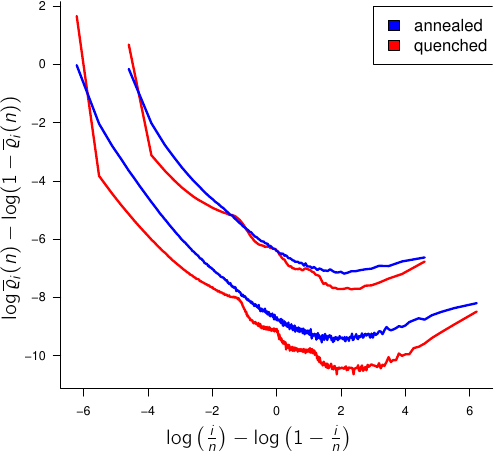}}\\
\subfloat[$\zeta(N)=2N, \alpha = 1.5$]{\includegraphics[angle=0,scale=.5]{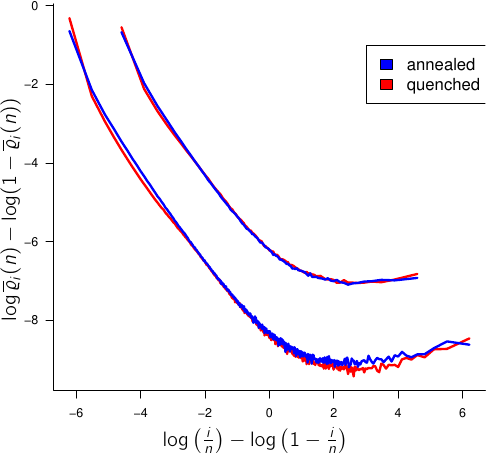}}
\subfloat[$\zeta(N)=4N^{2}, \alpha = 1.5$]{\includegraphics[angle=0,scale=.5]{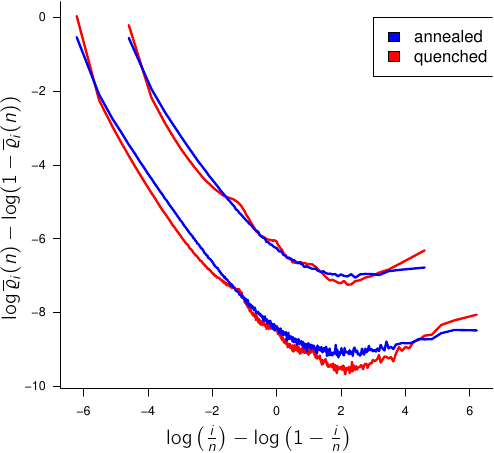}}
\caption{\breyta{Comparing $\overline\varrho_{i}^{N}$ and $\overline
\rho_{i}^{N}$.  Annealed  ($\overline \varrho_{i}(n)$, the empirical
mean of $R_{i}^{N}(n)$, recall~\eqref{eq:33},~\eqref{eq:estimates}
and~\eqref{eq:approxestimates}, blue lines)  and quenched
($\overline\rho_{i}^{N}(n)$, the empirical mean of 
 $\widetilde R_{i}^{N}(n)$, red lines) means}    when the
population evolves according to Definitions~\ref{dschwpop} and
~\ref{dfn:alpharandomall}, and~\eqref{eq:32},    for $N=125$
(population size $2N$ of diploid individuals), $\alpha$ and  upper bound $\zeta(N)$
as shown, $\varepsilon_{N} = 0.1$,  $\kappa = 2$; sample
size $n=50$ and $n=250$ diploid individuals; results from $10^{5}$
experiments }\label{fig:qaC}
\end{figure}

The algorithm here is different from the one for approximating 
$\EE{R_{i}^{N}(n)}$ described in Appendix~\ref{sec:estimatingERiN} in
that here we generate gene genealogies by recording ancestral
relations.   In each experiment the population evolves forward in time,
the ancestral relations are recorded and stored in $\mathds{A}^{(N,n)}$, and
every now and then we randomly sample $n$ diploid individuals ($2n$
gene copies);  this process is repeated until the sampled gene copies
are found to have common ancestor; the gene tree of the sampled gene
copies is then fixed and the branch lengths are read off the fixed
tree (a sample whose gene copies are without a common ancestor and so
with an incomplete gene tree is discarded).  This process is repeated
a given number of times, each time starting from scratch with a new
population. Averaging over population ancestries gives $\overline
\rho_{i}^{N}(n)$ (see Figures~\ref{fig:annealedA} and~\ref{fig:qaC}
for examples).

\begin{enumerate}
\item $\left( \mathfrak r_{i}^{N}(n), \ldots, \mathfrak r_{i}^{N}(n)  \right) \leftarrow (0,\ldots, 0)$
\item for each of $M$ experiments: 
\begin{enumerate}
\item initialise the ancestry $A_{\ell}(0) = \ell$ for $\ell = 1,2, \ldots, 4N$
\item $\left(  \ell_{1}^{N}(n), \ldots,  \ell_{2n-1}^{N}(n) \right) \leftarrow  (0,\ldots,0)$
\item {\bf until} a complete  sample tree is found:
\begin{enumerate}
\item add to $\mathds A^{(N,n)}$:
\begin{enumerate}
\item  sample  numbers of potential offspring $X_{1}, \ldots, X_{N}$
\item assign chromosomes to each potential offspring; the chromosomes
of family $i$ will occupy levels $4i$, $4i+1$, $4i+2$, $4i+3$
\item sample $2N$  surviving offspring
\end{enumerate}
\item sample uniformly at random and without replacement $2n$ levels 
\item check if the sample tree is complete 
\end{enumerate}
\item read the branch lengths $\ell_{i}^{N}(n)$   off the fixed complete  tree
\item update $\mathfrak r_{i}^{N}(n)$ for $i = 1,2,\ldots, 2n-1$
\begin{displaymath}
\mathfrak r_{i}^{N}(n) \leftarrow  \mathfrak r_{i}^{N}(n) +   \frac{  \ell_{i}^{N}(n)  }{  \ell_{1}^{N}(n)  + \cdots +   \ell_{2n-1}^{N}(n)     }
\end{displaymath}
\end{enumerate}
\item return an approximation  $\overline \rho_{i}^{N}\left(n \right) =
(1/M)  \mathfrak r_{i}^{N}(n) $  of $\EE{\widetilde R_{i}^{N}(n) }$
\end{enumerate}
Denote the $M$ independently generated ancestries by
$\mathds A_{1}^{(N,n)}, \ldots, \mathds A_{M}^{(N,n)}$ and write
$ \mathfrak r_{i}^{N}\left(n, \mathds A_{j}^{(N,n)} \right)$ for
relative branch lengths read off a  complete tree as  recorded in
$ \mathds A_{j}^{(N,n)}$.  Then for $i = 1,2, \ldots, 2n-1$ 
 \begin{displaymath}
\overline \rho_{i}^{N} \left(n \right)  = (1/M)\sum_{j=1}^{M} \mathfrak r_{i}^{N}\left(n, \mathds A_{j}^{(N,n)} \right)  
\end{displaymath}
In Figures~\ref{fig:qaC} and~\ref{fig:annealedA}  are examples
comparing $\overline \varrho_{i}^{N}(n)$ (see
\S~\ref{sec:estimatingERiN})  and $\overline\rho_{i}^{N}(n)$; the
graphs show the 
same pattern.

\bibliographystyle{plainnat}%
\bibliography{refs}%

@inbook{Freund2021,
  title = {Multiple-merger genealogies: Models,  consequences,  inference},
  ISBN = {9783985475056},
  ISSN = {2523-5168},
  url = {http://dx.doi.org/10.4171/ECR/17-1/9},
  DOI = {10.4171/ecr/17-1/9},
  booktitle = {Probabilistic Structures in Evolution},
  publisher = {EMS Press},
  author = {Freund,  Fabian},
  year = {2021},
  month = May,
  pages = {179–202}
}

@article{Milholland2017,
  title = {Differences between germline and somatic mutation rates in humans and mice},
  volume = {8},
  ISSN = {2041-1723},
  url = {http://dx.doi.org/10.1038/ncomms15183},
  DOI = {10.1038/ncomms15183},
  number = {1},
  journal = {Nature Communications},
  publisher = {Springer Science and Business Media LLC},
  author = {Milholland,  Brandon and Dong,  Xiao and Zhang,  Lei and Hao,  Xiaoxiao and Suh,  Yousin and Vijg,  Jan},
  year = {2017},
  month = May 
}

@article{Xue2009,
  title = {Human {Y} Chromosome Base-Substitution Mutation Rate Measured by Direct Sequencing in a Deep-Rooting Pedigree},
  volume = {19},
  ISSN = {0960-9822},
  url = {http://dx.doi.org/10.1016/j.cub.2009.07.032},
  DOI = {10.1016/j.cub.2009.07.032},
  number = {17},
  journal = {Current Biology},
  publisher = {Elsevier BV},
  author = {Xue,  Yali and Wang,  Qiuju and Long,  Quan and Ng,  Bee Ling and Swerdlow,  Harold and Burton,  John and Skuce,  Carl and Taylor,  Ruth and Abdellah,  Zahra and Zhao,  Yali and MacArthur,  Daniel G. and Quail,  Michael A. and Carter,  Nigel P. and Yang,  Huanming and Tyler-Smith,  Chris},
  year = {2009},
  month = Sept,
  pages = {1453–1457}
}

@article{Wooldridge2025,
  title = {Direct Measurement of the Mutation Rate and Its Evolutionary Consequences in a Critically Endangered Mollusk},
  volume = {42},
  ISSN = {1537-1719},
  url = {http://dx.doi.org/10.1093/molbev/msae266},
  DOI = {10.1093/molbev/msae266},
  number = {1},
  journal = {Molecular Biology and Evolution},
  publisher = {Oxford University Press (OUP)},
  author = {Wooldridge,  T Brock and Ford,  Sarah M and Conwell,  Holland C and Hyde,  John and Harris,  Kelley and Shapiro,  Beth},
  editor = {Agashe,  Deepa},
  year = {2025},
  month = Jan 
}

@article{Seplyarskiy2023,
  title = {A mutation rate model at the basepair resolution identifies the mutagenic effect of polymerase {III} transcription},
  volume = {55},
  ISSN = {1546-1718},
  url = {http://dx.doi.org/10.1038/s41588-023-01562-0},
  DOI = {10.1038/s41588-023-01562-0},
  number = {12},
  journal = {Nature Genetics},
  publisher = {Springer Science and Business Media LLC},
  author = {Seplyarskiy,  Vladimir and Koch,  Evan M. and Lee,  Daniel J. and Lichtman,  Joshua S. and Luan,  Harding H. and Sunyaev,  Shamil R.},
  year = {2023},
  month = Nov,
  pages = {2235–2242}
}

@article{Wang2026,
  title = {How precise are mutation rate estimates? Comparison of different approaches to estimate de novo mutation rates},
  volume = {135},
  ISSN = {1365-2540},
  url = {http://dx.doi.org/10.1038/s41437-026-00852-7},
  DOI = {10.1038/s41437-026-00852-7},
  number = {6},
  journal = {Heredity},
  publisher = {Springer Science and Business Media LLC},
  author = {Wang,  Xi and Zhang,  Chaowei and Wang,  Hongbo and Reid,  Kerry and Meril\"{a},  Juha},
  year = {2026},
  month = June,
  pages = {445–452}
}

@article{Kersting2014,
  title = {The evolving beta coalescent},
  volume = {19},
  ISSN = {1083-6489},
  url = {http://dx.doi.org/10.1214/EJP.v19-3332},
  DOI = {10.1214/ejp.v19-3332},
  number = {none},
  journal = {Electronic Journal of Probability},
  publisher = {Institute of Mathematical Statistics},
  author = {Kersting,  G\"{o}tz and Schweinsberg,  Jason and Wakolbinger,  Anton},
  year = {2014},
  month = Jan 
}

@article{Eldonhaploid2026,
  title = {Gene Genealogies in Haploid Populations Evolving According to Sweepstakes Reproduction},
  volume = {88},
  ISSN = {1522-9602},
  url = {http://dx.doi.org/10.1007/s11538-026-01736-x},
  DOI = {10.1007/s11538-026-01736-x},
  number = {9},
  journal = {Bulletin of Mathematical Biology},
  publisher = {Springer Science and Business Media LLC},
  author = {Eldon,  Bjarki},
  year = {2026},
  month = Aug 
}

@article{Matuszewski2018,
  title = {Coalescent Processes with Skewed Offspring Distributions and Nonequilibrium Demography},
  volume = {208},
  ISSN = {1943-2631},
  url = {http://dx.doi.org/10.1534/genetics.117.300499},
  DOI = {10.1534/genetics.117.300499},
  number = {1},
  journal = {Genetics},
  publisher = {Oxford University Press (OUP)},
  author = {Matuszewski,  Sebastian and Hildebrandt,  Marcel E and Achaz,  Guillaume and Jensen,  Jeffrey D},
  year = {2018},
  month = Jan,
  pages = {323–338}
}

@article{Eldon2026,
  title = {Beta-coalescents when sample size is large},
  url = {http://dx.doi.org/10.64898/2025.12.30.697022},
  DOI = {10.64898/2025.12.30.697022},
  journal={bioRxiv},
  publisher = {openRxiv},
  author = {Chetwynd-Diggle,  Jonathan A and Eldon, Bjarki},
  year = {2026},
  month = jan 
}

@article{Donnelly1995,
  title = {COALESCENTS AND GENEALOGICAL STRUCTURE UNDER NEUTRALITY},
  volume = {29},
  ISSN = {1545-2948},
  url = {http://dx.doi.org/10.1146/annurev.ge.29.120195.002153},
  DOI = {10.1146/annurev.ge.29.120195.002153},
  number = {1},
  journal = {Annual Review of Genetics},
  publisher = {Annual Reviews},
  author = {Donnelly,  Peter and Tavar{\'e},  Simon},
  year = {1995},
  month = dec,
  pages = {401–421}
}

@Article{ethier93:_flemin_viot,
  author = 	 {S Ethier and T Kurtz},
  title = 	 {{Fleming-Viot} processes in population genetics},
  journal = 	 {SIAM J Control Optim},
  year = 	 1993,
  volume = 	 31,
  doi={https://doi.org/10.1137/0331019},
  pages = 	 {345--86}}

@Article{fleming79:_some_markov,
  author = 	 {WH Fleming and M  Viot},
  title = 	 {Some measure-valued {Markov} processes in population genetics theory},
  journal = 	 {Indiana University Mathematics Journal},
  year = 	 1979,
  volume = 	 28,
  url = {https://www.jstor.org/stable/24892583},
  pages = 	 {817--43}}

@article{Hoban2013,
  title = {High variance in reproductive success generates a false signature of a genetic bottleneck in populations of constant size: a simulation study},
  volume = {14},
  ISSN = {1471-2105},
  url = {http://dx.doi.org/10.1186/1471-2105-14-309},
  DOI = {10.1186/1471-2105-14-309},
  number = {1},
  journal = {BMC Bioinformatics},
  publisher = {Springer Science and Business Media LLC},
  author = {Hoban,  Sean M and Mezzavilla,  Massimo and Gaggiotti,  Oscar E and Benazzo,  Andrea and van Oosterhout,  Cock and Bertorelle,  Giorgio},
  year = {2013},
  month = Oct 
}

@article{Eldon2016,
  title = {Current hypotheses to explain genetic chaos under the sea},
  volume = {62},
  ISSN = {2396-9814},
  url = {http://dx.doi.org/10.1093/cz/zow094},
  DOI = {10.1093/cz/zow094},
  number = {6},
  journal = {Current Zoology},
  publisher = {Oxford University Press (OUP)},
  author = {Eldon,  Bjarki and Riquet,  Florentine and Yearsley,  Jon and Jollivet,  Didier and Broquet,  Thomas},
  year = {2016},
  month = Sept,
  pages = {551–566}
}

@Book{ethier05:_markov,
  author = 	 {SN Ethier and TG Kurtz},
  title = 	 {Markov processes: characterization and convergence},
  publisher = 	 {Wiley},
  year = 	 2005,
  doi = {https://doi.org/10.1002/9780470316658},
  address = 	 {Hoboken, New Jersey}}

@article{Gnedin2014,
  doi = {10.1239/jap/1417528464},
  year = {2014},
  month = dec,
  publisher = {Cambridge University Press ({CUP})},
  volume = {51},
  number = {A},
  pages = {23--40},
  author = {Alexander Gnedin and Alexander Iksanov and Alexander Marynych},
  title = {{$\Lambda$}-coalescents: a survey},
  journal = {Journal of Applied Probability}
}

@article{DAHMER2014,
  doi = {10.1017/s0963548314000297},
  url = {https://doi.org/10.1017/s0963548314000297},
  year = {2014},
  month = jul,
  publisher = {Cambridge University Press ({CUP})},
  volume = {23},
  number = {6},
  pages = {1010--1027},
  author = {Iulia Dahmer and G\"{o}tz Kersting and Anton Wakolbinger},
  title = {The Total External Branch Length of Beta-Coalescents},
  journal = {Combinatorics,  Probability and Computing}
}

@article{freund2020cannings,
  title={Cannings models, population size changes and multiple-merger coalescents},
  author={Freund, Fabian},
  journal={Journal of mathematical biology},
  volume={80},
  number={5},
  pages={1497--1521},
  year={2020},
  doi={https://doi.org/10.1007/s00285-020-01470-5},
  publisher={Springer}
}

@article{Skorokhod1956,
  title = {Limit Theorems for Stochastic Processes},
  volume = {1},
  ISSN = {1095-7219},
  url = {https://doi.org/10.1137/1101022},
  DOI = {10.1137/1101022},
  number = {3},
  journal = {Theory of Probability \& Its Applications},
  publisher = {Society for Industrial \& Applied Mathematics (SIAM)},
  author = {Skorokhod,  A. V.},
  year = {1956},
  month = jan,
  pages = {261–290}
}

@Book{athreya06:_measur,
  author = 	 {KB Athreya and SN Lahiri},
  title = 	 {Measure theory and probability theory},
  publisher = 	 {Springer},
  doi = {https://doi.org/10.1007/978-0-387-35434-7},
  year = 	 2006}

@article{Diamantidis2024,
  title = {Bursts of coalescence within population pedigrees whenever big families occur},
  volume = {227},
  ISSN = {1943-2631},
  url = {http://dx.doi.org/10.1093/genetics/iyae030},
  DOI = {10.1093/genetics/iyae030},
  number = {1},
  journal = {GENETICS},
  publisher = {Oxford University Press (OUP)},
  author = {Diamantidis,  Dimitrios and Fan,  Wai-Tong (Louis) and Birkner,  Matthias and Wakeley,  John},
  editor = {Jain,  K},
  year = {2024},
  month = feb 
}

@article{PANOV2017379,
title = {Limit theorems for sums of random variables with mixture distribution},
journal = {Statistics \& Probability Letters},
volume = {129},
pages = {379-386},
year = {2017},
issn = {0167-7152},
doi = {https://doi.org/10.1016/j.spl.2017.06.017},
url = {https://www.sciencedirect.com/science/article/pii/S0167715217302213},
author = {Vladimir Panov}
}

@article{10.1214/aoms/1177700291,
author = {Bengt von Bahr and Carl-Gustav Esseen},
title = {{Inequalities for the $r$th Absolute Moment of a Sum of Random Variables, $1 \leqq r \leqq 2$}},
volume = {36},
journal = {The Annals of Mathematical Statistics},
number = {1},
publisher = {Institute of Mathematical Statistics},
pages = {299 -- 303},
year = {1965},
doi = {10.1214/aoms/1177700291},
URL = {https://doi.org/10.1214/aoms/1177700291}
}

@article{freund2023interpreting,
  doi = {10.1371/journal.pgen.1010677},
  url = {https://doi.org/10.1371/journal.pgen.1010677},
  year = {2023},
  month = mar,
  publisher = {Public Library of Science ({PLoS})},
  volume = {19},
  number = {3},
  pages = {e1010677},
  author = {Fabian Freund and Elise Kerdoncuff and Sebastian
                  Matuszewski and Marguerite Lapierre and Marcel
                  Hildebrandt and Jeffrey D. Jensen and Luca Ferretti
                  and Amaury Lambert and Timothy B. Sackton and
                  Guillaume Achaz},
  editor = {Lindi Wahl},
  title = {Interpreting the pervasive observation of {U-shaped} Site Frequency Spectra},
  journal = {{PLOS} Genetics}
}

@book{Bertoin2006,
  title = {Random Fragmentation and Coagulation Processes},
  ISBN = {9780511617768},
  url = {http://dx.doi.org/10.1017/CBO9780511617768},
  DOI = {10.1017/cbo9780511617768},
  publisher = {Cambridge University Press},
  author = {Bertoin,  Jean},
  year = {2006},
  month = Aug 
}

@article{Kingman1975,
  doi = {10.1111/j.2517-6161.1975.tb01024.x},
  url = {https://doi.org/10.1111/j.2517-6161.1975.tb01024.x},
  year = {1975},
  month = sep,
  publisher = {Wiley},
  volume = {37},
  number = {1},
  pages = {1--15},
  author = {J. F. C. Kingman},
  title = {Random Discrete Distributions},
  journal = {Journal of the Royal Statistical Society: Series B (Methodological)}
}

@article{Landim2015,
  doi = {10.1016/j.spa.2014.08.011},
  year = {2015},
  month = mar,
  publisher = {Elsevier {BV}},
  volume = {125},
  number = {3},
  pages = {1058--1088},
  author = {C. Landim},
  title = {A topology for limits of {Markov} chains},
  journal = {Stochastic Processes and their Applications}
}

@article{Etemadi1981,
  doi = {10.1007/bf01013465},
  year = {1981},
  month = feb,
  publisher = {Springer Science and Business Media {LLC}},
  volume = {55},
  number = {1},
  pages = {119--122},
  author = {N. Etemadi},
  title = {An elementary proof of the strong law of large numbers},
  journal = {Zeitschrift fuer Wahrscheinlichkeitstheorie und Verwandte Gebiete}
}

@article{Koskela2018,
  doi = {10.1515/sagmb-2017-0011},
  year  = {2018},
  month = {jun},
  publisher = {Walter de Gruyter {GmbH}},
  volume = {17},
  number = {3},
  author = {Jere Koskela},
  title = {Multi-locus data distinguishes between population growth and multiple merger coalescents},
  journal = {Statistical Applications in Genetics and Molecular Biology}
}

@article{Koskela2019,
  doi = {10.1016/j.mbs.2019.03.004},
  year = {2019},
  month = may,
  publisher = {Elsevier {BV}},
  volume = {311},
  pages = {1--12},
  author = {Jere Koskela and Maite Wilke Berenguer},
  title = {Robust model selection between population growth and multiple merger coalescents},
  journal = {Mathematical Biosciences}
}

@article{BLS15,
  doi = {10.1214/18-ejp175},
  year  = {2018},
  publisher = {Institute of Mathematical Statistics},
  volume = {23},
  number = {0},
  author = {Matthias Birkner and Huili Liu and Anja Sturm},
  title = {Coalescent results for diploid exchangeable population models},
  journal = {Electronic Journal of Probability}
}

@article{DS05,
  doi = {10.1016/j.spa.2005.04.009},
  year = {2005},
  month = oct,
  publisher = {Elsevier {BV}},
  volume = {115},
  number = {10},
  pages = {1628--1657},
  author = {Rick Durrett and Jason Schweinsberg},
  title = {A coalescent model for the effect of advantageous mutations on the genealogy of a population},
  journal = {Stochastic Processes and their Applications}
}

@article{Eldon2020,
  doi = {10.1146/annurev-genet-021920-095932},
  url = {https://doi.org/10.1146/annurev-genet-021920-095932},
  year = {2020},
  month = nov,
  publisher = {Annual Reviews},
  volume = {54},
  number = {1},
  pages = {213--236},
  author = {Bjarki Eldon},
  title = {Evolutionary Genomics of High Fecundity},
  journal = {Annual Review of Genetics}
}

@article {Arnasonsweepstakes2022,
article_type = {journal},
title = {Sweepstakes reproductive success via pervasive and recurrent selective sweeps},
author = {\'Arnason, Einar and Koskela, Jere and Halld\'orsd\'ottir, Katr\'in and Eldon, Bjarki},
editor = {Gagnaire, Pierre-Alexandre and Przeworski, Molly and Freund, Fabian and Gagnaire, Pierre-Alexandre},
volume = 12,
year = 2023,
month = {feb},
pub_date = {2023-02-20},
pages = {e80781},
citation = {eLife 2023;12:e80781},
doi = {10.7554/eLife.80781},
url = {https://doi.org/10.7554/eLife.80781},
journal = {eLife},
issn = {2050-084X},
publisher = {eLife Sciences Publications, Ltd},
}

@article{Eldon2023mec,
  doi = {10.1111/mec.16903},
  url = {https://doi.org/10.1111/mec.16903},
  year = {2023},
  month = mar,
  publisher = {Wiley},
  author = {Bjarki Eldon and Wolfgang Stephan},
  title = {Sweepstakes reproduction facilitates rapid adaptation in highly fecund populations},
  journal = {Molecular Ecology}
}

@article{K2000,
	author = {JFC Kingman},
	journal = {Genetics},
	pages = {1461--1463},
	title = {Origins of the coalescent: 1974--1982},
	volume = 156,
	doi={https://doi.org/10.1093/genetics/156.4.1461},
	year = 2000
}

@article{HM11b,
	author = {T Huillet and M M\"{o}hle},
	journal = {Stoch Models},
	pages = {521--554},
	title = {Population genetics models with skewed fertilities: forward and backward analysis},
	volume = 27,
	doi={https://doi.org/10.1080/15326349.2011.593411},
	year = 2011
}

@book{feng2010poisson,
  title={The {Poisson-Dirichlet} distribution and related topics: models and asymptotic behaviors},
  author={Feng, Shui},
  year={2010},
  doi={10.1007/978-3-642-11194-5},
  publisher={Springer Science \& Business Media}
}

@article{M98,
	author = {Martin M{\"{o}}hle},
	journal = {Adv Appl Prob},
	pages = {493--512},
	title = {A convergence theorem for {Markov} chains arising in population genetics and the coalescent with selfing},
	volume = 30,
	doi={https://doi.org/10.1239/aap/1035228080},
	year = 1998
}

@article{JFB:JFB13143,
author = {Waples, R. S.},
title = {Tiny estimates of the ${N_e}/{N}$ ratio in marine fishes: Are they real?},
journal = {Journal of Fish Biology},
volume = {89},
number = {6},
publisher = {Blackwell Publishing Ltd},
issn = {1095-8649},
doi = {10.1111/jfb.13143},
pages = {2479--2504},
year = {2016}
}

@article{schweinsberg03,
	author = {J Schweinsberg},
	journal = {Stoch Proc Appl},
	pages = {107--139},
	title = {Coalescent processes obtained from supercritical {G}alton-{W}atson processes},
	volume = 106,
	doi={10.1016/S0304-4149(03)00028-0},
	year = 2003
}

@article{S00,
	author = {J Schweinsberg},
	journal = {Electron J Probab},
	pages = {1--50},
	title = {Coalescents with simultaneous multiple collisions},
	volume = 5,
	doi={10.1214/EJP.v5-68},
	year = 2000
}

@article{MS03,
	author = {M M{\"{o}}hle and S Sagitov},
	journal = {J Math Biol},
	pages = {337--352},
	title = {Coalescent patterns in diploid exchangeable population models},
	volume = 47,
	doi={10.1007/s00285-003-0218-6},
	year = 2003
}

@article{S03,
	author = {S Sagitov},
	journal = {J Appl Probab},
	pages = {839--854},
	title = {Convergence to the coalescent with simultaneous multiple mergers},
	volume = 40,
	doi={10.1239/jap/1067436085},
	year = 2003
}

@article{Der2012,
  doi = {10.1534/genetics.112.140038},
  year  = {2012},
  month = {jun},
  publisher = {Genetics Society of America},
  volume = {191},
  number = {4},
  pages = {1331--1344},
  author = {R. Der and C. Epstein and J. B. Plotkin},
  title = {Dynamics of Neutral and Selected Alleles When the Offspring Distribution Is Skewed},
  journal = {Genetics}
}

@article{Der2014,
	author = {R. Der and J. B. Plotkin},
	doi = {10.1534/genetics.114.161422},
	journal = {Genetics},
	month = {jan},
	number = {4},
	pages = {1199--1216},
	publisher = {Genetics Society of America},
	title = {The Equilibrium Allele Frequency Distribution for a Population with Reproductive Skew},
	url = {http://dx.doi.org/10.1534/genetics.114.161422},
	volume = {196},
	year = {2014}
}

@article{Li1998,
	author = {Gang Li and Dennis Hedgecock},
	doi = {10.1139/f97-312},
	journal = {Can. J. Fish. Aquat. Sci.},
	month = {apr},
	number = {4},
	pages = {1025--1033},
	publisher = {Canadian Science Publishing},
	title = {Genetic heterogeneity, detected by {PCR}-{SSCP}, among samples of larval {P}acific oysters ( \emph{Crassostrea gigas} ) supports the hypothesis of large variance in reproductive success },
	volume = {55},
	year = {1998}
}

@article{K82,
title = {The coalescent},
journal = {Stochastic Processes and their Applications},
volume = {13},
number = {3},
pages = {235-248},
year = {1982},
issn = {0304-4149},
doi = {https://doi.org/10.1016/0304-4149(82)90011-4},
url = {https://www.sciencedirect.com/science/article/pii/0304414982900114},
author = {J.F.C. Kingman}
}

@article{K82b,
	author = {J F C Kingman},
	journal = {J App Probab},
	pages = {27--43},
	title = {On the genealogy of large populations},
	volume = {{19A}},
	doi={10.2307/3213548},
	year = 1982
}

@inproceedings{K82c,
	address = {Amsterdam},
	author = {J F C Kingman},
	booktitle = {Exchangeability in Probability and Statistics},
	editor = {G Koch and F Spizzichino},
	pages = {97--112},
	publisher = {North-Holland},
	title = {Exchangeability and the evolution of large populations},
	isbn = {0444864032},
	year = 1982
}

@article{K1978,
	author = {JFC Kingman},
	journal = {J London Math Soc},
	pages = {374--380},
	title = {The representation of partition structures},
	volume = 18,
	doi={https://doi.org/10.1112/jlms/s2-18.2.374},
	year = 1978
}

@article{H83b,
	author = {R R Hudson},
	journal = {Theor Popul Biol},
	pages = {183--201},
	title = {Properties of a neutral allele model with intragenic recombination},
	volume = 23,
	doi={https://doi.org/10.1016/0040-5809(83)90013-8},
	year = 1983
}

@article{T83,
	author = {F Tajima},
	journal = {Genetics},
	pages = {437--460},
	title = {Evolutionary relationships of {DNA} sequences in finite populations},
	volume = 105,
	doi={https://doi.org/10.1093/genetics/105.2.437},
	year = 1983
}

@article{Birkner2024,
  title = {The joint fluctuations of the lengths of the {Beta}({2 -  $\alpha$}, {$\alpha$})-coalescents},
  volume = {34},
  ISSN = {1050-5164},
  url = {http://dx.doi.org/10.1214/23-AAP1964},
  DOI = {10.1214/23-aap1964},
  number = {1A},
  journal = {The Annals of Applied Probability},
  publisher = {Institute of Mathematical Statistics},
  author = {Birkner,  Matthias and Dahmer,  Iulia and Diehl,  Christina S. and Kersting,  G\"{o}tz},
  year = {2024},
  month = feb 
}

@inproceedings{H94,
	address = {London},
	author = {D Hedgecock},
	booktitle = {Genetics and evolution of Aquatic Organisms},
	editor = {A Beaumont},
	pages = {1222--1344},
	publisher = {Chapman and Hall},
	title = {Does variance in reproductive success limit effective population sizes of marine organisms?},
	isbn = {978-0-412-49370-6},
	year = 1994
}

@article{Blath2016,
	author = {Jochen Blath and Mathias Christensen Cronj\"{a}ger and Bjarki Eldon and Matthias Hammer},
	doi = {10.1016/j.tpb.2016.04.002},
	journal = {Theoretical Population Biology},
	month = {aug},
	pages = {36--50},
	publisher = {Elsevier {BV}},
	title = {The site-frequency spectrum associated with {$\Xi$}-coalescents},
	url = {http://dx.doi.org/10.1016/j.tpb.2016.04.002},
	volume = {110},
	year = {2016}
}

@article{HP11,
	author = {D Hedgecock and A I Pudovkin},
	journal = {Bull Marine Science},
	pages = {971--1002},
	title = {Sweepstakes reproductive success in highly fecund marine fish and shellfish: a review and commentary},
	volume = 87,
	doi={10.5343/bms.2010.1051},
	year = 2011
}

@incollection{H82,
	address = {New York},
	author = {D Hedgecock and M Tracey and K Nelson},
	booktitle = {The {B}iology of {C}rustacea: {E}mbryology, {M}orphology, and {G}enetics},
	editor = {L G Abele},
	pages = {297--403},
	publisher = {Academic Press},
	title = {Genetics},
	volume = 2,
	isbn={0121064050},
	year = 1982
}

@incollection{B94,
	address = {New York},
	author = {A T Beckenbach},
	booktitle = {Non-neutral {E}volution},
	editor = {B Golding},
	pages = {188--198},
	publisher = {Chapman \& Hall},
	title = {Mitochondrial haplotype frequencies in oysters: neutral alternatives to selection models},
	doi={10.1007/978-1-4615-2383-3},
	year = 1994
}

@article{Vendrami2021,
  title = {Sweepstake reproductive success and collective dispersal produce chaotic genetic patchiness in a broadcast spawner},
  volume = {7},
  ISSN = {2375-2548},
  url = {http://dx.doi.org/10.1126/sciadv.abj4713},
  DOI = {10.1126/sciadv.abj4713},
  number = {37},
  journal = {Science Advances},
  publisher = {American Association for the Advancement of Science (AAAS)},
  author = {Vendrami,  David L. J. and Peck,  Lloyd S. and Clark,  Melody S. and Eldon,  Bjarki and Meredith,  Michael and Hoffman,  Joseph I.},
  year = {2021},
  month = sep 
}

@article{AH2015,
	author = {Einar \'{A}rnason and Katr\'{\i}n Halld\'{o}rsd\'{o}ttir},
	doi = {10.7717/peerj.786},
	journal = {{PeerJ}},
	pages = {e786},
	publisher = {{PeerJ}},
	title = {Nucleotide variation and balancing selection at the \emph{Ckma} gene in {A}tlantic cod: analysis with multiple merger coalescent models },
	volume = {3},
	year = {2015}
}

@article{A04,
	author = {E \'{A}rnason},
	journal = {Genetics},
	pages = {1871--1885},
	title = {Mitochondrial cytochrome \emph{b} variation in the high-fecundity {A}tlantic cod: trans-{A}tlantic clines and shallow gene genealogy},
	doi={https://doi.org/10.1534/genetics.166.4.1871},
	volume = 166,
	year = 2004
}

@article{EW06,
	author = {B Eldon and J Wakeley},
	journal = {Genetics},
	pages = {2621--2633},
	title = {Coalescent processes when the distribution of offspring number among individuals is highly skewed},
	volume = 172,
	doi={https://doi.org/10.1534/genetics.105.052175},
	year = 2006
}

@book{W07,
  author = {John Wakeley},
  title = {Coalescent Theory: An Introduction},
  year = {2009},
  publisher = {Roberts \& Company Publishers},
  address = {Greenwood Village},
  isbn = {0-9747077-5-9},
  language = {English},
}

@article{SW08,
	author = {O Sargsyan and J Wakeley},
	journal = {Theor Pop Biol},
	pages = {104--114},
	title = {A coalescent process with simultaneous multiple mergers for approximating the gene genealogies of many marine organisms},
	volume = 74,
	doi={10.1016/j.tpb.2008.04.009},
	year = 2008
}

@article{HM12,
	author = {T Huillet and M M\"{o}hle},
	journal = {Theor Popul Biol},
	pages = {5--14},
	title = {On the extended {M}oran model and its relation to coalescents with multiple collisions},
	volume = 87,
	doi={https://doi.org/10.1016/j.tpb.2011.09.004},
	year = 2013
}

@incollection{BB09,
	author = {M Birkner and J Blath},
	booktitle = {Trends in stochastic analysis},
	editor = {J Blath and P M\"{o}rters and M Scheutzow},
	pages = {329--363},
	publisher = {Cambridge University Press},
	title = {Measure-valued diffusions, general coalescents and population genetic inference},
	doi={https://doi.org/10.1017/CBO9781139107020},
	year = 2009
}

@article{SKS2016,
	author = {Jeffrey P. Spence and John A. Kamm and Yun S. Song},
	doi = {10.1534/genetics.115.184101},
	eprint = {http://www.genetics.org/content/202/4/1549.full.pdf},
	issn = {0016-6731},
	journal = {Genetics},
	number = {4},
	pages = {1549--1561},
	publisher = {Genetics},
	title = {The Site Frequency Spectrum for General Coalescents},
	url = {http://www.genetics.org/content/202/4/1549},
	volume = {202},
	year = {2016}
}

@article{BBE13,
	author = {M Birkner and J Blath and B Eldon},
	journal = {Genetics},
	pages = {255--290},
	title = {An ancestral recombination graph for diploid populations with skewed offspring distribution},
	volume = 193,
	doi={https://doi.org/10.1534/genetics.112.144329},
	year = 2013
}

@article{BBE2013a,
	author = {M Birkner and J Blath and B Eldon},
	journal = {Genetics},
	pages = {1037--1053},
	title = {Statistical properties of the site-frequency spectrum associated with {$\Lambda$}-coalescents},
	volume = 195,
	doi={https://doi.org/10.1534/genetics.113.156612},
	year = 2013
}

@article{BBMST09,
	author = {M Birkner and J Blath and M M\"{o}hle and M Steinr\"{u}cken and J Tams},
	journal = {ALEA Lat. Am. J. Probab. Math. Stat.},
	pages = {25--61},
	title = {A modified lookdown construction for the {X}i-{F}leming-{V}iot process with mutation and populations with recurrent bottlenecks},
	volume = 6,
	doi={https://doi.org/10.34657/1856},
	year = 2009
}

@article{Johri2022,
  title = {Recommendations for improving statistical inference in population genomics},
  volume = {20},
  ISSN = {1545-7885},
  url = {http://dx.doi.org/10.1371/journal.pbio.3001669},
  DOI = {10.1371/journal.pbio.3001669},
  number = {5},
  journal = {PLOS Biology},
  publisher = {Public Library of Science (PLoS)},
  author = {Johri,  Parul and Aquadro,  Charles F. and Beaumont,  Mark and Charlesworth,  Brian and Excoffier,  Laurent and Eyre-Walker,  Adam and Keightley,  Peter D. and Lynch,  Michael and McVean,  Gil and Payseur,  Bret A. and Pfeifer,  Susanne P. and Stephan,  Wolfgang and Jensen,  Jeffrey D.},
  year = {2022},
  month = May,
  pages = {e3001669}
}

@article{Fu2006,
	author = {{Y-X} Fu},
	journal = {Theor Popul Biol},
	pages = {385--394},
	title = {Exact coalescent for the {Wright-Fisher} model},
	volume = 69,
	doi={https://doi.org/10.1016/j.tpb.2005.11.005},
	year = 2006
}

@article{SD2005,
	author = {J Schweinsberg and R Durrett},
	journal = {Ann Appl Probab},
	title = {Random partitions approximating the coalescence of lineages during a selective sweep},
	volume = {1591--1651},
	doi={10.1214/105051605000000430},
	year = 2005
}

@article{Hedrick2005,
  doi = {10.1554/05-009},
  year  = {2005},
  publisher = {The Society for the Study of Evolution},
  volume = {59},
  number = {7},
  pages = {1596},
  author = {Philip Hedrick}, 
  title = {Large variance in reproductive success and the   ${N}_e/{N}$ ratio},
  journal = {Evolution}
}

@article{DK99,
	author = {P Donnelly and T G Kurtz},
	journal = {Ann Probab},
	pages = {166--205},
	title = {Particle Representations for Measure-Valued Population Models},
	volume = 27,
	doi={https://doi.org/10.1214/aop/1022677258},
	year = 1999
}

@article{P99,
	author = {J Pitman},
	journal = {Ann Probab},
	pages = {1870--1902},
	title = {Coalescents with multiple collisions},
	volume = 27,
	doi={10.1214/aop/1022874819},
	year = 1999
}

@article{S99,
	author = {S Sagitov},
	journal = {J Appl Probab},
	pages = {1116--1125},
	title = {The general coalescent with asynchronous mergers of ancestral lines},
	volume = 36,
	doi={10.1239/jap/1032374759},
	year = 1999
}

@article{TV09,
	author = {JE Taylor and A V\'{e}ber},
	journal = {Electron J Probab},
	pages = {242--288},
	title = {Coalescent processes in subdivided populations subject to recurrent mass extinctions},
	volume = 14,
	doi={https://doi.org/10.1214/EJP.v14-595},
	year = 2009
}

@article{MS01,
	author = {M M\"{o}hle and S Sagitov},
	journal = {Ann Probab},
	pages = {1547--1562},
	title = {A classification of coalescent processes for haploid exchangeable population models},
	volume = 29,
	doi={https://doi.org/10.1214/aop/1015345761},
	year = 2001
}

@article{BBC05,
	author = {M Birkner and J Blath and M Capaldo and A M Etheridge and M M\"{o}hle and J Schweinsberg and A Wakolbinger},
	journal = {Electron. J. Probab},
	pages = {303--325},
	title = {Alpha-stable branching and {Beta}-coalescents},
	volume = 10,
	doi={https://doi.org/10.1214/EJP.v10-241},
	year = 2005
}

@article{B09,
	author = {N Berestycki},
	journal = {Ensaios Math\'{e}maticos},
	pages = {1--193},
	title = {Recent progress in Coalescent Theory},
	volume = 16,
	doi={10.21711/217504322009/em161},
	year = 2009
}

@article{wright31:_evolut_mendel,
    author = {Wright, Sewall},
    title = {EVOLUTION IN MENDELIAN POPULATIONS},
    journal = {Genetics},
    volume = {16},
    number = {2},
    pages = {97-159},
    year = {1931},
    month = {03},
    issn = {1943-2631},
    doi = {10.1093/genetics/16.2.97},
    url = {https://doi.org/10.1093/genetics/16.2.97},
    eprint = {https://academic.oup.com/genetics/article-pdf/16/2/97/35081059/genetics0097.pdf},
}

@Article{fisher22,
  author = 	 {RA Fisher},
  title = 	 {On the dominance ratio},
  journal = 	 {Proc Royal Society Edinburgh},
  year = 	 1923,
  volume = 	 42,
  doi={10.1017/S0370164600023993},
  pages = 	 {321--341}}

@article{A88,
	author = {J C Avise and R M Ball and J Arnold},
	journal = {Mol Biol Evol},
	pages = {331--344},
	title = {Current versus historical population sizes in vertebrate species with high gene flow: a comparison based on mitochondrial {DNA} lineages and inbreeding theory for neutral mutations},
	volume = 5,
	year = 1988
}

@article{PW90,
	author = {S R Palumbi and A C Wilson},
	journal = {Evolution},
	pages = {403--415},
	title = {Mitochondrial {DNA} diversity in the sea-urchins \emph{{S}trongylocentrotus purpuratus} and \emph{{S}trongylocentrotus droebachiensis}},
	volume = 44,
	year = 1990
}

@article{BBS08,
	author = {J Berestycki and N Berestycki and J Schweinsberg},
	journal = {Ann Inst H Poincar\'{e} Probab Statist},
	pages = {214--238},
	title = {Small-time behavior of {beta} coalescents},
	volume = 44,
	doi={https://doi.org/10.1214/07-AIHP103},
	year = 2008
}

@article{BBS07,
	author = {J Berestycki and N Berestycki and J Schweinsberg},
	journal = {Ann Probab},
	pages = {1835--1887},
	title = {Beta-coalescents and continuous stable random trees},
	doi={https://doi.org/10.1214/009117906000001114},
	volume = 35,
	year = 2007
}

@article{mohle2000total,
  title={Total variation distances and rates of convergence for ancestral coalescent processes in exchangeable population models},
  author={M{\"o}hle, Martin},
  journal={Advances in Applied Probability},
  pages={983--993},
  year={2000},
  doi={10.1017/S0001867800010417},
  publisher={JSTOR}
}
\end{document}